%% file: main.tex
\documentclass[longauth]{aa} 
\usepackage{graphicx}
\usepackage{txfonts}
\usepackage{amsmath,amsfonts,amssymb} 
\usepackage{natbib}
\usepackage{multicol,caption}
\usepackage{lipsum}
\usepackage{booktabs}
\usepackage{threeparttable}
\usepackage{multirow}
\usepackage{here}
\usepackage{lscape}

\bibpunct{(}{)}{;}{a}{}{,}
\usepackage[colorlinks=true,urlcolor=blue,linkcolor=blue,citecolor=blue]{hyperref}

\newcommand{\pyttv}{\texttt{PyTTV}\xspace}
\newcommand{\pytransit}{\texttt{PyTransit}\xspace}
\newcommand{\emcee}{\texttt{emcee}\xspace}
\newcommand{\rebound}{\texttt{REBOUND}\xspace}

\newcommand{\tess}{TESS\xspace}
\newcommand{\maroonx}{MAROON-X\xspace}

\newcommand{\NP}{\mathcal{N}}
\newcommand{\UP}{\mathcal{U}}

\usepackage{threeparttable}
\usepackage{booktabs}
\usepackage{longtable}
\usepackage{ragged2e}

\usepackage{tikz,xcolor,hyperref}

\definecolor{lime}{HTML}{A6CE39}
\DeclareRobustCommand{\orcidicon}{%
	\hspace{-1.5mm}
	\begin{tikzpicture}
	\draw[lime, fill=lime] (0,0) 
	circle [radius=0.16] 
	node[white] {{\fontfamily{qag}\selectfont \tiny ID}};
	\draw[white, fill=white] (-0.0625,0.095) 
	circle [radius=0.007];
	\end{tikzpicture}
	\hspace{-2.5mm}
}
\foreach \x in {A, ..., Z}{%
	\expandafter\xdef\csname orcid\x\endcsname{\noexpand\href{https://orcid.org/\csname orcidauthor\x\endcsname}{\noexpand\orcidicon}}
}

\foreach \x in {a, ..., z}{%
	\expandafter\xdef\csname orcid\x\endcsname{\noexpand\href{https://orcid.org/\csname orcidauthor\x\endcsname}{\noexpand\orcidicon}}
}

\begin{document}

\title{TOI-2015 b: a sub-Neptune in strong gravitational interaction with an outer non-transiting planet}

\include{coauthors_list}

\date{Received/accepted}
\titlerunning{TOI-2015\,b and TOI-2015\,c}\authorrunning{K. Barkaoui et al.}	 

\abstract{
TOI-2015 is a known exoplanetary system around an M4 dwarf star, consisting of a transiting sub-Neptune planet in a 3.35-day orbital period, TOI-2015\,b, accompanied by a non-transiting companion, TOI-2015\,c.
High-precision radial velocity measurements were taken with the \maroonx\ spectrograph, and high-precision photometric data were collected, primarily using the SPECULOOS, MUSCAT, TRAPPIST and LCOGT networks. We collected 63 transit light curves and 49 different transit epochs for TOI-2015\,b.
We re-characterize the target star by combining optical spectra obtained by the \maroonx, Shane/KAST and IRTF/SpeX  spectrographs, Bayesian Model
Averaging (BMA) and 
Spectral Energy Distribution (SED) analysis. 
The TOI-2015 host star is a $K=10.3$\,mag M4-type dwarf  with a sub-solar metallicity of [e/H]$= -0.31 \pm 0.16$, and an effective temperature of $T_{\rm eff} \approx 3200$\,K.
Our photodynamical analysis of the system strongly favors the 5:3 mean motion resonance and in this scenario the planet b (TOI-2015\,b) has an orbital period of $P_b=3.34$~days, a mass of $M_p=9.02^{+0.32}_{-0.36}$\,$M_\oplus$, a radius of  $R_p=3.309^{+0.013}_{-0.011}$\,$R_\oplus$,
resulting in a density  of  $\rho_p = 0.25 \pm 0.01$\,$\rho_\oplus= 1.40 \pm 0.06$~g~cm$^{-3}$, indicative of a Neptune-like composition. Its transits exhibit large ($>1$\,hr) timing variations indicative of an outer perturber in the system. We performed a global analysis of the high-resolution radial-velocity measurements, the photometric data, and the TTVs, and inferred that  TOI-2015 hosts a second planet, TOI-2015\,c, in a non-transiting configuration. Our analysis places it near a 5:3 resonance with an orbital period of $P_c = 5.583$\,days and a mass of $M_p = 8.91^{+0.38}_{-0.40}$\,$M_\oplus$.
The dynamical configuration of TOI-2015\,b and TOI-2015\,c can be used to constrain the system's planetary formation and migration history.
Based on the mass-radius composition models, TOI-2015\,b is a water-rich or rocky planet with a hydrogen-helium envelope.
Moreover, TOI-2015\,b has a high transmission spectroscopic metric (TSM$=149$), making it a favorable target for future transmission spectroscopic observations with \emph{JWST}  to constrain the atmospheric composition of the planet. Such observations would also help to break the degeneracies in theoretical models of the planet's interior structure.}


\maketitle

\section{Introduction}

The Milky Way  is dominated by M dwarf stars \citep{Henry_1994AAS,Henry_2006}, which are intriguing  targets for searching and characterizing small planets. M-dwarf systems offer a unique opportunity to explore their physical properties thanks to their small sizes and low masses. The relatively small size of the host star leads to deep transits, large radial-velocity (RV) and large TTVs signals. This allows us to explore the interior structure and atmospheric composition of the planets, by measuring their sizes and masses (e.g. \citet{Dorn_2017A&A}).

Transit timing variations (TTVs, \citealt{Agol_2005MNRAS,Holman_2005Sci}) can be used to search for additional non-transiting exoplanets and to estimate their physical parameters (orbital parameters and planetary masses) thanks to the planets gravitational perturbation.
The TTV amplitude depends on the mass of the perturbing planet as well as the orbital eccentricities and longitudes of the pericenter of each planet (see, e.g., \citealt{Deck_2016ApJ}).
Transiting planets in a multi-planet system with strong gravitational influences offer us an excellent  opportunities to explore the interior composition of exoplanets. Moreover, the combination of planet sizes measured from transit depths and masses measured from TTVs and RVs yields densities and clues to the interior composition of planets. Near-resonant multi-planet systems, which exhibit nearly-exact integer ratios of their orbital periods, offer special opportunities to derive the formation and evolution mechanisms of the systems.
Currently, there are about 374 planetary systems that show TTVs, of which only $\sim 20$ orbit M dwarfs, including the well-characterized  TRAPPIST-1 system, a resonant system with seven transiting rocky worlds orbiting a nearby late M dwarf star \citep{Gillon_2016Natur,Gillon_2017Natur,Agol_2021}.

TOI-2015 has been confirmed by \cite{Jones_2024} using the Transiting Exoplanet Survey Satellite data, WIRO-2.3m, RBO-0.6m and ARC-3.5m photometric data,  and radial velocities collected with the Habitable-zone Planet Finder Spectrograph. The \cite{Jones_2024}'s analysis placed the non-transiting planet TOI-2015\,c within near 2:1 resonance. 
The authors found that TOI-2015\,b has a radius of $R_b=3.37^{+0.15}_{-0.20}\,R_\oplus$ and a mass of $M_b = 13.3^{+4.7}_{-4.5}\,M_\oplus$, and TOI-2015\,c has a mass of $M_c = 6.8^{+3.5}_{-2.3}\,M_\oplus$. However, other possible two-planet solutions--such as 3:2 and 4:3 near-resonances could not be conclusively excluded without complementary photometric and radial velocity observations.
In this paper, we extended their analysis including further period ratios (5:3 and 5:2 resonances) to justify and set the stage for the justification for the paper. 
In this context, we present a new photodynamical analysis of the TOI-2015 system using \emph{TESS} data, new ground-based photometric data and radial velocity measurements collected with the \maroonx\ spectrograph. We collected a total of 63 transit light curves of the transiting planet, TOI-2015\,b and 28 radial velocity measurements. This enabled us to enhance the accuracy of the physical parameters of both planets, and better constrain composition and origin.

The paper is structured as follows. Section~\ref{photometric_observation} describes the \emph{TESS} and ground-based follow-up (photometric, spectroscopic, and high angular resolution imaging) observations used to characterise the system. 
Section~\ref{stellar_carac} describes the stellar characterization of TOI-2015 using spectral energy distributions (SEDs), stellar evolution models, spectroscopic observations, and Spectral and photometric fitting using Bayesian Model Averaging (BMA).
The validation of the planet transit signals in the photometric data is presented in Section~\ref{Validate_planet}.
Photodynamical analysis of photometric data, RVs and TTV measurements is presented in Section~\ref{Global_modelling}, and an independent TTVs analysis is presented in Section~\ref{indep_fit}, allowing us to characterize the system.
The planet searches and detection limits are presented in Section~\ref{sec:sherlock}.
Finally, our discussion and conclusions are presented in Sections~\ref{discuss_results} and \ref{conclusion}, respectively.

\section{Observation and data reduction} \label{photometric_observation}

\subsection{\emph{TESS} photometry}
\label{sec:tess_photom}

The host star TIC\,368287008 (TOI-2015) was observed by \emph{TESS} \citep{Ricker_2015JATIS_TESS} in Sectors 24, 51 and 78 for 27 days each with 2-min cadence. Observation dates are presented in Table~\ref{TESS_obs_table}. The Sector\,51 campaign started on 2022\,Apr\,22 and ended on 2022\,May\,18. The two gaps during Sector\,51, were caused by scattered light and high background levels\footnote{{\tt TESS Report for Sector\,51: }\href{https://archive.stsci.edu/missions/tess/doc/tess_drn/tess_sector_51_drn74_v02.pdf}{tess\_sector\_51\_drn74\_v02.pdf}}, and we ignored them. The Sector\,78 campaign started on 2024\,Apr\,23 and ended on 2024\,May\,21. The target was observed in Camera\,1 CCD\,3. The first gap in the data is due to the Safe Hold mode of the spacecraft, while the second gap is due to the scattered light as Camera\,1 was too close to the Earth between BJD$_{\rm TDB}$ = 2460444 and BJD$_{\rm TDB}$ = 2460449 (see TESS observation report)\footnote{{\tt TESS Report for Sector\,78: }\href{https://archive.stsci.edu/missions/tess/doc/tess_drn/tess_sector_78_drn110_v01.pdf}{tess\_sector\_78\_drn110\_v01.pdf}}.

\begin{figure}[!]
	\centering
     \includegraphics[scale=0.5]{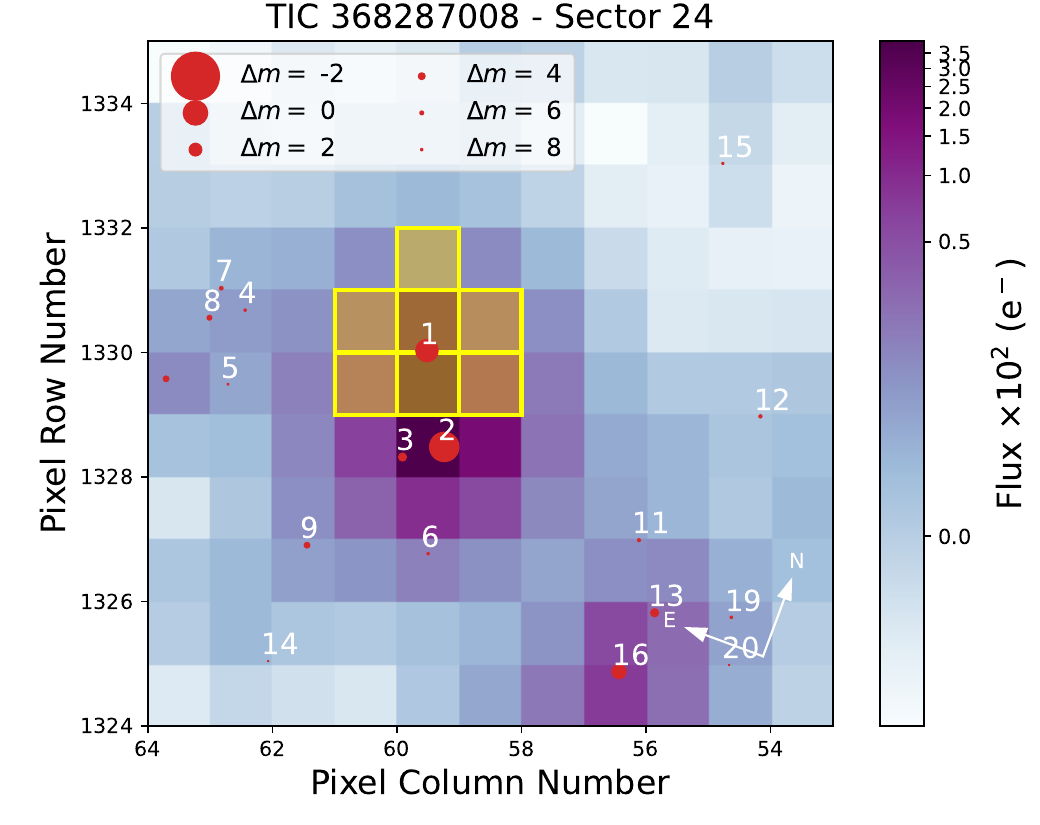} \\
     \includegraphics[scale=0.5]{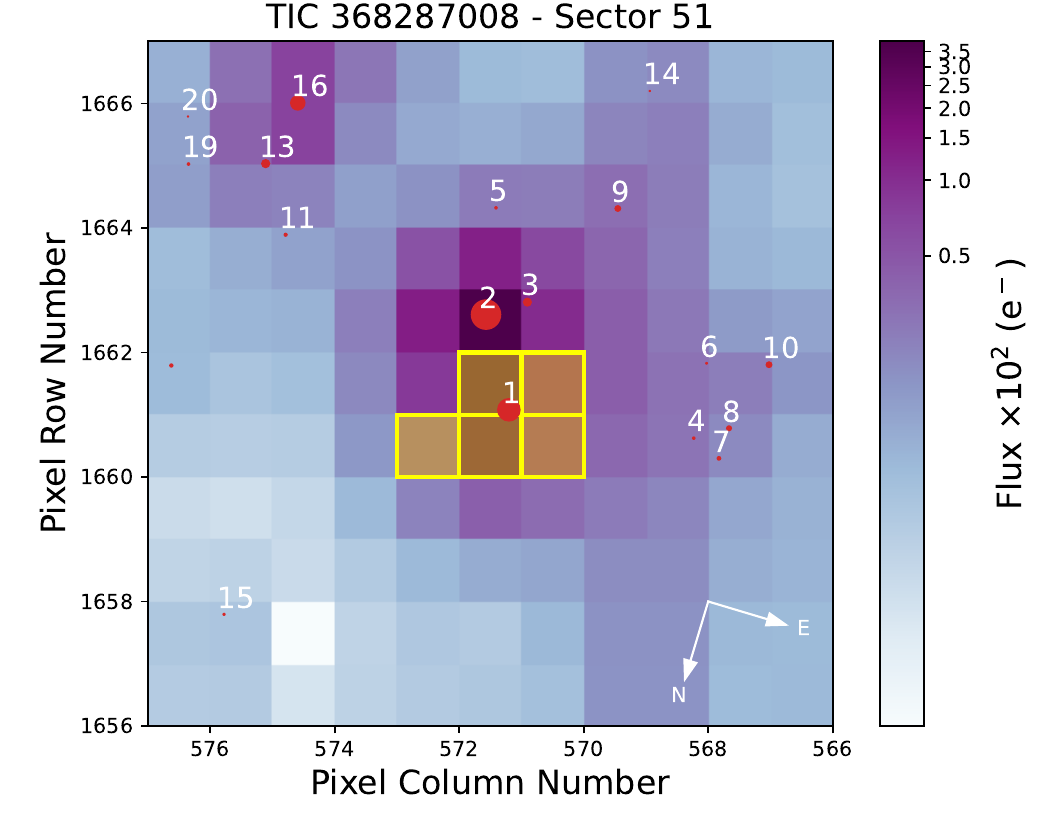}\\
     \includegraphics[scale=0.5]{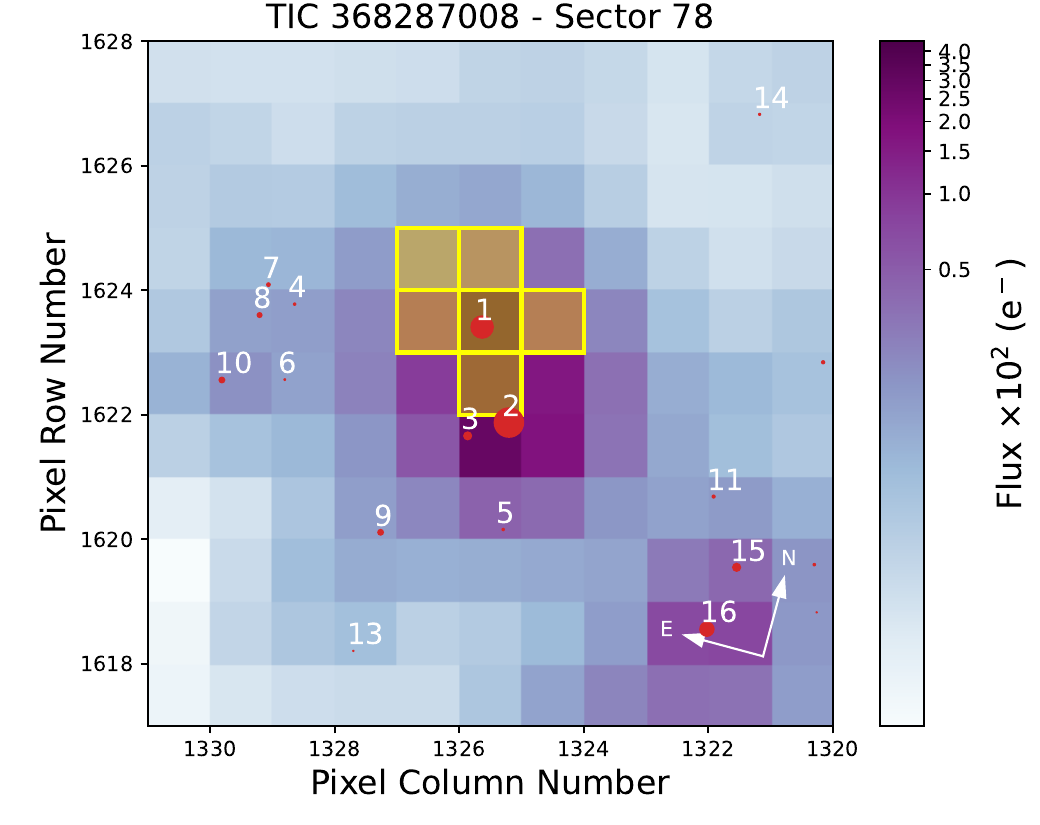}
	\caption{\emph{TESS} target pixel file images of TOI-2015 observed in Sectors 24 (top), 51 (middle), and 78 (bottom) made by \href{https://github.com/jlillo/tpfplotter}{\tt tpfplotter} \citep{Aller_2020AA}. Red dots show the location of Gaia DR3 sources, and the yellow shaded regions show the photometric apertures used to extract the photometric measurements.} 
	\label{Target_pixel}
\end{figure}

\begin{figure*}[!]
	\centering
	\includegraphics[scale=0.3]{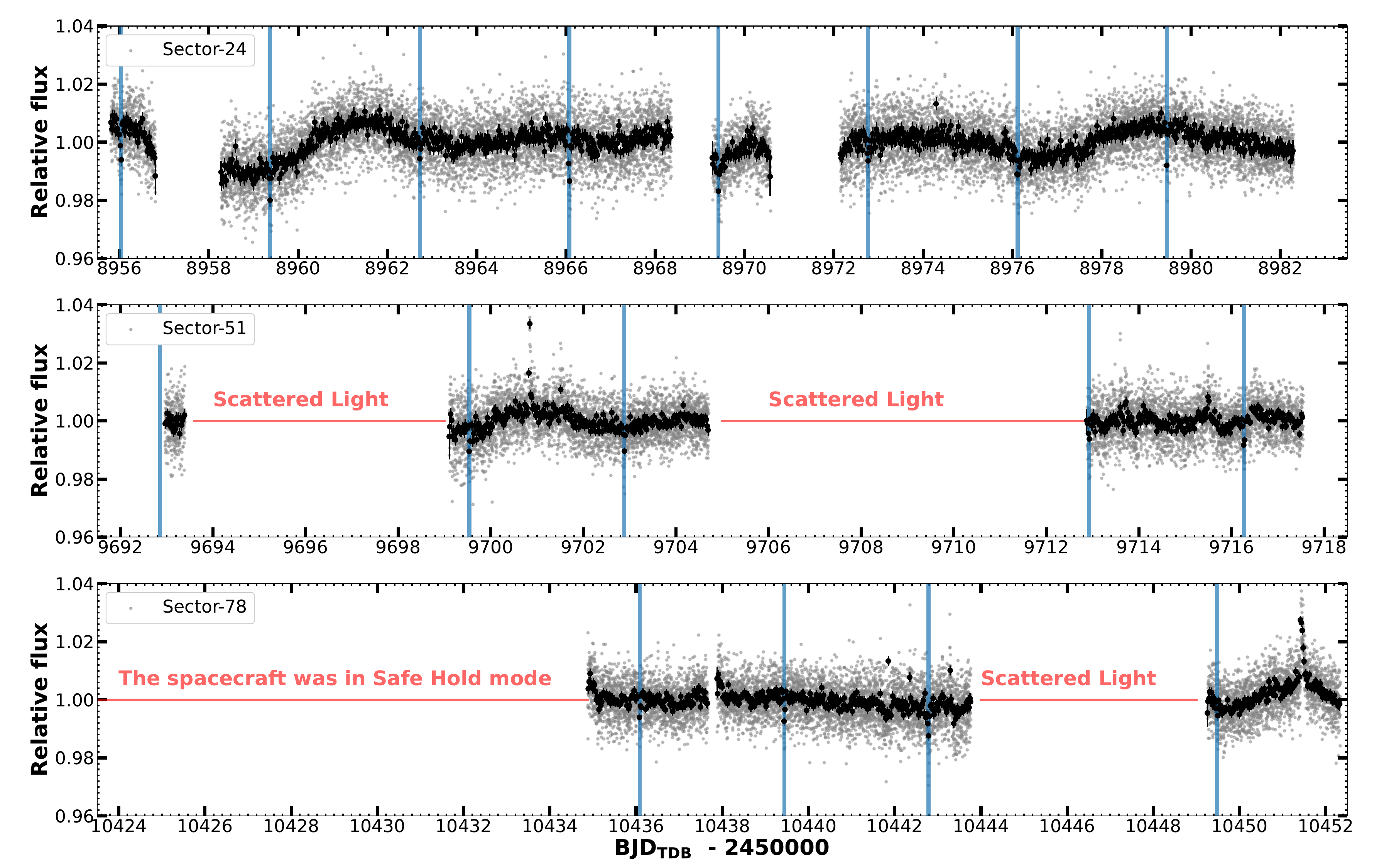}
	\caption{\emph{TESS} PDC-SAP flux extracted from the 2-min cadence data of TOI-2015. The target was observed in Sectors 24 (top), 51 (middle) and 78 (bottom). The light gray points show the 2-min cadence data, and the black points show the flux in 30-min bins. The transit locations of TOI-2015\,b are shown with vertical blue lines.} 
	\label{tess_lcs_pdcsap}
\end{figure*}

To analyse the \emph{TESS} photometric data, we used the Presearch Data Conditioning light curves (PCD-SAP) extracted from the Mikulski Archive for Space Telescopes \citep{Stumpe_2012PASP,Smith_2012PASP,Stumpe_2014} constructed by the \emph{TESS} Science Processing Operations Center \citep[SPOC,][]{SPOC_Jenkins_2016SPIE} at Ames Research Center.
PDC-SAP light curves have been corrected for any crowding and  instrument systematics effects.
The signature of TOI-2015\,b was first detected by the TESS SPOC pipeline in Sectors 24 and 51.
The TOI-2015 FOV including the  location of nearby Gaia DR3 sources \citep{Gaia_Collaboration_2021A&A} and photometric apertures are presented in Figure\,\ref{Target_pixel}.
\emph{TESS} transit light curves for TOI-2015\,b are presented in Figure\,\ref{tess_lcs_pdcsap}.

\subsection{Ground-based photometric follow-up} \label{ground_based_photo}

\begin{figure*}[!]
	\includegraphics[scale=0.3]{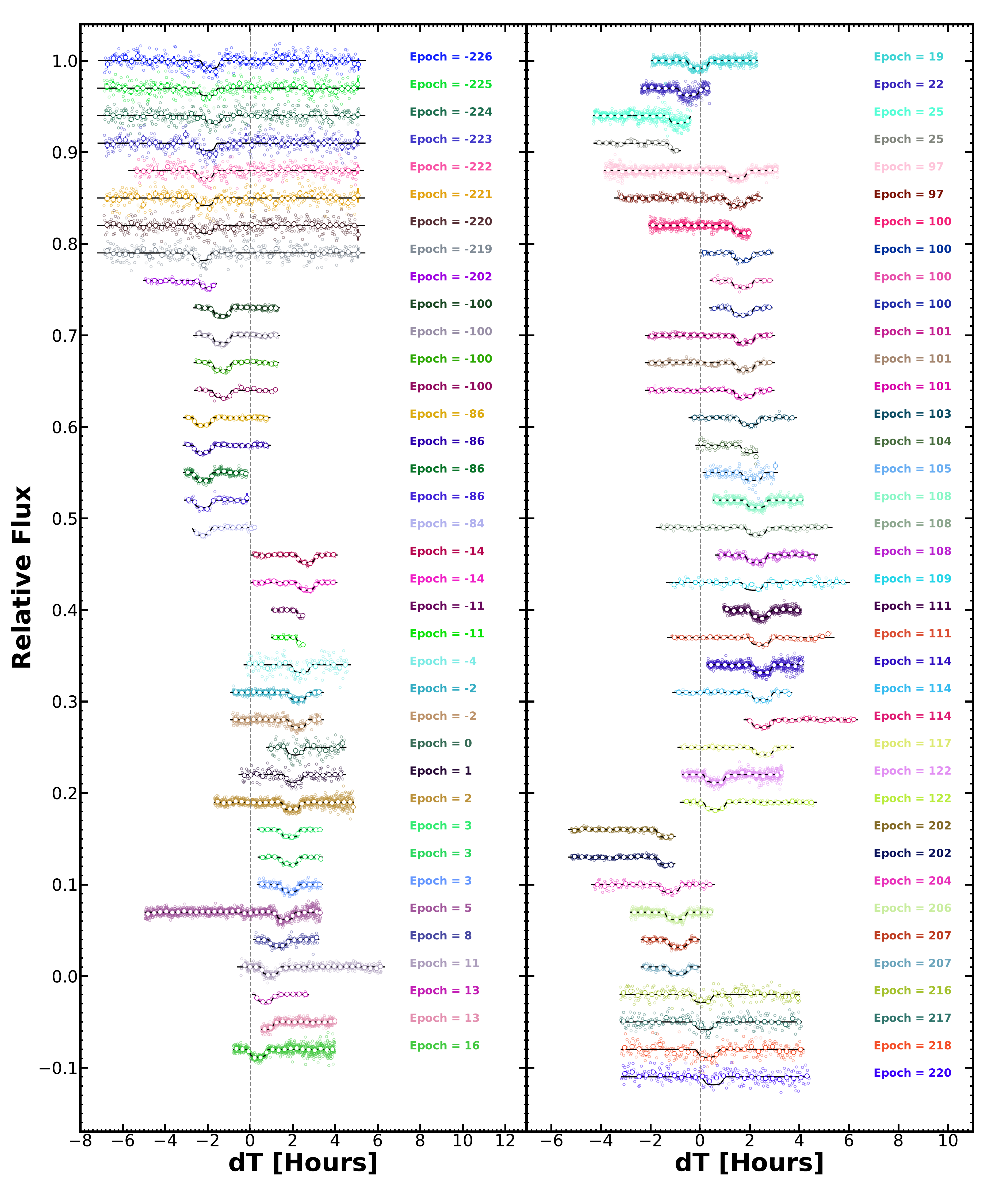}
	\caption{\emph{TESS} and ground-based transit light curves for TOI-2015\,b, plotted with arbitrary vertical offsets for clarity. The colored data points show the relative flux and the black lines show the best-fitting transit model superimposed. The light curves are shifted along the x-axis according to the TTVs and along y-axis for the visibility. The corresponding transit epoch is shown on the right of each transit light curve.} 
	\label{TOI2015b_lcs}
\end{figure*}

We used the {\tt TESS Transit Finder} tool in order to schedule the photometric observations. It is a  customized version of the {\tt Tapir} software package \citep{jensen2013}. These are summarized in the following sections, and the resulting transit light curves are presented in Figure\,\ref{TOI2015b_lcs}. The observation log is presented in Table\,\ref{obs_table}.  The transit timing measurements are presented in Table\,\ref{timing_table}.

\subsubsection{SPECULOOS-North}
We used SPECULOOS-North/Artemis \citep[Search for habitable Planets EClipsing ULtra-cOOl Stars,][]{Burdanov2022} at Teide observatory to observe the transits of TOI-2015\,b. Artemis is a 1.0-m Ritchey-Chr\'etien telescope equipped with a thermoelectrically cooled 2K$\times$2K Andor iKon-L BEX2-DD CCD camera with a pixel scale of $0.35\arcsec$/pixel and a total FOV of $12\arcmin\times12\arcmin$. SPECULOOS-North is a twin of the SPECULOOS-South \citep{Jehin2018Msngr,Delrez2018,Sebastian_2021AA} and SAINT-EX \citep[Search And characterIsatioN of Transiting EXoplanets, ][]{Demory_AA_SAINTEX_2020} telescopes. We observed a total of 15 (full and partial) transits in the $I+z$ filter with an exposure time of 10s. The observation dates are presented in Table\,\ref{obs_table}.
Data calibration and  photometric  extraction were performed using the {\tt PROSE}\footnote{{\tt Prose:} \url{https://github.com/lgrcia/prose}} pipeline \citep{prose_2022}.

\subsubsection{SAINT-EX}

We observed one full transit of TOI-2015\,b with SAINT-EX  on UT 2022\,Jun\,19 in the $I+z'$ filter with an exposure time of 10\,s. SAINT-EX \citep{Demory_AA_SAINTEX_2020}  is a 1.0-m $f/8$ Ritchey-Chr\'etien telescope located at the Sierra de San Pedro M\'artir in Baja California, M\'exico.  SAINT-EX is equipped with a thermoelectrically cooled 2K$\times$2K Andor iKon-L CCD camera, with a FOV of 12\arcmin$\times$12\arcmin\ and  a pixel scale of $0.35\arcsec$/pixel. The data calibration  and  photometric  extraction were performed using the {\tt PROSE} pipeline \citep{prose_2022}.

\subsubsection{TRAPPIST-South}

One full transit was observed with the TRAPPIST-South  \citep[TRAnsiting Planets and PlanetesImals Small Telescope,][]{Jehin2011,Gillon2011} telescope on UT 2022\,May\,29 in the $I+z'$ filter with an exposure time of 55\,s. It is a 60-cm Ritchey-Chr\'etien telescope located at ESO-La Silla Observatory in Chile, which is the twin of TRAPPIST-North. It is equipped with a 2K$\times$2K FLI Proline CCD camera with a FOV of $22\arcmin\times22\arcmin$ and a pixel-scale of 0.65\arcsec/pixel. The data calibration and photometric  extraction were performed using the {\tt PROSE} pipeline \citep{prose_2022}.

\subsubsection{TRAPPIST-North}

TRAPPIST-North \citep{Barkaoui2019_TN} observed three transits of TOI-2015\,b in the $I+z'$ filter with an exposure time of 50s. 
The telescope is a 60-cm Ritchey-Chr\'etien telescope located at Oukaimeden Observatory, and it is equipped with a thermoelectrically cooled 2K$\times$2K Andor iKon-L BEX2-DD CCD camera with a pixel scale of 0.6\arcsec\, resulting a FOV of $20\arcmin\times20\arcmin$. 
The data calibration and  photometric  extraction were performed using the {\tt PROSE} pipeline \citep{prose_2022}. The observation dates are given in Table\,\ref{obs_table}.

\subsubsection{MuSCAT}


We observed one full transit of TOI-2015\,b on UT 2022\,May\,06 with MuSCAT, a simultaneous multi-band camera installed on the 188-cm telescope in Okayama, Japan \citep{Narita_2015JATIS_Muscat1}. MuSCAT has three optical channels of $g'$, $r'$, and $z_s$- bands with a pixel scale of 0.358\arcsec/pixel and $6'.1 \times 6'.1$ field of view.
The reduction and the aperture photometry were conducted by the custom pipeline described in \citet{Fukui_2011PASJ}. The optimal aperture radius and set of comparison stars were selected for each band to minimize the scatter in the light curves.

\subsubsection{MuSCAT2}

Two full transits of TOI-2015\,b were observed on UT 2022\,May\,19 and 2022\,May\,29 with the MuSCAT2 multicolor imager \citep{Narita2018} mounted on the 1.52m-Telescopio Carlos S\'anchez (TCS) at the Teide Observatory in Tenerife (Canary Islands, Spain). 
Both transits were carried out simultaneously in the Sloan-$g$, -$r$, -$i$, and $z_\mathrm{s}$.
The photometric measurements were extracted using an uncontaminated photometric aperture (see Table\,\ref{obs_table}). 
The data calibration and photometric analysis were performed using the MuSCAT2 photometry pipeline \citep{Parviainen2020}. 

\subsubsection{LCOGT-2m0/MuSCAT3}

We used the Las Cumbres Observatory Global Telescope (LCOGT; \citealt{Brown_2013}) 2.0-m
Faulkes Telescope North at Haleakala Observatory in Hawaii to observe a total of six transits of TOI-2015\,b simultaneously in Sloan-$g',r',i'$ and $z_s$ filters.  The observation dates are given in Table\,\ref{obs_table}. The telescope is equipped with the MuSCAT3 multi-band imager \citep{Narita_2020SPIE11447E}. The data calibration was performed using the standard LCOGT {\tt BANZAI} pipeline \citep{McCully_2018SPIE10707E}, and photometric measurements were extracted using {\tt AstroImageJ}\footnote{{\tt AstroImageJ:}~\url{https://www.astro.louisville.edu/software/astroimagej/}} \citep{Collins_2017}.

\subsubsection{LCOGT-1.0-m}

We used the Las Cumbres Observatory Global Telescope (LCOGT; \citealt{Brown_2013}) 1.0-m network to observe a total of 15  transits (full and partial) of TOI-2015\,b in the Sloan-$i'$ filter. The observation dates are given in Table\,\ref{obs_table}.
Each telescope is equipped with 4096$\times$4096 SINISTRO Cameras, having an image scale of $0.389\arcsec$ per pixel and a total FOV of $26^{\prime} \times 26^{\prime}$. The data calibration was performed using the standard LCOGT {\tt BANZAI} pipeline \citep{McCully_2018SPIE10707E} and photometric measurements were extracted using {\tt AstroImageJ} \citep{Collins_2017}.

\subsubsection{OSN-1.5m}

We observed one partial transit of TOI-2015\,b on UT 2022\,Jun\,25 in the Johnson-Cousin $I$ filter  using the T150 at the Sierra Nevada Observatory in Granada (Spain). The T150 is equipped with a 2K$\times$2K Andor iKon-L BEX2DD CCD camera with a pixel scale of 0.232\arcsec, resulting in a field-of-view of 7.9\arcmin$\times$7.9\arcmin. The data calibration and photometric extraction were performed using {\tt AstroImageJ} \citep{Collins_2017}.

\subsubsection{IAC80}

We observed a full transit of TOI-2015\,b on UT 2024\,May\,09 in the Sloan-$r'$ filter using the IAC80 telescope located at Teide Observatory in Tenerife (Canary Island, Spain). The IAC80 is an 82\,cm telescope equipped with a 4K$\times$4K CCD camera with a plate scale of 0.32\arcsec/pixel, resulting in a field-of-view of 21.98\arcmin$\times$22.06\arcmin. The data calibration and photometric extraction were performed using {\tt AstroImageJ} \citep{Collins_2017}.

\subsubsection{T100}
One full transit was observed with the 1-m Turkish telescope T100, located at the T\"urkiye National Observatories Bak{\i}rl{\i}tepe Campus at an altitude of 2500 meters on UT 2023\,Jun\,25. The telescope is equipped with a cryo-cooled SI\,1100 CCD that has $4096\times4096$ pixels, providing an effective field of view (FoV) of $21^{\prime}\times21^{\prime}$. We used the CCD in $2\times2$ binning mode to decrease the readout time from 45\,s to 15\,s. We slightly defocused the telescope to increase the precision \citep{2015ASPC..496..370B} and observed without a filter, to capture all wavelengths of light and thus maximize the precision. Calibration of the raw images, aperture photometry with respect to an ensemble of comparison stars, and airmass detrending were performed using {\tt AstroImageJ}.

\subsubsection{CAHA-1.23m}

One full transit of TOI-2015\,b was observed on UT 2023\,Jun\,15 with the Zeiss 1.23\,m telescope at the Observatory of Calar Alto in Spain. The telescope is equipped with an iKon-XL\,230 camera, with $4096 \times 4108$\,pixels of size $15\,\mu$m. The pixel scale is 0.32\,arcsec\,pixel$^{-1}$ and the FOV is 21.4\,arcmin $\times$ 21.5\,arcmin. The transit was monitored through the special uncoated GG-495 glass long-pass filter (transparent at $>500$\,nm). Observations were performed by slightly defocusing the telescopes, in order to increase the photometric precision (e.g., \citealp{southworth2012,mancini2013}), and using autoguiding. An exposure time between 110 and 120 seconds was adopted. The science data were calibrated by adopting the same procedure as in \citet{mancini2017}.
Standard aperture photometry was used to extract the light curves of the planetary transit. This was done by placing the usual three apertures on the target and on five good comparison stars and running the APER routine \citep{southworth2010}. The sizes of the apertures were decided after several attempts, by selecting those having the lowest scatter when compared with a fitted model. The resulting light curve was normalized to zero magnitude by a quadratic fit to the out-of-transit data versus the comparison stars.

\subsection{Spectroscopy}  \label{Spectroscopy}

\subsubsection{Spectroscopic follow-up using \maroonx} 
\label{sec:maroonx_obs}

We observed TOI-2015 15 times with the \maroonx\ spectrograph \citep{Seifahrt18, Seifahrt22} on Gemini-North between UT 2021\,Apr and UT 2023\,Jul.  \maroonx\ is an extreme precision radial velocity (EPRV) spectrograph with a wavelength coverage suitable for M\,dwarf observations.  \maroonx\ has two CCDs encompassing different wavelength ranges, with a ``red'' channel at $650-920$\,nm and a ``blue'' channel at $500-670$\,nm.  Both of these CCDs were exposed simultaneously during an observation of TOI-2015, meaning that we have 15 red-channel RVs and 15 blue-channel RVs.  As these two channels are independent of one another and encompass different wavelength ranges, we consider them to be two different instruments for the purposes of our analysis, as they may capture different chromatic stellar signals.

The \maroonx\ data were reduced using a custom \texttt{Python3} pipeline using tools developed for the CRIRES instrument \citep{Bean10}.  We calculated RVs from the reduced spectrum using a modified version of \texttt{serval} \citep{Zechmeister20} customized to work with \maroonx\ data.  \texttt{serval} calculates RVs by co-adding all of the available spectra for the target to produce a high-SNR template, which is then compared to each individual spectrum in order to calculate the relative RV shift.  

\begin{figure*}[!]
	\centering
	\includegraphics[scale=0.31]{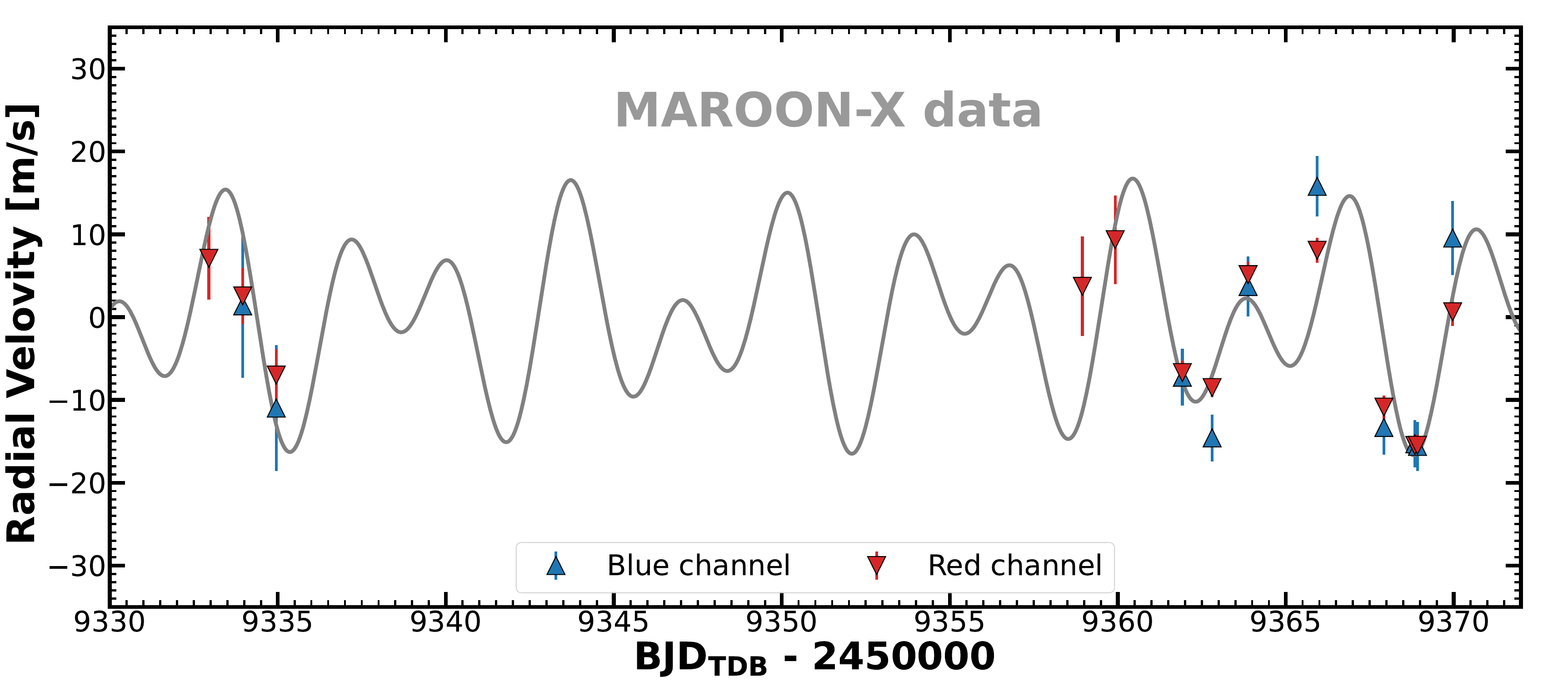}
     \includegraphics[scale=0.31]{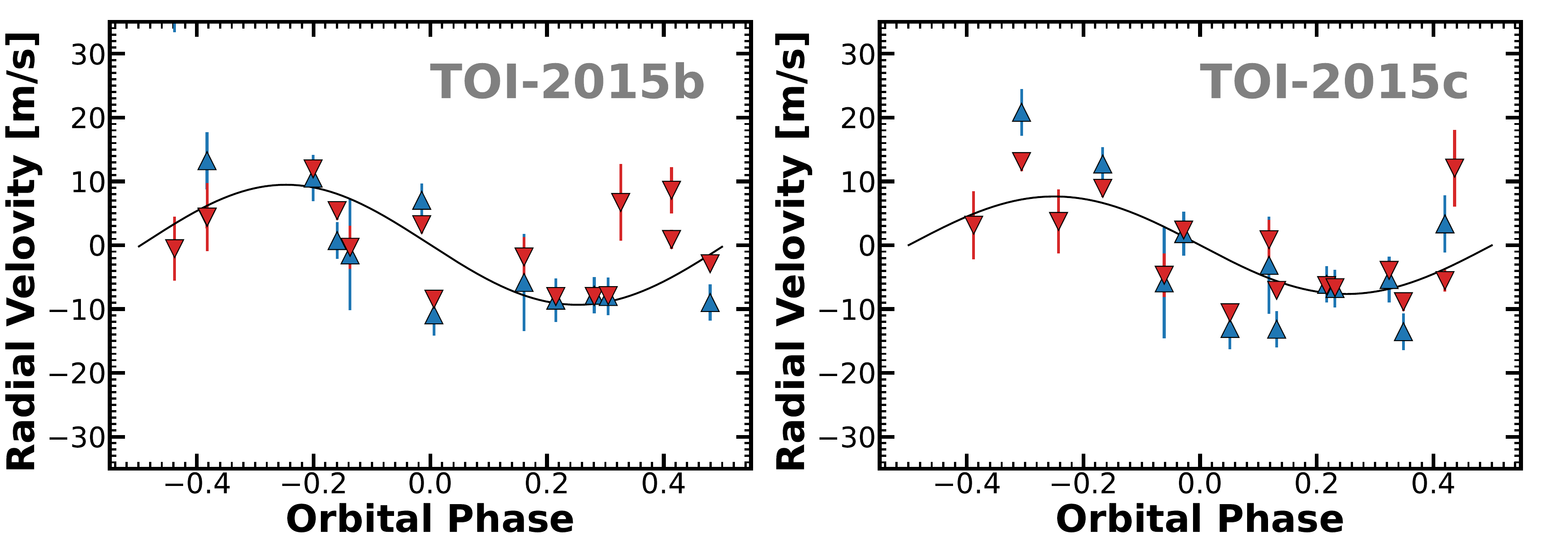}
	\caption{Radial velocity measurements in "red" (red triangles) and "blue" (blue triangles) channels with error bars) of the two-planets TOI-2015\,b and TOI-2015\,c collected with the \maroonx\  spectrograph. The gray solid line shows the best-fit radial velocity data for the 5:3 scenario.} 
	\label{TOI2015b_rvs_mx}
\end{figure*}

Our data have exposure times ranging from $520$ to $1800$\,s, with one exposure at 2400\,s.  In general, we found that the spectra with 520\,s exposure times had very large RV errors due to the faint nature of the host star.  We increased our exposure times later in the survey in order to improve the RV precision (and reduce time lost to telescope overhead).  Overall, we have ten exposures with the longer exposure times. Two of the short exposures in the blue channel had SNR$\,<\,10$ and were thus not included in the \texttt{serval} results.  Omitting the 520\,s exposures, the red-channel spectra have a median SNR of 94 and a resulting median RV error of 1.4\,m\,s$^{-1}$.  The blue-channel spectra have a median SNR of 33 and a median RV error of 3.1\,m\,s$^{-1}$.  The resulting RV measurements and curve are presented in Table\,\ref{table_Maroonx_RVs_TOI2015} and Figure\,\ref{TOI2015b_rvs_mx}.

\begin{table}[!]
\centering
\caption{Radial velocity measurements for TOI-2015 obtained by \maroonx\ in the "red" and "blue" channels.}
	{\renewcommand{\arraystretch}{1.5}
 \resizebox{0.48\textwidth}{!}{
		\begin{tabular}{l|c|c}
		\hline
     & Red channel & Blue channel \\
     \hline
     BJD$_{\rm TDB}$     & RV$\pm$ $\sigma_{\rm RV}$ [m/s] & RV$\pm$ $\sigma_{\rm RV}$ [m/s]      \\
        \hline
        2459332.94938  & $-2.581    \pm 5.003 $ & $36.052 \pm 13.656$  \\
        2459333.95573  & $-6.697 	 \pm 3.404$ & $-16.506 \pm 8.681$ \\
        2459334.95363	& $-15.835  \pm 3.113$ & $-28.145 \pm 7.595$ \\
        2459358.94624	& $4.576    \pm 6.018$ & $- $ \\
        2459359.92272	& $10.594	 \pm 5.348$ & $- $ \\
        2459361.92241	& $-4.669   \pm 1.427$ & $-5.860 \pm 3.417 $  \\
        2459362.80767	& $-6.133   \pm 1.153$ & $-12.592 \pm 2.828 $ \\
        2459363.87855	& $7.915	 \pm 1.514$ & $6.392 \pm 3.623 $ \\
        2459365.93460	& $11.654   \pm 1.494$ & $19.844 \pm 3.651 $ \\
        2459367.91986	& $-6.518   \pm 1.391$ & $-7.968 \pm 3.255 $ \\
        2459368.83965	& $-10.798  \pm 1.239$ & $-9.334 \pm 2.871 $  \\
        2459368.91956	& $-10.754  \pm 1.270$ & $-9.599 \pm 2.950 $ \\
        2459369.96677	& $5.781	 \pm 1.705$ & $16.241 \pm 4.477 $ \\
        2460127.91889	& $20.109   \pm 1.351$ & $26.724 \pm 2.611 $ \\
        2460130.78194	& $8.402    \pm 1.450$ & $5.732 \pm 2.868 $ \\
		\hline
		\end{tabular}} }
		\label{table_Maroonx_RVs_TOI2015}
\end{table}

\subsubsection{IRTF/SpeX spectroscopy}

We collected a medium-resolution near-infrared spectrum of TOI-2015 with the SpeX spectrograph \citep{Rayner2003} on the 3.2-m NASA Infrared Telescope Facility (IRTF) on UT 2022\,Apr\,19.
Thin cirrus was present, and seeing was 1$\farcs$0.
Using the short-wavelength cross-dispersed (SXD) mode with the $0.3'' \times 15''$ slit aligned to the parallactic angle, we gathered spectra covering 0.80--2.42\,$\mu$m with a resolving power of $R{\sim}2000$.
Nodding in an ABBA pattern, we collected six exposures of 114.9\,s each, totaling 689.4\,s on source.
We collected a standard set of SXD flat-field and arc-lamp exposures after the science observations, followed by a set of six, 3.7-s exposures of the A0\,V star HD\,140729 ($V{=}6.1$).
Data calibration was performed using {\tt Spextool v4.1} \citep{Cushing2004}, following the instructions for standard usage in the Spextool User's Manual\footnote{Available at \url{http://irtfweb.ifa.hawaii.edu/~spex/observer/}.}. 
The final spectrum has a median SNR per pixel of 83 with peaks in the $J$, $H$, and $K$ bands of 108, 119, and 110, respectively, and an average of 2.7\,pixels per resolution element.

\subsubsection{Shane/Kast optical spectroscopy}

We observed TOI-2015 with the Kast double spectrograph \citep{kastspectrograph} mounted on the 3m Shane telescope at Lick Observatory on UT 2022\,Jul\,02. Conditions were mostly clear with scattered clouds and 1$\arcsec$ seeing. We used the 1$\arcsec$ slit aligned to the parallactic angle to obtain blue and red optical spectra split at 5700~{\AA} by the d57 dichroic, and dispersed by the 600/4310 grism and 600/7500 grating, respectively, for a common resolution of $\lambda/\Delta\lambda$ $\approx2000$. We obtained a single 1200\,s exposure in the blue channel and two 600\,s exposures in the red channel at an average airmass of 1.07. The G2~V star HD 104385 ($V=8.57$) was observed immediately before TOI-2015 for telluric absorption calibration, and the spectrophotometric calibrator Feige\,66 \citep{1992PASP..104..533H,1994PASP..106..566H} was observed during the night for flux calibration. We obtained HeHgCd and HeNeArHg arc lamp exposures at the start of the night to wavelength calibrate our blue and red data, respectively; flat-field lamp exposures for pixel response calibration were also obtained. Data were reduced using the kastredux code\footnote{\url{https://github.com/aburgasser/kastredux}.} using standard settings. We focused on our analysis on the higher-quality red optical data which span 6000-9000~{\AA} and have a median signal-to-noise = 130  at 7350~{\AA}.

\subsection{High-resolution imaging}  \label{high-res}

High-angular-resolution imaging is required to check for nearby sources that could contaminate the \emph{TESS} photometry, resulting in an underestimated radius of the occulting object, or that can be the source of astrophysical false positives, such as blended eclipsing binaries.

\subsubsection{4.1-m\,SOAR observations of TOI-2015}
We searched for stellar companions to TOI-2015 using  speckle imaging installed on the 4.1m Southern Astrophysical Research (SOAR) telescope \citep{Tokovinin_2018PASP} on UT 2021\,Apr\,25, observing in the Cousins-I filter, a similar bandpass as \emph{TESS}. This observation was sensitive enough to detect a 4.2-mag fainter star at an angular distance of 1\,arcsec from the target. Further details of the SOAR observations are available in \cite{Ziegler_2020AJ}. 
Figure\,\ref{SHANE_SOAR_highres} shows the speckle auto-correlation functions and the 5$\sigma$ detection sensitivity. No nearby sources have been detected within 3\arcsec of TOI-2015 in the SOAR data.

\subsubsection{3.0m-Shane observations of TOI-2015}
\begin{figure*}[!]
	\includegraphics[scale=0.7]{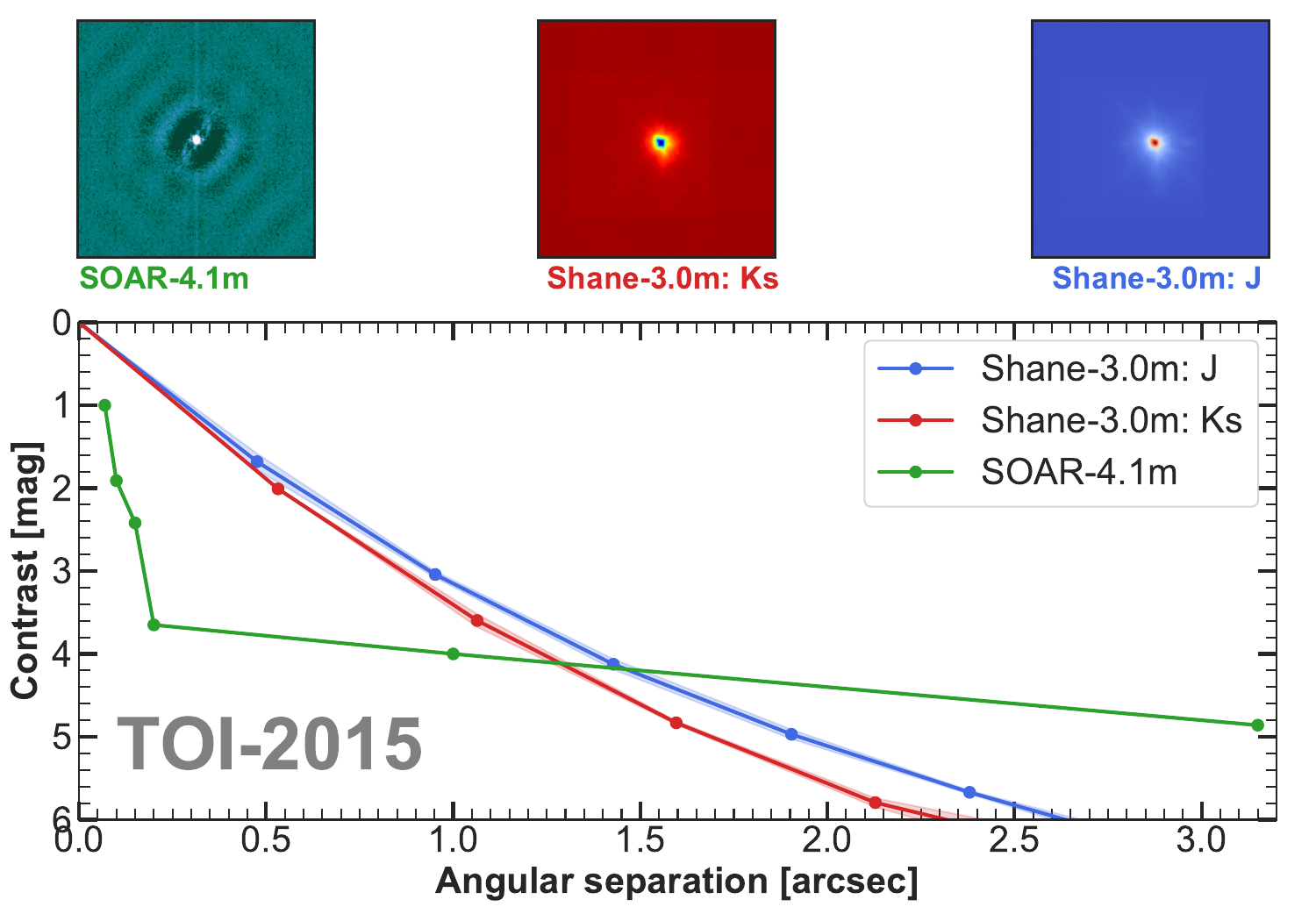}
	\caption{High angular resolution imaging of TOI-2015 from the 4.1-m SOAR telescope on UT 2021\,Apr\,25 in the $K_s$ filter (green panel), and  the 3.0-m SHANE telescope on UT 2021\,Mar\,04 in the $K$ (red panel) and  $J$ (blue panel) bands. No stellar companions were found within detection limits.}
	\label{SHANE_SOAR_highres}
\end{figure*}

We observed TOI-2015 with the ShARCS camera on the Shane 3.0-m telescope located at Lick Observatory \citep{2012SPIE.8447E..3GK, 2014SPIE.9148E..05G, 2014SPIE.9148E..3AM} on UT 2021\,Mar\,04. Observations were taken with the Shane adaptive optics system in natural guide star mode to search for nearby unresolved stellar companions. Sequences of observations were collected using the $K_s$ and $J$  filters. The data reduction was performed using the publicly available \texttt{SImMER}\footnote{{\tt SImMER}: https://github.com/arjunsavel/SImMER} pipeline \citep{2020AJ....160..287S, 2022PASP..134l4501S}. No nearby stellar sources have been detected within detection limits (see Figure\,\ref{SHANE_SOAR_highres}). 

\section{Stellar characterization} \label{stellar_carac}

\subsection{SED analysis and evolutionary models} \label{SED_evol_analys}

\begin{figure}[!]
	\centering
	\includegraphics[scale=0.3]{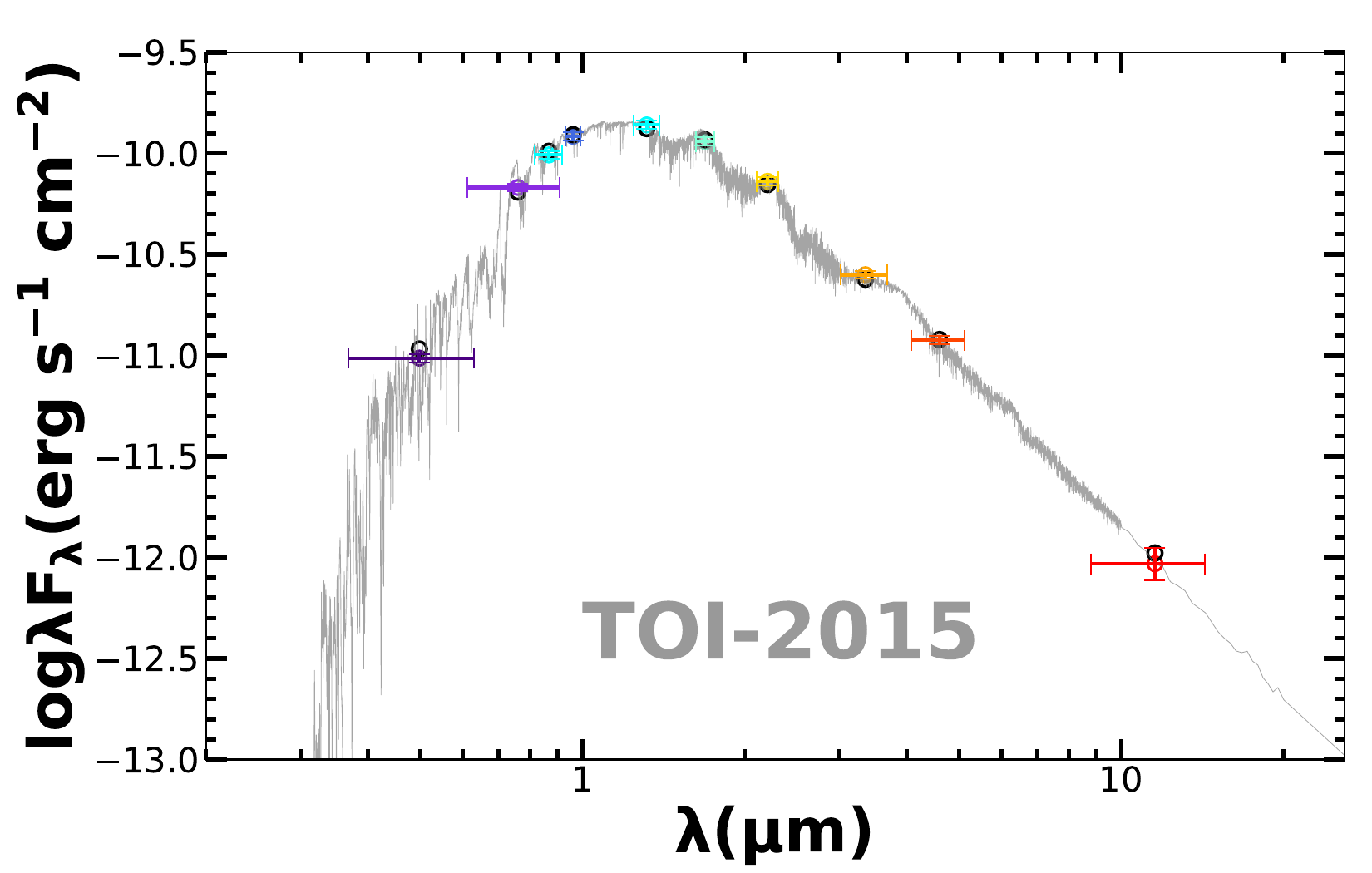}
	\caption{ Spectral Energy Distribution (SED) fit of TOI-2015. The gray curve is the best-fitting NextGen atmosphere model, the black circles show the model fluxes, while the colored circles with error bars show the observed fluxes.}
    	\label{SED_plots}
\end{figure}

As a first determination of the basic stellar parameters, we performed an analysis of the broadband SED of the star together with the {\it Gaia\/} DR3 parallax \citep[with no systematic offset applied; see, e.g.,][]{StassunTorres:2021}, in order to derive an empirical measurement of the stellar radius, following the procedures described in \citet{Stassun:2016,Stassun:2017,Stassun:2018}. We pulled the $JHK_S$ magnitudes from {\it 2MASS}, the W1--W3 magnitudes from {\it WISE}, the $G_{\rm BP} G_{\rm RP}$ magnitudes from {\it Gaia}, and the $zy$ magnitudes from {\it Pan-STARRS}. Together, the available photometry spans the full stellar SED over the wavelength range $0.4-10\,\mu$m (see Figure\,\ref{SED_plots}).

We performed a fit using NextGen stellar atmosphere models, with the free parameters being the effective temperature ($T_{\rm eff}$) and metallicity ([Fe/H]); for the latter we adopted the spectroscopically determined value using \maroonx\ data. In addition, we fixed the extinction $A_V \equiv 0$ due to the proximity of the system to the Earth. The resulting fit (Figure\,\ref{SED_plots}) has a best-fit $T_{\rm eff} = 3200 \pm 75$~K, with a reduced $\chi^2$ of 1.3. Integrating the model SED gives the bolometric flux at Earth, $F_{\rm bol} = 1.552 \pm 0.018 \times 10^{-10}$ erg~s$^{-1}$~cm$^{-2}$, which with the {\it Gaia\/} parallax gives the luminosity, $L_{\rm bol} = 0.010837 \pm 0.000063$~L$_\odot$. Taking the $L_{\rm bol}$ and $T_{\rm eff}$ together gives the stellar radius, $R_\star = 0.339 \pm 0.016$~R$_\odot$. In addition, we derived  the stellar mass from the empirical $M_K$ relations of \citet{Mann:2019}, giving $M_\star = 0.33 \pm 0.02$~M$_\odot$. 

We also checked stellar parameters as obtained from evolutionary modeling using CLES models for low-mass stars \citep{Scuflaire_2008Ap,Fernandes_2019}. We used as inputs the luminosity derived just above, the metallicity [Fe/H]$=-0.31\pm0.16$ from \maroonx, and assuming an age $>1$~Gyr. We derived $M_\star=0.28 \pm 0.06~M_\odot$,  within 1-sigma agreement with the stellar mass found above. Other stellar parameters (stellar radius, effective temperature, etc.) are also found to be within $1\,\sigma$ of the values derived above.

\subsection{Spectroscopic analysis} \label{spec_analysis}

\begin{figure}[!]
    \centering
    \includegraphics[width=\columnwidth, keepaspectratio]{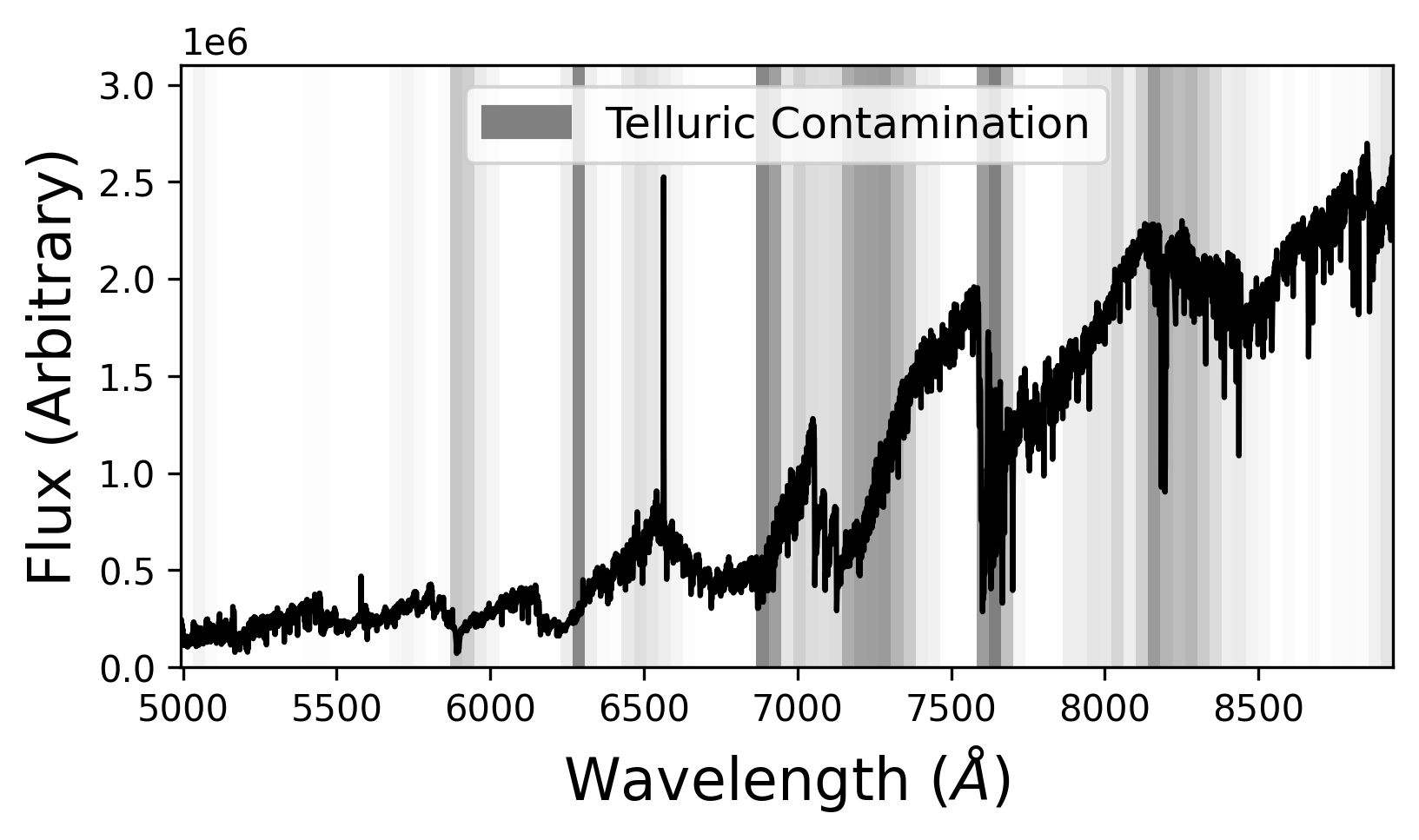}
    \caption{
    MAROON-X spectrum of TOI-2015. The spectrum of TOI-2015 is shown in black. The gray regions are regions with telluric features in the spectrum, with darker gray indicating more tellurics.
    }
    \label{fig:maroonx}
\end{figure}

We determined stellar parameters from the red part of the \maroonx\ spectrum (see Figure~\ref{fig:maroonx}) following the method described in \citet{Passegger_2020} using the PHOENIX-ACES model grid \citep{Husser_2013}, and assuming a stellar age of 5~Gyr for the evolutionary models (PARSEC, \citet{Bressan_2012MNRAS}; \citet{Chen_2014MNRAS,Chen_2015MNRAS}; \citet{Tang_2014MNRAS}). With this, we derive parameters of $T_{\rm eff}= 3211 \pm 51$K, $\log{g_\star}= 5.04 \pm 0.04$, and [Fe/H]$= -0.31\pm0.16$.
The derived metallicity of [Fe/H]$= -0.31 \pm 0.16$ is significantly different from the values determined from SpeX and Kast spectra. We note that these spectrographs have lower spectral resolutions, and therefore different determination techniques have been used. As shown by
\citet{Passegger_2022}, who compared different stellar parameter determination techniques, significant differences in the metallicity can be found by different methods, even when used on the same high-resolution spectra. \\

An independent analysis ran with the same methods as \cite{Brady_2024} recovered similar values for $T_{\rm eff}$ ($3237 \pm 82$\,K) and [Fe/H] ($-0.22 \pm 0.16$), which makes sense as both techniques utilize comparisons to PHOENIX model spectra to recover the stellar parameters.  However, we note that the \cite{Brady_2024} technique tends to perform unreliably for M dwarfs cooler than 3200\,K, and the recovered temperature for TOI-2015 falls very close to this limit. 

We also followed methods similar to those in \cite{Brady_2024} to estimate the stellar $v\sin{i}$ by comparing the width of its cross correlation function with an artificially-broadened \maroonx\ spectrum of Barnard's Star, which, as an M\,4 star \citep{Kirkpatrick_1991}, has a similar spectral type to TOI-2015.  Overall, we recovered a $v\,\sin{i}$ value of $2.1 \pm 0.4$\,km\,s$^{-1}$.  However, we note that, due to \maroonx's resolution, this method cannot  measure $v$sin$i$ values lower than 2\,km\,s$^{-1}$, meaning that it is possible that TOI-2015 is rotating more slowly than the detection limit of \maroonx.    This value implies that TOI-2015 may be rotating more slowly than the $v\,\sin{i}=3.2\pm 0.6$\,km\,s$^{-1}$ quoted in \cite{Jones_2024}, which was close to the resolution limit of HPF.  This discrepancy can be explained by the fact that \maroonx\ has a higher resolution than HPF, allowing it to probe longer rotation periods.  To be conservative, we thus quote a 2\,$\sigma$ upper limit of $v\,\sin{i}<2.9$\,km\,s$^{-1}$ for TOI-2015 from the \maroonx\ data.  This is in agreement with the 8.7\,d photometric rotation period quoted by \cite{Jones_2024}, which predicts a $v\,\sin{i} \lesssim 2$\,km\,s$^{-1}$.\\

\begin{figure}[!]
    \centering
    \includegraphics[width=\columnwidth, keepaspectratio]{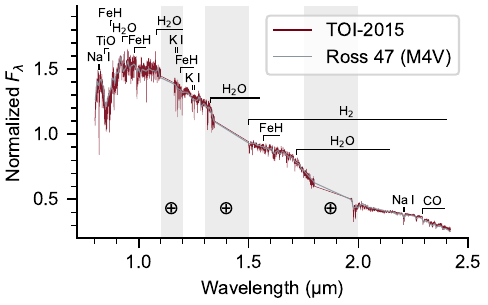}
    \caption{
    SpeX SXD spectrum of TOI-2015.
        The spectrum of TOI-2015 is shown in red and the SpeX Prism spectrum of the M4 standard Ross 47 is shown in grey for comparison.
        Strong spectral features of M dwarfs are indicated, and regions of high telluric absorption are shaded.
    }
    \label{fig:spex}
\end{figure}

Figure\,\ref{fig:spex} shows the SpeX SXD spectrum of TOI-2015.
We used the SpeX Prism Library Analysis Toolkit \citep[SPLAT, ][]{splat} to compare the spectrum to those of SpeX Prism standards \citep{Kirkpatrick2010}.
We found the best spectral match to be the M4 standard Ross 47 (Gl\,213), and we adopted an infrared spectral type of M4.0$\pm$0.5 for TOI-2015.
Using SPLAT, we measured the equivalent widths of the $K$-band Na\,\textsc{i} and Ca\,\textsc{i} doublets and the H$_2$O--K2 index \citep{Rojas-Ayala2012}.
Following the \citet{Mann2013} relation between these observables and stellar metallicity, and propagating uncertainties using a Monte Carlo approach (see \citealt{Delrez_2022_A&A}), we estimated an iron abundance of [Fe/H]$=+0.29\pm0.13$.\\

\begin{figure}[!]
	\centering
	\includegraphics[scale=0.5]{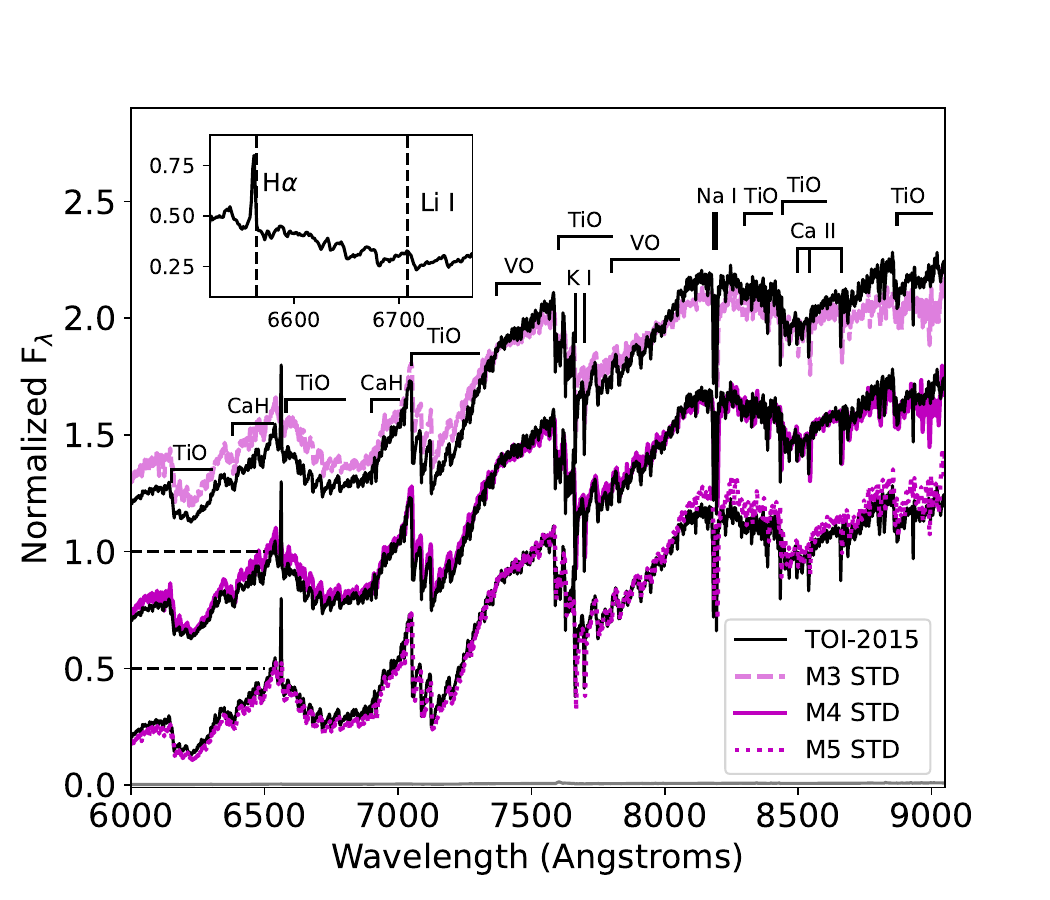}
	\caption{Normalized Shane/Kast optical spectrum of TOI-2015 (black lines) compared to three SDSS spectral templates from \citet[magenta lines]{2007AJ....133..531B}. Characteristic spectral features of mid-type M dwarfs are labeled. The inset box shows the 6520--6770~{\AA} region encompassing H$\alpha$ (in emission) and Li~I (not detected).}
	\label{fig:kast}
\end{figure}

Figure\,\ref{fig:kast} compares the reduced Shane/Kast red optical spectrum of TOI-2015 to the M3, M4, and M5 dwarf spectral templates from 
\citet{2007AJ....133..531B}. The spectral morphology of TOI-2015 is intermediate between the M4 and M5 templates, suggesting an M4.5$\pm$0.5 optical classification, consistent with the near-infrared type. Both template-based classifications are also consistent with optical index-based classifications from \citet{1995AJ....110.1838R,1997AJ....113..806G,1999AJ....118.2466M,2003AJ....125.1598L}; and \citet{2007MNRAS.381.1067R}, which span M3.5-M4.5.
We detected strong H$\alpha$ emission at 6563~{\AA} with an equivalent width of $-3.09\pm0.08$~{\AA}, corresponding to $\log{\left(L_{{\rm H}\alpha}/L_{\rm bol}\right)} = -4.02\pm0.06$ using the $\chi$ factor relation of \citet{2014ApJ...795..161D}. 
This strong emission implies an activity age of no more than $5-6$~Gyr \citep{2008AJ....135..785W}, while the absence of detectable Li\,\textsc{i} absorption at 6708~{\AA} rules out an age younger than $\sim30$\,Myr. We measure the metallicity index $\zeta=1.057\pm0.002$  \citep{2013AJ....145..102L} which corresponds to a slightly super-solar metallicity of [Fe/H]$=0.08\pm0.20$ using the \citet{Mann2013} calibration. This is formally consistent with the super-solar metallicity inferred from the SpeX data.\\

\subsection{Spectral and photometric fitting using Bayesian Model Averaging} \label{BMA}
The SpeX SXD spectrum of TOI-2015, alongside photometric data from {\it 2MASS} ($JHK_S$), {\it ALLWISE} ($W1,W2$), and {\it Gaia} DR3 ($G, G_{\rm RP}$), was used for a SED fit using four synthetic spectral model grids. The SpeX SXD spectrum was convolved to a wavelength resolution of $R = 200$ to ensure consistent analysis across the different grids. Absolute flux calibration was achieved by scaling the optical part of the SpeX SXD spectrum to match the Gaia XP spectrum \citep{GAIADR3}. For the spectral and photometric fitting, we employed four synthetic spectral libraries specifically designed to model the M dwarf regime: BT-Settl CIFIST \citep{BTSETTLCIFIST}, BT-Settl AGSS \citep{BTSETTLAGSS}, Phoenix ACES \citep{Husser_2013}, and SPHINX \citep{SPHINX}. These libraries offer theoretical spectra and atmospheric models that account for various physical conditions in M dwarfs, making them appropriate for this analysis.
The fitting process was carried out using the \texttt{species} toolkit \citep{SPECIES}, which incorporates the nested sampling algorithm from the UltraNest package \citep{ULTRANEST} to efficiently explore parameter space and estimate posterior distributions as well as Bayesian evidence.
We adopted empirical relations for the $T_\mathrm{eff}$, radius, and mass of M dwarfs from \citet[][]{Mann_2015} as priors in this analysis, using {\it Gaia} DR3 magnitudes, parallax, and {\it 2MASS} $K_s$ magnitude as constraints. The model grids cover [Fe/H] from $-1.0$ to $+0.5$\,dex. A flat prior was used for extinction ($A_\mathrm{v}$), with potential values constrained between 0 and 0.2, given the star’s proximity. The parameters fitted included $T_\mathrm{eff}$, [Fe/H], $\log{g_\star}$, parallax, radius, and extinction. The carbon-to-oxygen (C/O) ratio was also fitted for the SPHINX library without any prior. Furthermore, stellar mass and luminosity estimates were derived from relations involving $\log{g_\star}$, radius, parallax, and $T_\mathrm{eff}$.

\begin{figure}[!]
    \centering
    \includegraphics[width=\columnwidth, keepaspectratio]{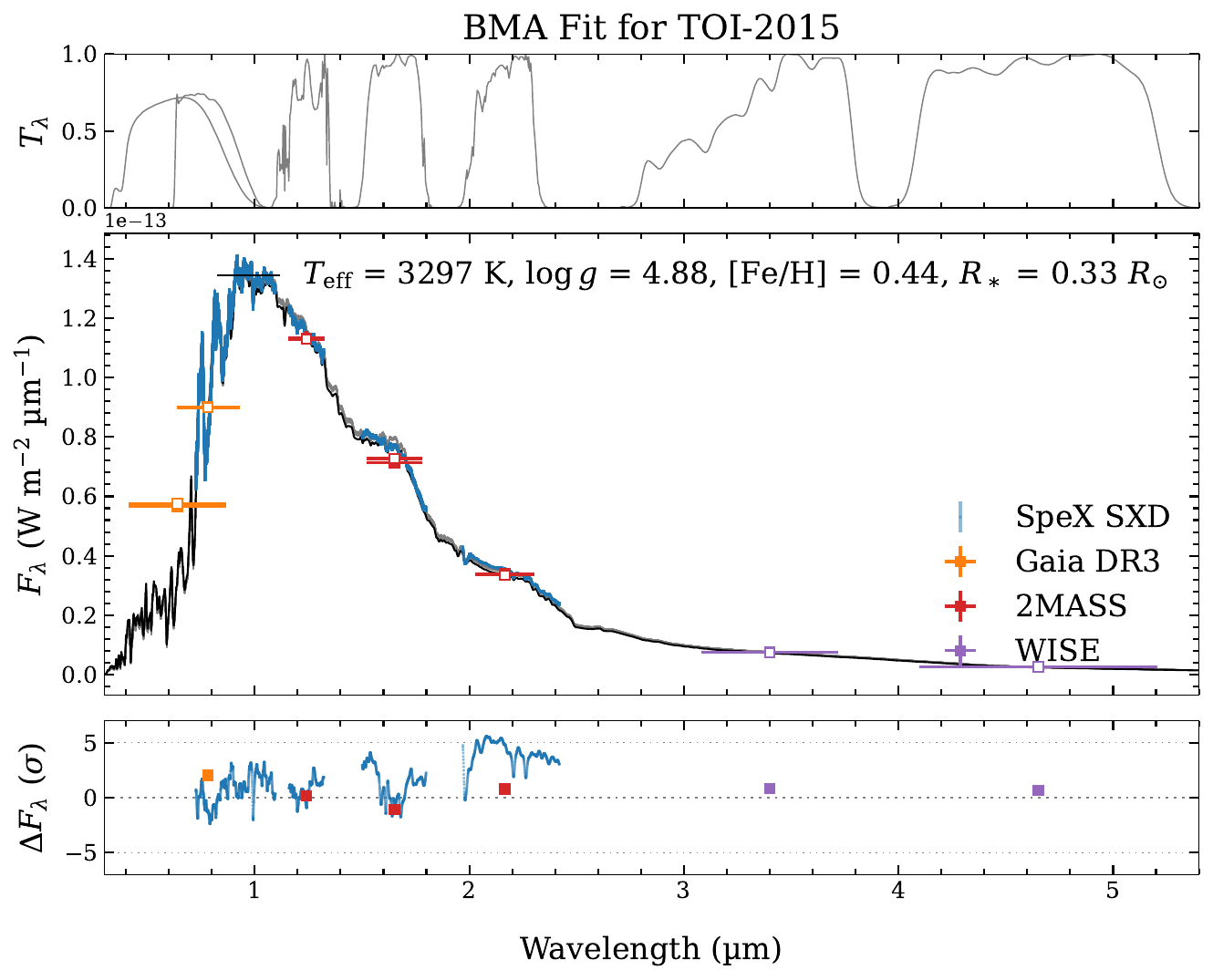}
    \caption{BMA fit for TOI-2015. The top panel shows the transmission curves of the filters considered, and the bottom panel displays the residuals. The observed SpeX SXD spectrum is shown in blue, while the black line represents the BT-Settl CIFIST synthetic spectrum using the BMA-derived stellar parameters, as shown in Table 1. The fluxes related to the photometry are shown in colors: orange for Gaia DR3, red for 2MASS, and purple for WISE.}
    \label{fig:BMAFIT}
\end{figure}

We applied a Bayesian Model Averaging (BMA) approach to combine the posterior distributions from each synthetic library. This method provides a robust estimation of stellar parameters by accounting for uncertainties between different models, yielding our final estimates for $T_\mathrm{eff}$, [Fe/H], $\log{g_\star}$, parallax, radius, extinction, and, in the case of SPHINX, the C/O ratio. The best-fitted synthetic spectrum and the residuals are shown in Figure\,\ref{fig:BMAFIT}, with the BMA values listed in Table\,\ref{stellarpar}. The final values obtained through BMA closely matched the best-fit results from the BT-Settl CIFIST library, since it is the grid with the highest evidence. All grids provide a supersolar metallicity for TOI-2015 and the largest discrepancies in the posteriors distributions are found for $T_\mathrm{eff}$ (median values from 3170\,K to 3315\,K), and therefore, for the stellar radius (median values from 0.3262 to 0.3548~$R_{\odot}$).

\begin{table*}[!]
\caption{Astrometry, photometry, and spectroscopy stellar properties of TOI-2015.}
\centering
	{\renewcommand{\arraystretch}{0.9}
     \resizebox{0.8\textwidth}{!}{
		\begin{tabular}{lcc}
			\hline
			\hline
			\multicolumn{3}{c}{  Star information}   \\
			\hline
			{\it Identifying information:}  & \\
                        & \multicolumn{2}{c}{TOI 2015}    \\
			            & \multicolumn{2}{c}{TIC 368287008}    \\
			            & \multicolumn{2}{c}{DR3 ID 1270814787767025408}  \\
			            & \multicolumn{2}{c}{2MASS J15283191+2721388}    \\
                        & \multicolumn{2}{c}{WISE J152831.86+272139.5}\\
			\hline
   			\hline
			Parameter & Value &  Source   \\
			\hline
			{\it Parallax  and distance:} &   \\
			RA [J2000]     &  15:28:31.84  &  (1) \\
			Dec [J2000]    & +27:21:39.86  &  (1)\\
			Parallax [$mas$] & $21.13 \pm 0.02$ &   (1)\\
                $\mu_{\rm RA}$ [mas\,yr$^{-1}$] & $-56.244 \pm 0.010$ & (1) \\
                $\mu_{\rm Dec}$ [mas\,yr$^{-1}$] & $63.807 \pm 0.014$ & (1) \\
			Distance [pc]  & $47.32 \pm 0.05$ & (1)\\
			\hline
			{\it Photometric properties:} & \\
			\emph{TESS}$_{\rm mag}$           &  $12.839 \pm 0.007$  & (2)  \\
			$V_{\rm mag}$ [UCAC4]       & $16.11 \pm 0.20$  & (3) \\
			$B_{\rm mag}$ [UCAC4]       & $17.13 \pm 0.10$  &  (3) \\
			$J_{\rm mag}$ [2MASS]       & $11.12 \pm 0.02$  &  (4) \\
			$H_{\rm mag}$ [2MASS]       & $10.52 \pm 0.02$  &    (4) \\
			$K_{\rm mag}$ [2MASS]       & $10.26 \pm 0.02$  &   (4) \\			
			$G_{\rm mag}$ [Gaia DR3]   & $14.176 \pm 0.001$  &  (1)  \\
			$W1_{\rm mag}$ [WISE]       & $10.098 \pm 0.022$ &  (5) \\
			$W2_{\rm mag}$ [WISE]       & $9.93 \pm 0.02$  &   (5) \\
			$W3_{\rm mag}$ [WISE]       & $9.77 \pm 0.04$  &   (5)\\
            $W4_{\rm mag}$ [WISE]       & $9.32$  &   (5)\\
			\hline
			\multicolumn{2}{l}{\it Spectroscopic and derived parameters}
\\ $T_{\rm eff}$ [K]
    & $\bf 3297_{-28}^{+14}$  &   [BAM] this work
\\  & $3211 \pm 51$  &   [\maroonx] this work
\\  & $ 3200 \pm 75$  &  [SED] this work
\\ $\log{g_\star}$ [dex]           
    & $\bf 5.04 \pm 0.04$  &  [\maroonx] this work
\\  & $5.01 \pm 0.02$  &  [SED] this work
\\ ${\rm [Fe/H]}$ [dex] 
    &  $+0.29\pm0.13$  &  [optical] this work
\\  & $\bf -0.31\pm0.16$  &  [\maroonx] this work
\\ $M_\star$  [$M_\odot$]
    & $ 0.33 \pm 0.02$  &   [SED] this work
\\  & $   0.3043 \pm 0.0067$  &     $^a$this work
\\  & $ \bf  0.3041_{-0.0312}^{+0.0353}$ &  [BMA] this work
\\ $R_\star$ [$R_\odot$]
    & $ 0.339 \pm 0.016$  &  [SED] this work
\\  & $  0.3240 \pm 0.0094$  &    $^a$this work
\\  & $ \bf  0.3273^{+0.0029}_{-0.0051}$ & [BMA] this work
\\ $F_{\rm bol}$  [erg\,cm$^{-2}$s$^{-1}$]
    & $ 1.552 \pm 0.018\times 10^{-11}$   & [SED] this work
\\ $A_V$ [mag]
    & $0.02 \pm 0.02$    & [SED] this work
\\ $L_\star$  [$L_\odot$]
    & $0.0322_{-0.0025}^{+0.0028}$  & [SED] this work
\\  $\rho_\star$  [$\rho_\odot$]
    & $8.37^{+1.03}_{-0.98}$  & [BMA] this work
\\ Age  [Gyr]
    & $\lesssim$5--6  & [H$\alpha$] this work
\\ $v\,\sin{i}$ [km\,s$^{-1}$]
    & $\lesssim$2  &  [\maroonx] this work
\\ Spectral type
    &   M4$\pm$0.5 & [NIR] this work
\\  &   M4.5$\pm$0.5 & [optical] this work
\\ \hline
	\end{tabular} }}
	 \tablefoot{ 
	{\bf (1):} Gaia EDR3 \cite{Gaia_Collaboration_2021A&A}; 
	{\bf (2)} \emph{TESS} Input Catalog \cite{Stassun_2018AJ_TESS_Catalog}; 
	{\bf (3)} UCAC4 \cite{Zacharias_2012yCat.1322};
	{\bf (4)} 2MASS \cite{Skrutskie_2006AJ_2MASS};
	{\bf (5)} WISE \cite{Cutri_2014yCat.2328}. $^a$ Stellar mass and radius values are computed from \cite{Mann_2015, Mann:2019}. Parameters in bold are the stellar parameters used in priors for photodynamical modeling presented in Section~\ref{Global_modelling}.
	}
    \label{stellarpar}
\end{table*}

\section{Planet validation} \label{Validate_planet}

\subsection{\emph{TESS} data validation}

The SPOC performed a transit search of Sector 24 on 23 May 2020 \citep{jenkins2002,jenkins2010,Jenkins2020}, which yielded a candidate with a 3.35~days period at a signal to noise ratio of $S/N = 13.6$. The TESS Science Office reviewed the vetting information and issued an alert on 19 June 2020 \citep{guerrero2021}. The transit signal was also recovered in Sectors 51 and 78, and the transit signature passed all the diagnostic tests presented in the Data Validation reports.

The transit depth found was $9753\pm735$~ppm, corresponding to a planet radius of $3.4\pm0.4$\,R$_\oplus$, and with a period of $3.34916\pm0.00001$\,days. 
A comparison of the odd and even transits' depths led to a 2-$\sigma$ agreement. The target is quite isolated and no neighboring star was included in the \emph{TESS} aperture  (see Figure\,\ref{Target_pixel}), although TOI-2015 was identified as the likely source of the events. 

According to the difference image centroiding test \citep{Twicken2018} for Sector 24, the host star is located within $5.058 \pm 2.9$~arcsec of the transit source. This result was then tightened up to $0.806 \pm 3.0$~arcsec in the sector 24-51 search. The transit source location is thus consistent with the host star.

\subsection{Ground-based photometric follow-up}

\begin{figure}[!]
	\centering
	\includegraphics[scale=0.3]{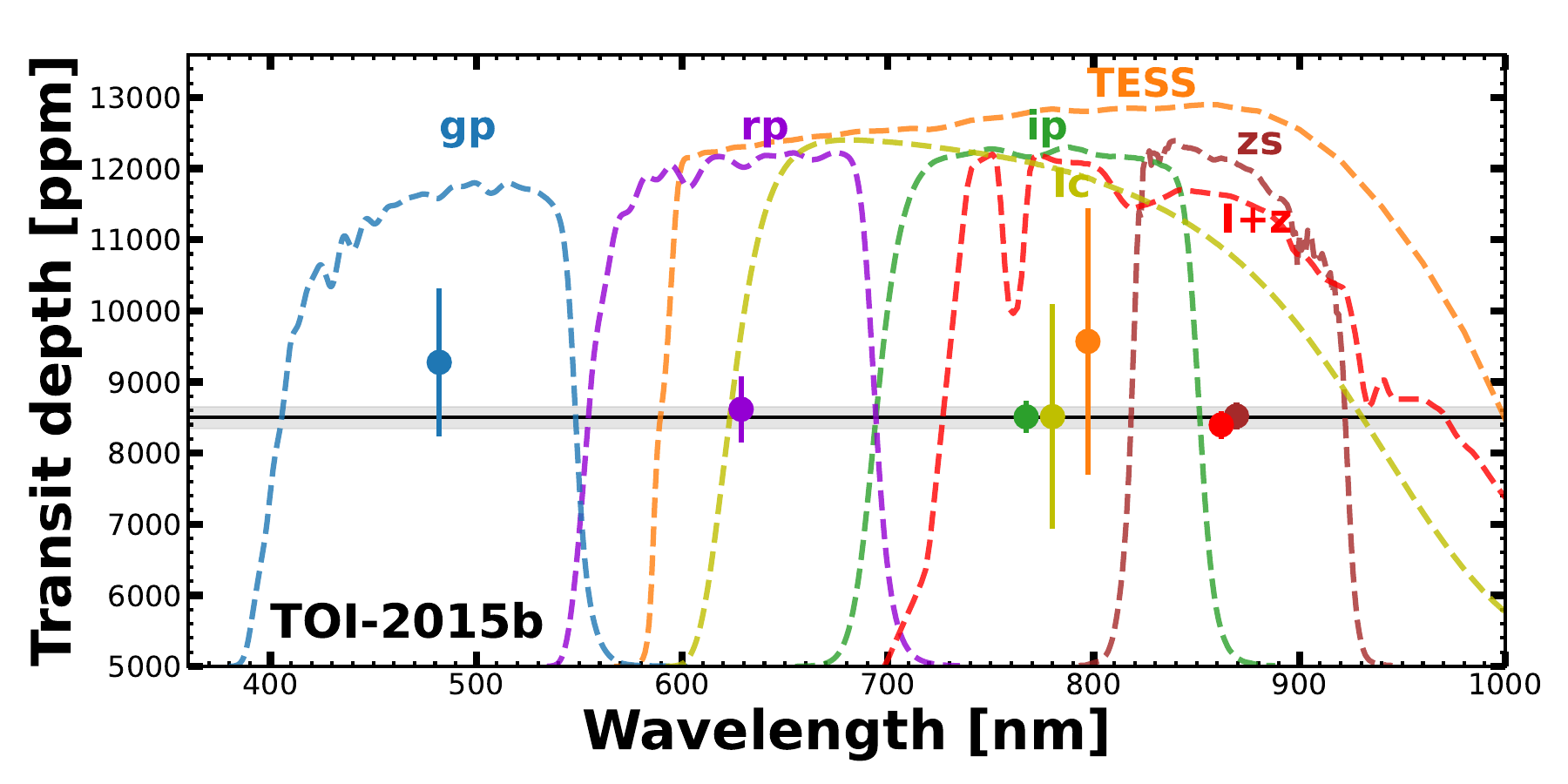}
	\caption{Measured transit depths in different bands (colored dots with error bars) obtained in the global analysis for  TOI-2015\,b. The horizontal black  line  corresponds to the depth obtained from the achromatic fit with a $1\sigma$ error bar (shaded region). All measurements agree with the common transit depth at $1\sigma$. Colored dashed lines show the transmission for each filter.}
	\label{toi2015b_depths}
\end{figure}

We used the  ground-based photometric observations to {\it i}) confirm the transit event on the target, {\it ii)} measure the transit timing variations, and {\it iii}) check for the chromaticity for the transit depth in different wavelengths. 
Two closest neighboring stars to TOI-2015  are TIC~368287010 ($T_{\rm mag} = 12.97 $, $\Delta T_{\rm mag} = 0.18 $) at $33.3\arcsec$ and TIC~368287012 ($T_{\rm mag} = 16.49 $, $\Delta T_{\rm mag} = 3.69 $) at $37.0\arcsec$.
We collected the observations in the $I+z$, Johnson-$I_c$, Sloan-$g'$, -$r'$, -$i'$ and $z_\mathrm{s}$ filter, covering a range  400 to 1000\,nm. This resulted  in a non-chromatic dependence in different bands. The measured transit depths are presented in Figure\,\ref{toi2015b_depths}.

\subsection{Archival imaging}

\begin{figure*}[!]
	\centering
	\includegraphics[scale=0.7]{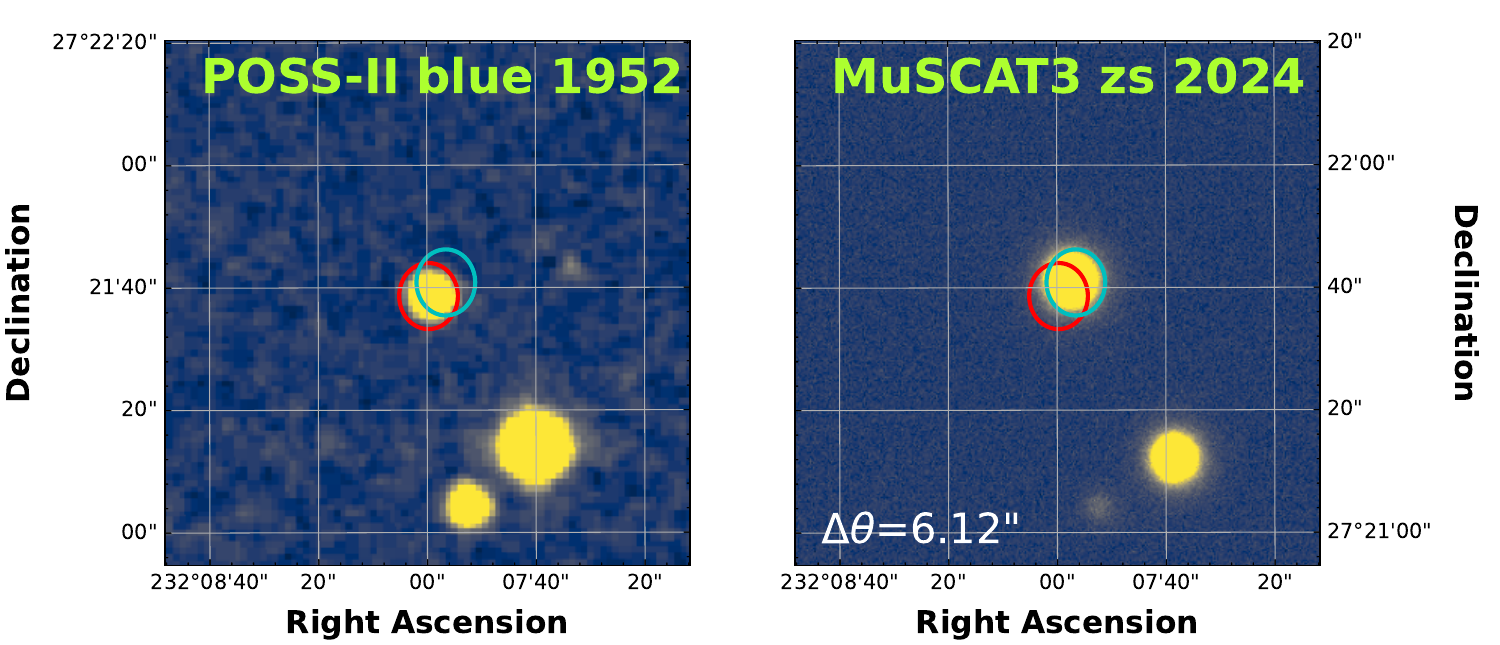}
	\caption{Evolution of the TOI-2015's position.  The {\it left panel} shows the an archival image of TOI-2015 taken using a photographic plate on the Palomar Schmidt Telescope in the blue filter. The {\it right planet} shows the \textit{z$_s$} image from LCO-HAL-2m0/MuSCAT3 taken in 2024.}
	\label{archival_images}
\end{figure*}

We used archival science images of TOI-2015 to exclude the background stellar objects that could be blended with our target in its current position. This kind of object might introduce the same transit event that we observed in our data and skew the physical properties of the system that we obtained from our photodynamical analysis.
TOI-2015 has a relatively low proper motion of 64~mas/yr. We used images from POSS-II/DSS \citep{1963POSS-I} in 1952 in the blue filter and LCO-HAL-2m0/MuSCAT3 in 2024 in the\textit{z$_s$} filter, and they span 72 years. 
The target has moved by only 6.12\arcsec\ from 1952 to 2024. There is no stellar background source at the present-day position of TOI-2015 (see Figure\,\ref{archival_images}).

\subsection{Statistical validation}

We used the {\tt TRICERATOPS}\footnote{{\tt TRICERATOPS:}~\url{https://github.com/stevengiacalone/triceratops}} (Tool for Rating Interesting Candidate Exoplanets and Reliability Analysis of Transits Originating from Proximate Stars; \citealt{Giacalone_2021AJ}) package to calculate the False Positive Probability, which allows us to identify whether a given candidate is a planet or a nearby false positive. {\tt TRICERATOPS} returns two parameters, which are the FPP (False Positive Probability) and the NFPP (Nearby False Positive Probability).
{\tt TRICERATOPS} uses the phase-folded \emph{TESS} light curves of the candidate and runs a Bayesian fit of several different possible astrophysical scenarios. It also allows us to implement high-contrast imaging observations in order to improve our results. In our case, we used the 4.1-m\,SOAR contrast curves described in Section~\ref{high-res}.
Using the \emph{TESS} light curves of TOI-2015\,b phase-folded on the orbital periods obtained from the photometric fit (see Section~\ref{Global_modelling}). Using ground-based observations (see Section~\ref{ground_based_photo}), the transit event is detected on the target, and thus we excluded other nearby sources, which means that NFPP = 0 for TOI-2015\,b. Typically, $i$) a given candidate is considered statistically validated when  FPP$<$0.015 and NFPP$<$0.001, $ii$) a given candidate is likely a planet if FPP$<$0.5 and NFPP$<$0.001, and $iii$) a given candidate is a nearby false positive if NFPP$>$0.1.
We ran {\tt TRICERATOPS} using the TOI-2015\,b phase-folded \emph{TESS} light curves and the 4.1-m SOAR high contrast imaging observations. We obtained NFPP=0 and FPP$= 0.0055 \pm 0.0011 $. TOI-2015\,b is likely a planet.

\section{Global modelling: Photometric, TTVs and RVs} \label{Global_modelling}

We performed a set of photodynamical analyses modeling the \tess data, ground-based photometric observations, and the \maroonx radial velocity measurements jointly using \pyttv \citep{Korth2023,korth2024ApJ}. The code models the photometry and the radial velocities simultaneously using \rebound \citep{Rein2012, Rein2015, Tamayo2020} for dynamical integration and \pytransit \citep{Parviainen2015, Parviainen2020a, Parviainen2020b} for transit modeling, and provides posterior densities for the model parameters estimated using MCMC sampling in a standard Bayesian parameter estimation framework.

Table\,\ref{table:pyttv_parameters} lists the model parameters and their priors. All planetary parameters, except for the $\log_{10}$ mass and radius ratio, are defined at a reference time of $t_\mathrm{ref}=2459424.785$~BJD$_{\rm TDB}$. We set uniform priors on the logarithmic masses of the two planets, with ranges designed to aid optimization without constraining the posteriors. For the inner planet's radius ratio, we apply a loosely informative prior based on the visible transit depth, while the outer planet's radius ratio is assigned a dummy normal prior, as the data cannot constrain it.
We set a normal prior on the inner planet's transit center and an uniform prior on the outer planet's mean anomaly (the planets are parameterized differently since one transits and the other does not). A loosely informative prior is set on the inner planet's impact parameter based on the transit fit, and a loosely informative normal prior is used for the outer planet's impact parameter. We do not constrain the outer planet to non-transiting geometries because we are interested in seeing the preferred solutions without informing the analysis that the outer planet does not transit. 
Additionally, we set zero-centered half-normal priors on the eccentricities, $\NP(0.0, 0.083)$, allowing for eccentric orbits while biasing against high eccentricities. Finally, we apply a normal prior on the stellar density, $\NP(13, 1.3)$~g~cm$^{-3}$, based on stellar characterization.

Since the period of the second planet is unknown, we must repeat the photodynamical analysis for all plausible period commensurability scenarios that could lead to the observed TTVs. We ignore the scenarios with an inner non-transiting planet and carry out analysis for scenarios where the other planet's orbital period is close to the 2:1, 5:3, 5:2, 4:3 and 3:2 period commensurabilities. For each scenario, we set the prior for the outer planet's period to $\NP(r \times 3.3491408, 0.02)$\,days, where $r$ is the period ratio corresponding to the scenario, and the 3.3491408~days period is derived from the linear ephemeris model.

The analysis starts with a global optimization using the differential evolution method \citep{Storn1997a, Price2005} implemented in \pytransit \citep{Parviainen2015}. The optimizer starts with a population of parameter vectors drawn from the model prior and clumps the population close to the global posterior mode. After the optimization, we use the clumped parameter vector population to initialize the \emcee sampler \citep{Foreman-Mackey2012}, which we then use for MCMC sampling to obtain a sample from the parameter posterior. Finally, we test the stability of the posterior solution by integrating a subset of posterior samples over 10,000 years.

The best-fit solutions yield differential Bayesian Information Criterion (BIC - BIC$_{5:3}$) values of 150, 0, -365, 517, and 183 for the 2:1, 5:3, 5:2, 4:3, and 3:2 scenarios, respectively. These results indicate that the BIC  favors the 5:2 period commensurability scenario, with the 5:3 scenario emerging as the second-most preferred. However, stability tests reveal that the solutions for the 5:2 and 3:2 scenarios are unstable,  leading to the ejection of one of the planets from the system within 10,000 years. Conversely, the solutions for the 2:1, 5:3, and 4:3 scenarios remain stable over the tested period. Among these, the 5:3 scenario is strongly favored over the 2:1 and 4:3 scenarios and also exhibits the smallest orbital inclination differences compared to the other configurations.

\begin{figure*}[!]
	\centering
	\includegraphics[scale=0.32]{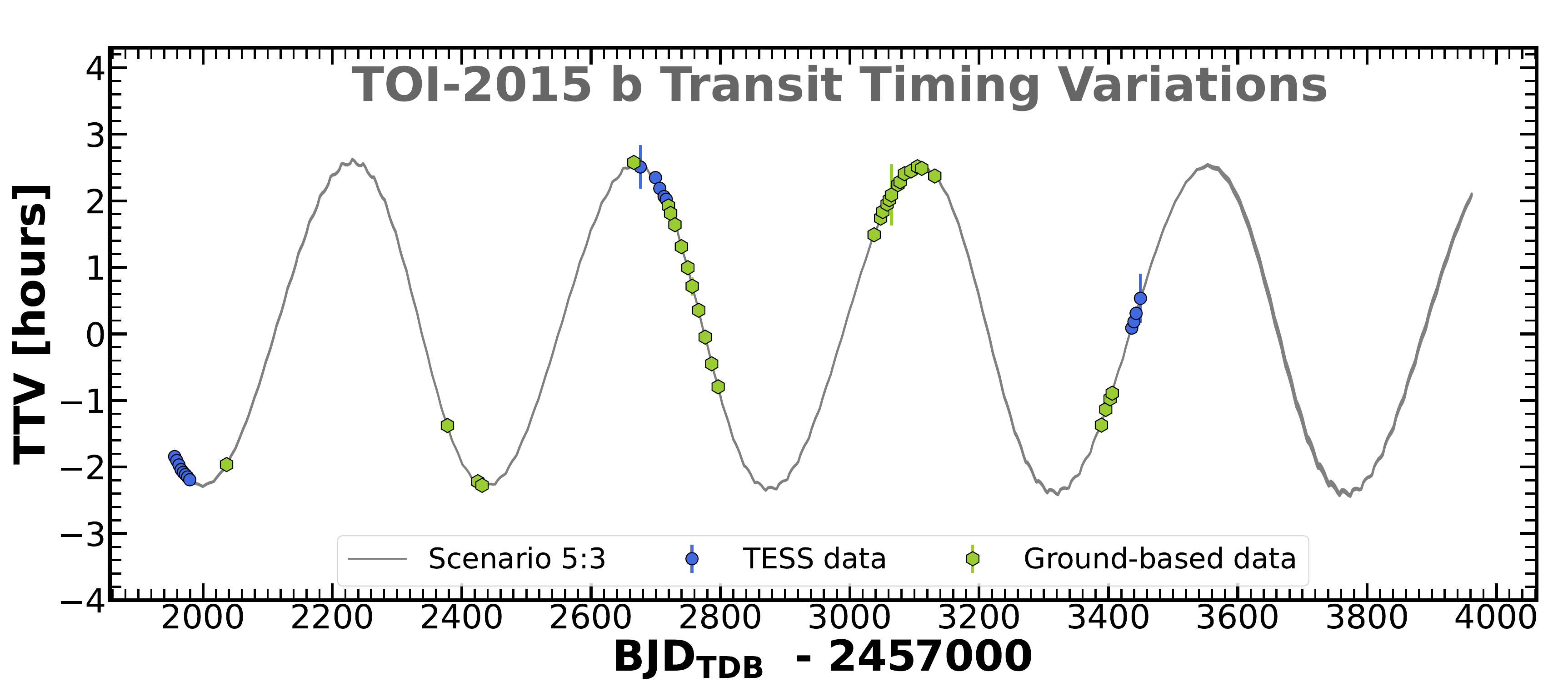}
	\caption{ Transit timing variation (TTV) measurements for TOI-2015\,b from \emph{TESS} (blue points) and ground-based (green points) facilities. The gray line is the best-fit transit-timing model for the 5:3  scenario.} 
	\label{fig_TOI2015b_ttvs}
\end{figure*}

Adopting the 5:3 scenario, we find that TOI-2015\,b is a sub-Neptune with a radius of $R_p = 3.309^{+0.013}_{-0.011}\,R_\oplus$, a mass of $M_b = 9.20^{+0.32}_{-0.36}\,M_\oplus$, and an eccentricity of $e_b = 0.0789^{+0.0018}_{-0.0016}$. The non-transiting planet, TOI-2015\,c has a mass of $M_c = 9.52^{+0.42}_{-0.36}\,M_\oplus$, an orbital period of $P_c = 5.5829$\,days, and an eccentricity of $e_c = 0.0004^{+0.0002}_{-0.0001}$ (see Table\,\ref{tois_mcmc_params_53}). Figure\,\ref{fig_TOI2015b_ttvs} shows the TTVs data with fit for the 5:3 near resonance scenario.
We also present the results for the 2:1 and 5:2 scenarios in Tables~\ref{tois_mcmc_params_21} and \ref{tois_mcmc_params_52}. \\

We also investigated the orbital stability of the 5:2, 5:3, and 2:1 MMR scenarios. To proceed, we extracted 200 configurations from each model posterior, and computed the dynamical evolution of the 3$\times$200 configurations over 300kyr. For these simulations, we used the \texttt{WHFast} integrator \citep{Rein2015WHFast} from the \texttt{rebound} software package \citep{Rein2012}, with an integration timestep of $\sim$ $1/70~P_b$ and a symplectic corrector of order 17. The number of system configurations that survived after 300kyr (no escape, no close encounter) in the 5:2, 5:3, and 2:1 MMR scenarios are 159/200, 200/200, and 172/200, respectively. From these results, the 5:3 MMR scenario contains the largest number of stable configurations. Instead of the small orbital eccentricities, this higher stability rate is largely due to the low mutual inclination of this scenario. Indeed, we also ran simulations on the 5:2MMR scenario with the hypothesis of co-planarity, and also obtained a higher stability rate. In conclusion, while the 5:3 MMR scenario is more plausible due to the larger number of stable systems, we cannot firmly exclude any of the other scenarios from pure orbital stability considerations.

\begin{table*}[!]
\caption{Physical parameters of the TOI-2015 system for the 5:3 near resonance scenario.}
	\begin{center}
		{\renewcommand{\arraystretch}{1.5}
				\resizebox{0.9\textwidth}{!}{
			\begin{tabular}{lllc}
				\hline
			        Parameter & Unit &  TOI-2015\,b  & TOI-2015\,c    \\
           \hline
            Orbital period $P$ & days             & $3.348237^{+0.000035}_{-0.000030}$ & $5.582796^{+0.000042}_{-0.000041}$ \\ 
            Transit-timing $T_0$ & BJD$_{\rm TDB}$ &  $2459424.78570^{+0.00016}_{-0.00015}$  &   --   \\
            Orbital semi-major axis $a$ & au      & $0.029322 \pm 0.000003$ &  $0.041232 \pm 0.000004$   \\
            Impact parameter $b$ & $R_\star$      & $0.74346^{+0.0029}_{-0.0030}$  &  $2.02^{+0.11}_{-0.10}$  \\
			Transit duration $W$ & hour            &  $0.9876^{+0.0029}_{-0.0034}$  &  --  \\
            eccentricity $e$     & --  & $0.0789^{+0.0018}_{-0.0016}$   & $0.00033^{+0.0003}_{-0.0002}$ \\
            $\sqrt{e}\cos(w)$    &  -- &  $ 0.2624^{+0.0047}_{-0.0042}$  &  $0.01597^{+0.0060}_{-0.0059}$ \\
            $\sqrt{e}\sin(w)$    &  -- &  $-0.1008^{+0.0012}_{-0.0082}$  &  $0.0087^{+0.0023}_{-0.0032}$ \\
            Mean anomaly  $M$    & deg &  --  &   $3.547^{+0.019}_{-0.025}$     \\
			Orbital inclination $i$ & deg         &  $87.61 \pm 0.01$ &  $85.70^{+0.27}_{-0.24}$  \\
			Radius ratio $R_p /R_\star $ & -- & $0.09278^{+0.00035}_{-0.00030}$ & -- \\
            Scaled semi-major axis $a/R_\star$          &  --   &   $19.2811^{+0.0061}_{-0.0060}$   &   $ 27.1122 \pm 0.0098 $   \\
            Planet radius $R_p$ & $R_\oplus $    &  $3.309^{+0.013}_{-0.011}$ &  -- \\
            Planet Mass $M_p$   & $M_\oplus$       &  $9.20^{+0.32}_{-0.35}$  &  $ 8.91^{+0.38}_{-0.40} $ \\
            Planet density $\rho_p$  & g/cm$^3$       & $ 1.400^{+0.052}_{-0.056} $   &   -- \\
			Planet irradiation $S_p$  & $S_\oplus$     &  $ 13.185 \pm 0.387 $ &  $6.68\pm 0.19$\\
            Equilibrium temperature $T_{\rm eq}$ & K    &  $530.9^{+2.9}_{-3.9}$ &  $448 \pm 3$   \\
            Planet surface gravity $\log (g_p [\mathrm{cm/s}^2]) $ &   -- &   $ 2.915 \pm 0.105 $   &   --   \\
            $^a$ TSM & -- & $149.4 \pm 5.6$ & -- \\
            $^a$ ESM & -- & $10.8 \pm 0.3$ & --  \\
   \hline
		\end{tabular}}}
	\end{center}
	\tablefoot{$^a$ TSM and ESM values are computed from \cite{kem}.}
	\label{tois_mcmc_params_53}
\end{table*}

\section{Independent analysis of TTV data} \label{indep_fit}


An independent analysis of the transit-timing variations of the TOI-2015\,b system was conducted using the transit-timing code \texttt{NbodyGradient} \citep{Agol2021b}. 
The fits were initialized near several mean-motion resonances with period ratios near 4:3, 3:2, 5:3, 2:1 and 5:2.  We found that each near-resonance gave fits to the transit times of similar quality, indicating that it is difficult to distinguish between these resonances based solely on transit-timing data, as found by \citet{Jones_2024}. 

However, based on these fits, we find that the eccentricities are much lower in the 5:3 scenario than in the other cases ($\sim 0.01$ vs $\sim 0.1$).  In addition, we find that the masses in the 5:3 solution are very similar to those from our RV fit to the \maroonx data with 2-planets on circular orbits, indicating that this solution is, in fact, preferred over the others. The eccentricities in this fit are held to zero for convenience, but this turns out to be consistent with the low eccentricity found in the best-fit TTV model near the 5:3 period ratio.

\section{Planet searches and detection limits from \emph{TESS}} 
\label{sec:sherlock}

We processed the available 120\,s \emph{TESS} data using the \texttt{SHERLOCK} package \citep{pozuelos2020,Demory_AA_SAINTEX_2020} to recover the original signal corresponding to TOI-2015\,b detected by SPOC and search for other potential transiting planets that might remain unnoticed due to detection thresholds. We explored orbital periods from 0.5 to 25\,d using ten detrended scenarios corresponding to window sizes ranging from 0.1 to 1.3\,d; we refer the reader to \citet{pozuelos2023} for further details about different searching strategies and to \cite{devora2024} for a comprehensive update of all the \texttt{SHERLOCK}'s capabilities. 

In the first run, we found the signal corresponding to the TOI-2015\,b, which allowed us to confirm the detectability of this candidate independently. In the subsequent runs, \texttt{SHERLOCK} did not find any interesting detection, all the signals found being either attributable to intrinsic noise in the light curve not fully decorrelated by our detrending or to spurious detections.  

\begin{figure}[!]
	\centering
	\includegraphics[width=\columnwidth, keepaspectratio]{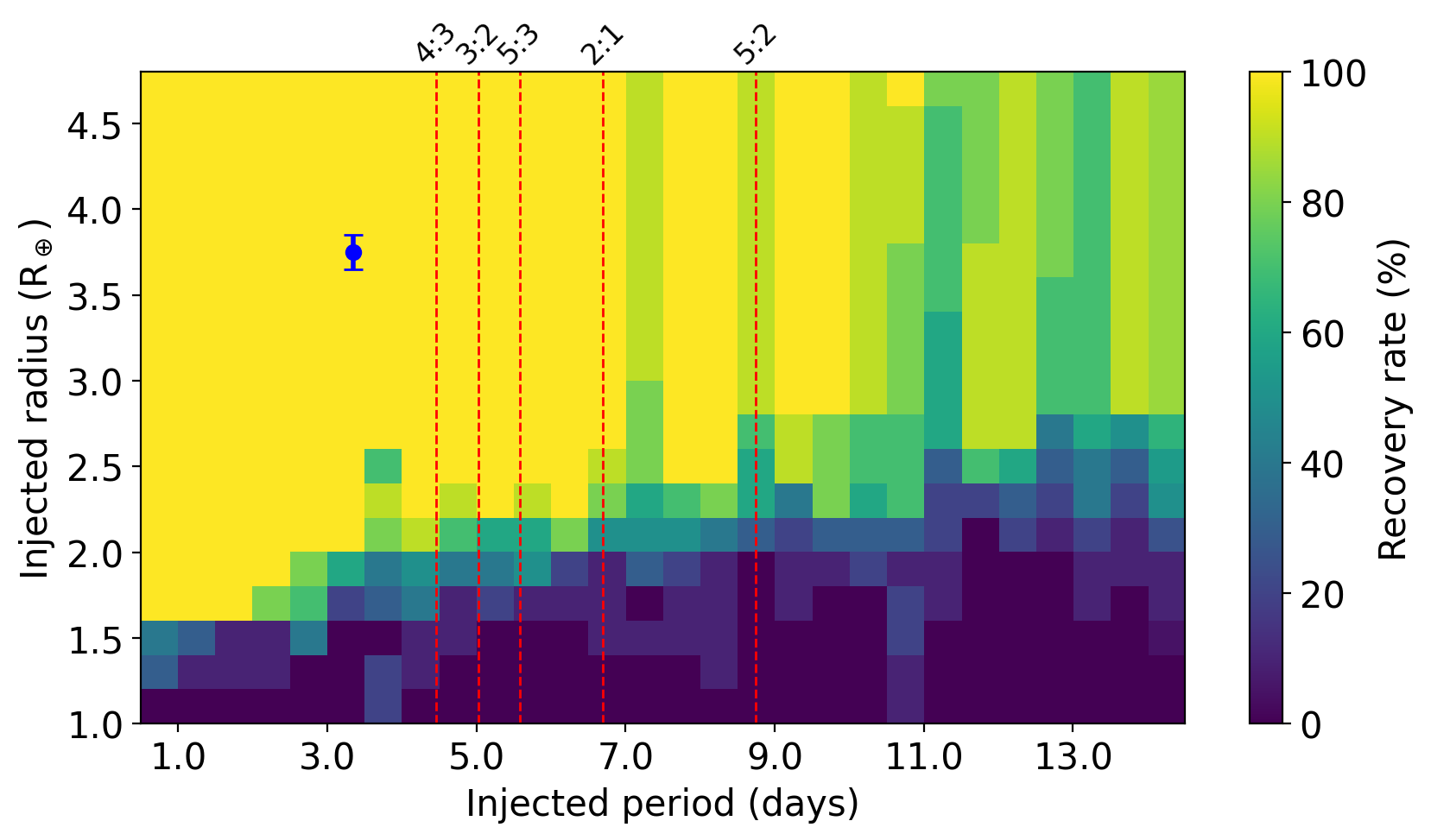}
 	\caption{Injection-and-recovery test performed to check the detectability of additional planets in the TOI-2015 system. We explored a total of 6000 different scenarios with the orbital period ranging from 0.5 to 15\,days and radius from 1 to 5\,R$_{\oplus}$. Larger recovery rates are presented in green and yellow colors, while lower recovery rates are shown in blue and darker hues. Planets larger than 3.0\,R$_{\oplus}$ with $P \lesssim 10$\,d would be easily detectable. Planets below $1.5\,R_\oplus$ would be undetectable, with recovery rates lower than 20\% in all cases. The dashed red lines show the MMR possibilities for the non-transiting planet. The blue dot refers to the planet TOI-2015\,b.} 
	\label{fig:matrix}
\end{figure}

Once we explored the data in the search for extra transiting planets, and since we know that there is at least a second planet in the system, we wanted to establish detection limits with the current data set. To this end we employed the \texttt{MATRIX} package \citep[see, e.g.,][]{matrix,Delrez_2022_A&A}, which injects synthetic planets into the data using a range of orbital periods, planetary radii, and orbital phases, and tries to recover them, mimicking the procedure conducted by \texttt{SHERLOCK}. The purpose of this analysis is to determine the range of planetary sizes and orbital periods that could be reliably detected in our data set. In particular, we generated 6000 scenarios with the orbital period ranging from 0.5 to 15\,d with steps of 0.5\,d, radius from 1 to 5\,R$_{\oplus}$ with steps of 0.2\,R$_{\oplus}$. Each radius--period pair was evaluated at ten orbital phases. The results are shown in the Figure\,\ref{fig:matrix}. We found that the planet corresponding to TOI-2015\,b falls in a region with a 100\% recovery rate. Moreover, we found that any planet larger than 3.0\,R$_{\oplus}$ would be easily detectable, with recovery rates higher than 70\% at any orbital period explored in this study. In contrast, planets with radii below 1.5\,R$_{\oplus}$ would be undetectable, with recovery rates lower than 20\% in all cases. Planets with radii between 1.5 and 3.0\,$R_{\oplus}$ represent the transition region with recovery rates from 50 to 100\%, with the shorter orbital period being the easier to detect and vice-versa. While the favored solution from the global fit presented in Section~\ref{Global_modelling} corresponds to the MMR 5:3 between planets TOI-2015\,b, and c, still other solutions might be possible, such as 4:3, 3:2, 2:1 and 5:2 (see Section~\ref{indep_fit}). We highlighted all these possibilities in Figure\,\ref{fig:matrix}, and we conclude that in all the cases, any transiting planet larger than 2.0\,R$_{\oplus}$ should be easily detectable, while smaller transiting planets might remain undetected in the current data set, and hence we cannot entirely rule out the transiting nature of planet~c. 

\section{Discussion} \label{discuss_results}


\subsection{Mass-radius and composition}
\label{sec:mr_comp}

\begin{figure}[!]
	\centering
	\includegraphics[scale=0.18]{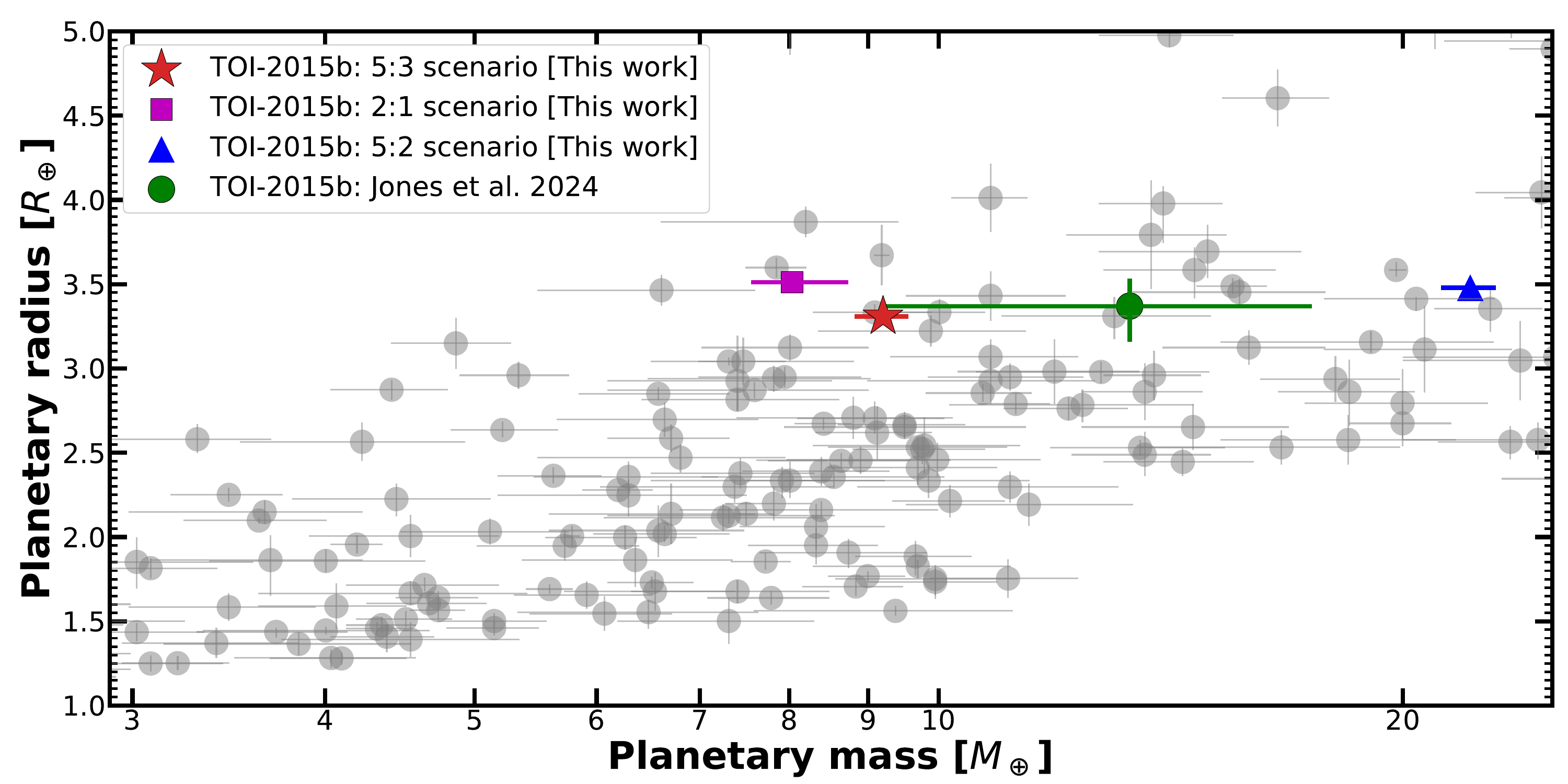}
    \includegraphics[scale=0.18]{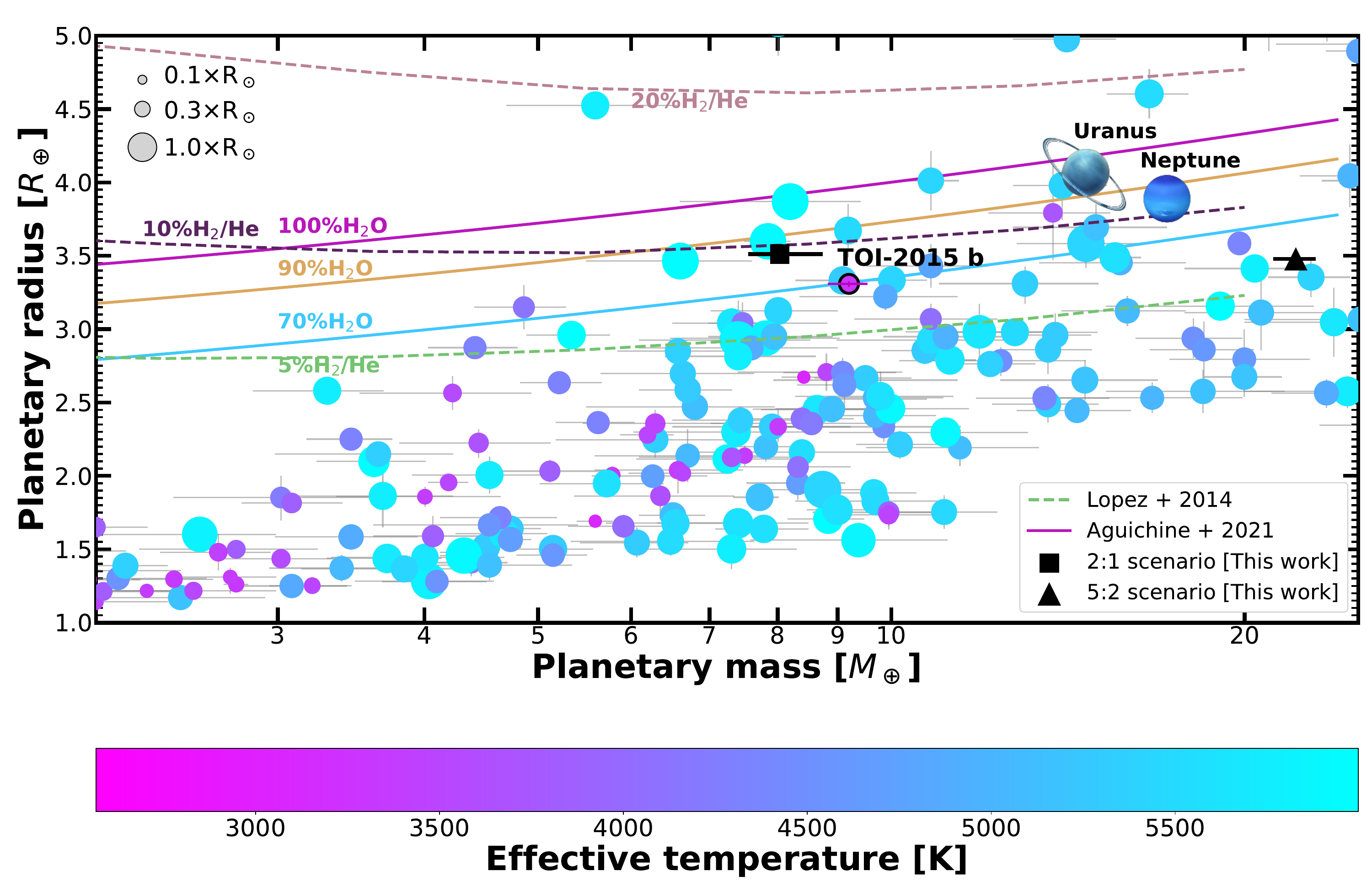}
	\caption{Planetary radius as a function of the planetary mass of known transiting exoplanets well characterized, with radius and mass precisions better than 10\% and 20\%, respectively. Data are extracted from {\tt TEPCat} \citep{Southworth_2011MNRAS}. {\it Top panel} shows the comparison between the planetary parameters for TOI-2015\,b. The red star shows our updated measurements, while the green dot with error bars the planetary parameters derived by \cite{Jones_2024}. We also highlighted the planetary parameters for other scenarios 2:1 (magenta square) and 5:2 (blue triangle).
    {\it Bottom panel} shows the comparison between TOI-2015\,b to other transiting planetary systems.
    The systems are colored according to the stellar effective temperature. The size of the points is scaled according to the stellar radius. Dashed lines show the mass-radius composition models from \cite{Lopez_2014}.  We display mass-radius curves for hydrogen-helium compositions of 5\%, 10\% and 20\% H$_2$/He. We assumed a planet with an incident flux of  $S_p=10\,S_\oplus$, and an age of $>1$~Gyr. The solid lines present the mass-radius composition models from  \cite{Aguichine_2021ApJ}. We display mass-radius curves for water-rich compositions of 70\%, 90\% and 100\%~H$_2$O. We assumed a planetary equilibrium temperature of $T_{\rm eq} = 500K$ and a core mass fraction of $x_{\rm core} = 0.3$. We also highlighted the planetary parameters for other scenarios 2:1 (black square) and 5:2 (black triangle). Two solar system planets (Uranus and Neptune) are also displayed.} 
	\label{Rp_Mp_diagram}
\end{figure}

Analyzing the observations from the \emph{TESS} mission together with ground-based photometry and radial velocity measurements collected with the \maroonx\ spectrograph, we confirm the planetary nature of the transiting planet TOI-2015\,b around its M4-dwarf host star.
We find that the transiting planet TOI-2015\,b has a radius and a mass of $R_p = 3.309^{+0.013}_{-0.011}\,R_\oplus$ and $M_p = 9.20^{+0.32}_{-0.36}\,M_\oplus$, respectively. This results in a mean density of $\rho_p = 1.400^{+0.052}_{-0.056}$~g~cm$^{-3}$,  indicative of a Neptune-like composition.
We present a comparative analysis of the mass and radius of TOI-2015\,b with other transiting exoplanets, and composition models from \cite{Aguichine_2021ApJ} and \cite{Lopez_2014}.
\cite{Aguichine_2021ApJ}'s models considered the mass-radius relation inferred from the water-rich composition models. These models assume a H$_2$O-dominated atmosphere on the top of a high-pressure water layer.
We computed the mass-radius relationship  from 
\cite{Aguichine_2021ApJ}, for different water fractions, and for a planetary equilibrium temperature of $T_{\rm eq} = 500K$ (i.e.\ the equilibrium temperature closest to the one for TOI-2015\,b). We assumed a core mass fraction of $x_{\rm core} = 0.3$ (i.e. Earth-like interior composition). 
\begin{equation}
    \log_{10}(R_p) = a\log_{10}(M_p) + \exp[-d\times(\log_{10}(M_p)+c)] + b,
\end{equation}
where $R_p$ is the planet radius in $R_\oplus$, $M_p$ is the planet mass in $M_\oplus$, and $a$, $b$, $c$ and $d$ are coefficients obtained by \citet{Aguichine_2021ApJ} fits. 
The results are shown in Figure\,\ref{Rp_Mp_diagram} (colored solid lines). Based on this preliminary comparison, TOI-2015\,b is compatible with a high water mass fraction of 70\%.

We also performed a similar analysis assuming a composition based on a rocky core surrounded by a hydrogen and helium envelope \citep{Lopez_2014}. These models are generated for planets with masses from 1 to 20~$M_\oplus$, ages from 100~Myr to 10~Gyr, incident flux from 0.1 to 1000~$F_\oplus$, and envelope fractions from 0.01 to 20\%. We used the tabulated mass-radius relations from the interior models of \citet{Lopez_2014} for different hydrogen-helium (H$_2$/He) fractions, assuming a planet incident flux of $S_p = 10\,S_\oplus$ (i.e. incident flux closest to the one for TOI-2015\,b) and systems older than 1~Gyr. Figure\,\ref{Rp_Mp_diagram} shows the mass-radius models (colored dashed lines). In this case, we find that the properties of TOI-2015\,b are consistent with a gaseous envelope with a mass fraction $f_{\rm env}\approx$10$\%$ of the mass of the planet. 

Our results show that models with water-rich and hydrogen-helium envelopes provide equally good matches to the planet density. A future measurement of the transmission spectrum of TOI-2015\,b with \emph{JWST} might help break this degeneracy by providing a direct atmospheric composition of the upper layers of the outer envelope of the planet's atmosphere.

\subsection{Location in the period-radius diagram}

\begin{figure}[!]
	\centering
	\includegraphics[scale=0.23]{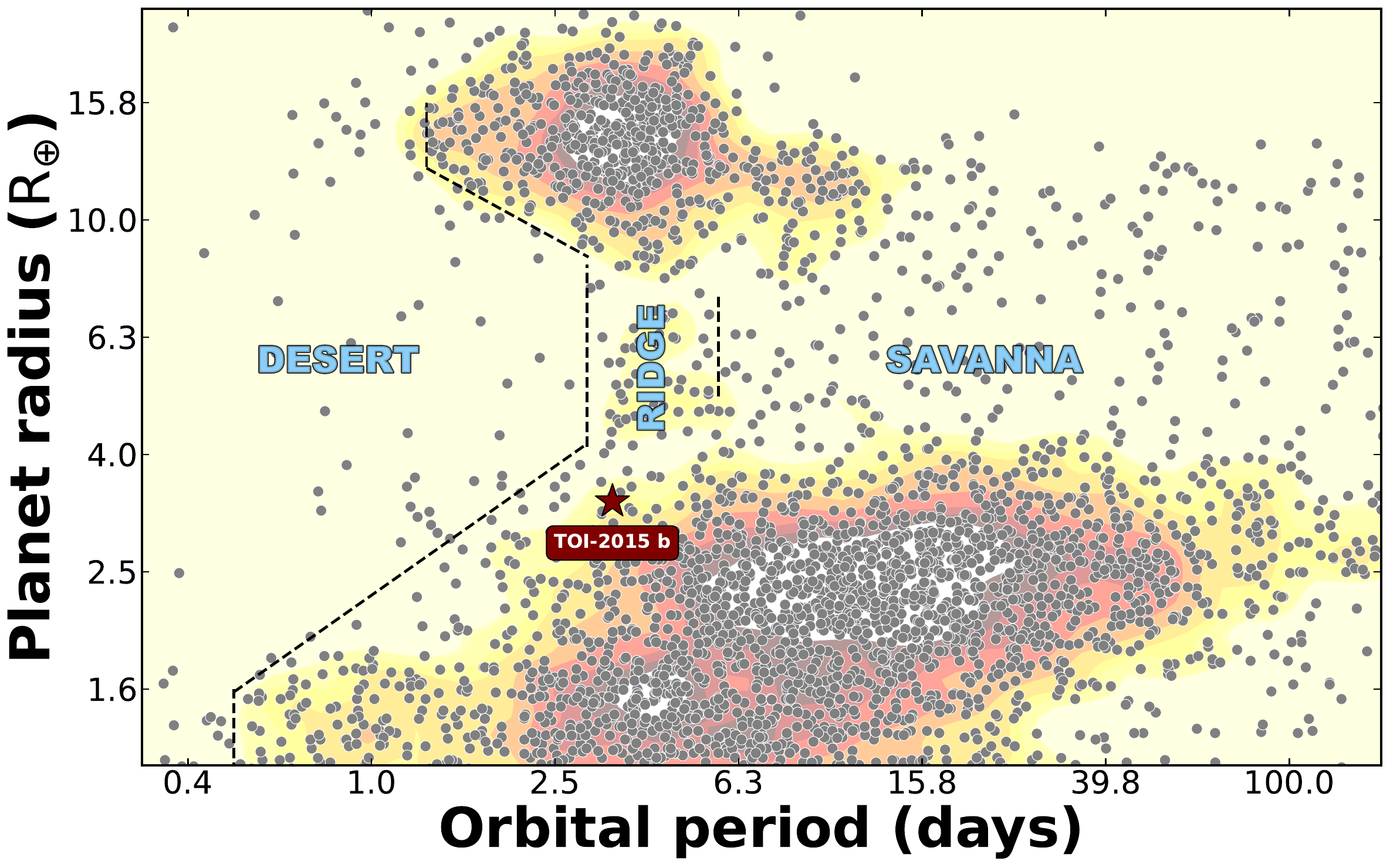}
	\caption{Planetary radius versus orbital period diagram of known transiting exoplanets. Data are extracted from  {\tt NASA Archive of Exoplanets.}  The location of the Neptunian desert, ridge, and savanna derived by \citet{Gonzalez_2024AA} are highlighted. TOI-2015\,b is well placed in the Neptunian-ridge region. This plot is made using {\tt nep-des} (\url{https://github.com/castro-gzlz/nep-des}).} 
	\label{Rp_Period_diagrams}
\end{figure}

Based on our photodynamical analysis of the system, TOI-2015\,b has a period of $P_b = 3.35$\,days. 
In Figure\,\ref{Rp_Period_diagrams}, we plotted the planet radius as a function of the orbital period of transiting exoplanets, along with the boundaries of the Neptune-desert.
TOI-2015\,b is located within a region between the "Neptunian-desert" and "savanna", defined as the "Neptunian-ridge" defined by \cite{Gonzalez_2024AA}. This work suggests that the evolutionary mechanisms that bring planets to the Neptunian ridge might be similar to those that bring larger exoplanets to the hot-Jupiter ($\simeq3-5$~days) region.
TOI-2015\,b has a low eccentricity of $e_b = 0.0770^{+0.0016}_{-0.0018}$. This eccentricity is not high enough to start the HEM (high-eccentricity migration)\citep{Fortney_2021JGRE,Bourrier_2023AA}.

To account for the large-amplitude transit timing variations of TOI-2015\,b, we identify an outer non-transiting companion likely to be in a 5:3 near resonance motion, TOI-2015\,c. Our TTV fit 
(Figure\,\ref{fig_TOI2015b_ttvs}) implies an orbital period of $P_c = 5.582904^{+0.000044}_{-0.000043}$\,days, a mass of $M_c = 8.91^{+0.38}_{-0.40}\,M_\oplus$, and an  eccentricity of $e_c = 0.00033^{+0.0003}_{-0.0002}$. The presence of this companion in the system might be partly responsible for inward migration of TOI-2015\,b.
The location of TOI-2015\,b in the radius-period diagram (Figure\,\ref{Rp_Period_diagrams}), together with the presence of the companion, and future  spin-orbit angle measurements will provide us some more insights on the formation and evolution history of the system.

\subsection{Potential for atmospheric characterization for TOI-2015\,b}

\begin{figure}[!]
	\centering
	\includegraphics[scale=0.22]{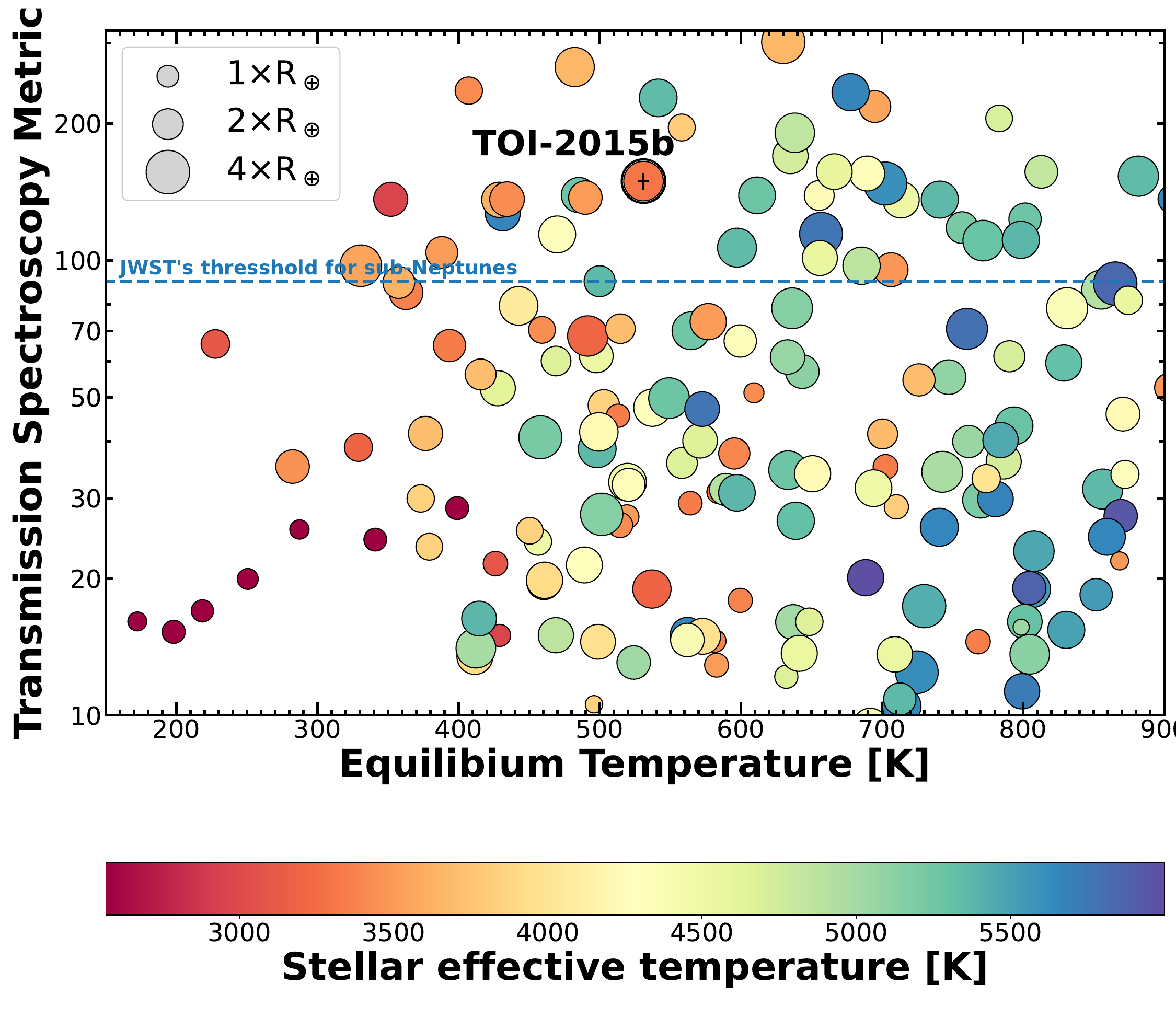}
	\caption{Feasibility of TOI-2015\,b for the transmission spectroscopy studies. Transmission Spectroscopy Metric (TSM; \cite{kem}) as a function of the planetary equilibrium temperature of all known transiting exoplanets with mass measurement. Data are extracted from {\tt NASA Exoplanets Archive.} The size of the points scales according to the planetary radius. The points are colored according to the stellar effective temperature. TOI-2015\,b is highlighted by the black circle and error bars.} 
	\label{TSM_diagram}
\end{figure}

To quantify the suitability of transiting exoplanets for atmospheric characterization through transmission spectroscopy, we used the Transmission Spectroscopy Metric (TSM) derived by \cite{kem}. By combining the planetary radius, mass and equilibrium temperature, and the infrared stellar brightness, we find that TOI-2015\,b has a TSM of $149.4\pm 5.6$. Figure\,\ref{TSM_diagram} plots the TSM as a function of the planetary equilibrium temperature for transiting exoplanets with mass measurements. This shows that TOI-2015\,b is one of the most best sub-Neptune planets for atmospheric exploration with \emph{JWST}.

We further explored the potential of TOI-2015\,b for transmission spectroscopy with \emph{JWST} through spectral simulations for a suite of atmospheric scenarios. We adopted \texttt{TauREx 3} \citep{Al-refaie2021} to simulate the synthetic transmission spectra. TOI-2015\,b could retain an H/He-dominated and/or a water-rich atmosphere (Sect\,\ref{sec:mr_comp}). We modeled H/He atmospheres with 1$\times$ and 100$\times$ scaled solar abundances using the atmospheric chemical equilibrium module of \citet{Agundez2012}, including collision-induced absorption by H$_2$–H$_2$ and H$_2$–He \citep{Abel2011,Abel2012,Fletcher2018}. For each chemical setup, we considered the cases of clear and hazy atmospheres. The haze was modeled using Mie scattering with the formalism of \cite{Lee2013}, assuming the same haze parameters as in previous studies \citep{Orell-miquel2023,Goffo2024}. 
We note that a super-solar metallicity is typically expected for sub-Neptune-sized planets (e.g., \citealp{Fortney2013,Thorngren2016}), while the planetary equilibrium temperature ($T_{\rm eq}$) within $T_{\rm eq}\approx$400--600\,K also points to a high degree of haziness due to inefficient haze removal \citep{Gao2020,Ohno2021,Yu2021}.
Moreover, we modeled the case of a pure H$_2$O atmosphere.

The \texttt{ExoTETHyS} \citep{Morello2021} package has been used to simulate the corresponding \textit{JWST} spectra with the NIRISS-SOSS ($\lambda=$[0.6\,$\mu$m--2.8\,$\mu$m]), NIRSpec-G395H ($\lambda=$[2.88\,$\mu$m--5.20\,$\mu$m]), and MIRI-LRS ($\lambda=$[5\,$\mu$m--12\,$\mu$m]) instrumental modes. The \texttt{ExoTETHyS} code has been cross-validated against the Exoplanet Characterization Toolkit (ExoCTK, \citealp{Bourque2021}) and \texttt{PandExo} \citep{Batalha2017} in a series of previous studies \citep{Murgas2021,Espinoza2022,Luque2022a,Luque2022b,Chaturvedi2022,Lillo-box2023,Orell-miquel2023,Palle2023,Goffo2024}. We conservatively increased the uncertainty estimates by 20\%. We considered wavelength bins with a spectral resolution of $R\sim$100 for NIRISS and NIRSpec, and a constant bin size of 0.25\,$\mu$m for MIRI-LRS observations, following the recommendations from recent \textit{JWST} Early Release Science papers \citep{Carter2024,Powell2024}.

\begin{figure}
\centering
\includegraphics[width=0.49\textwidth]{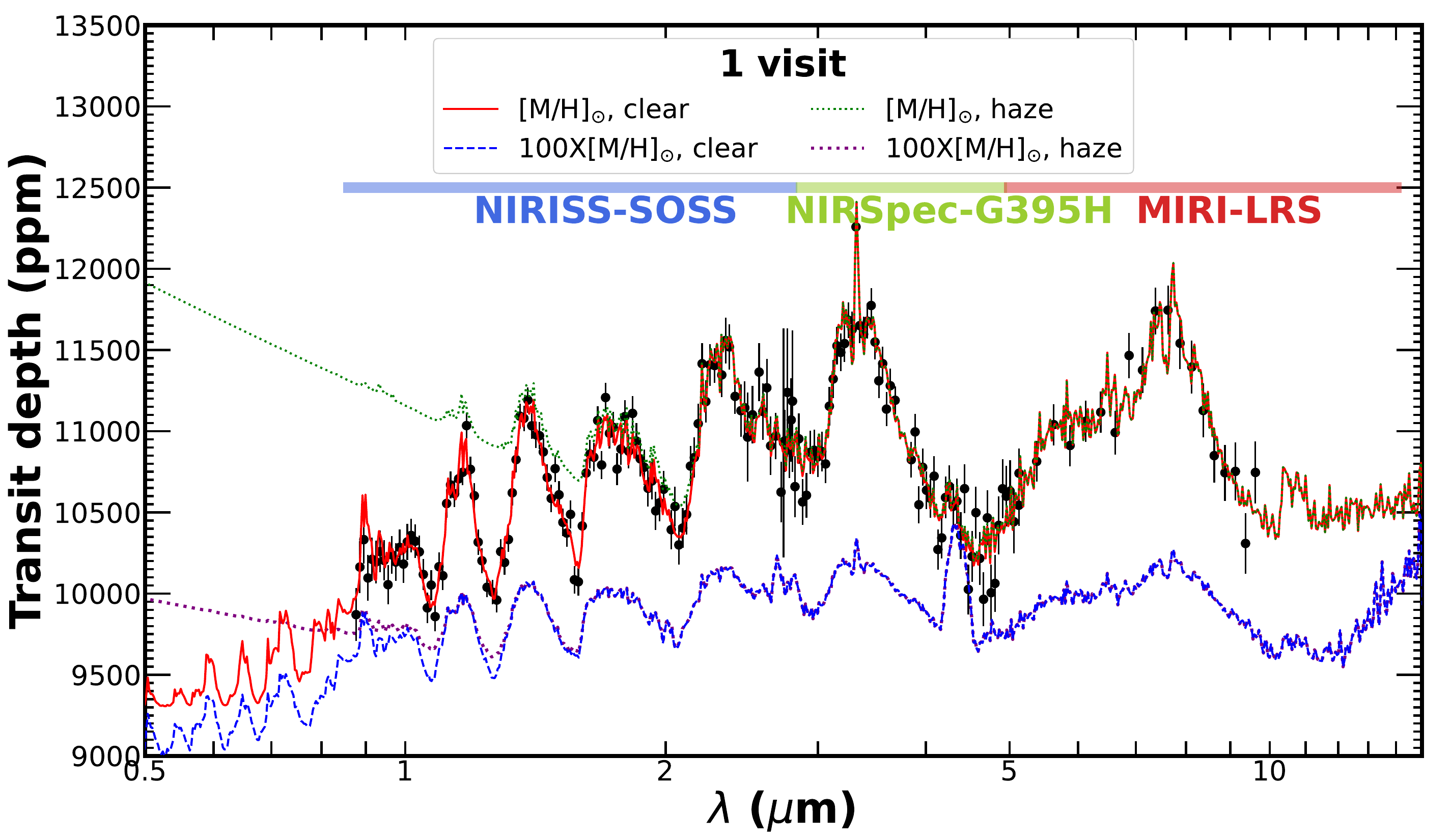}\\
\includegraphics[width=0.49\textwidth]{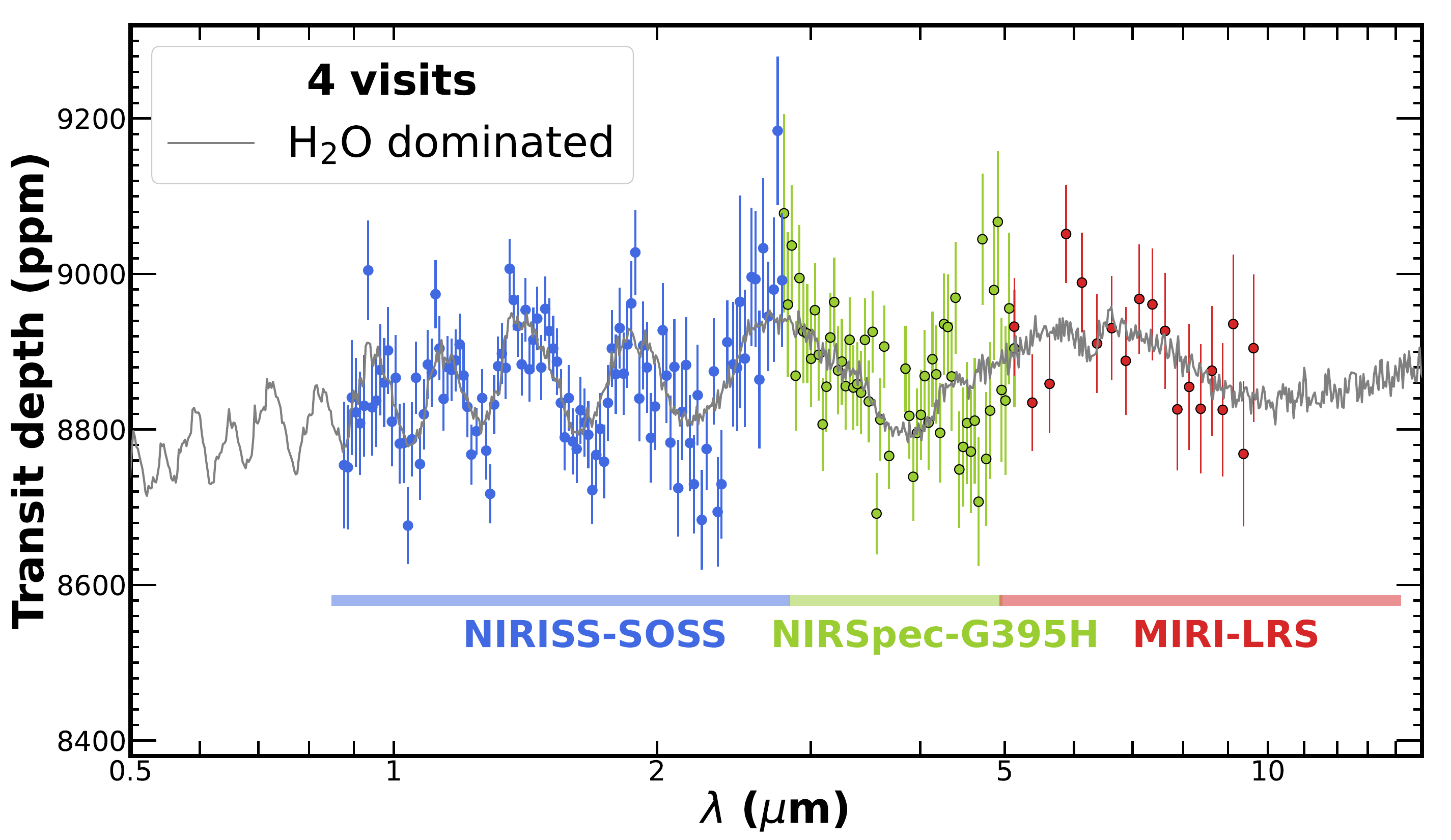}
\caption{Simulated synthetic atmospheric transmission spectra of TOI-2015\,b. \textit{Top:} fiducial models for clear or hazy H/He atmospheres with scaled solar abundances. \textit{Bottom:} model for a steam H$_2$O atmosphere. 
Simulated measurements with error bars are shown for the observation of one (\textit{top}) or four (\textit{bottom}) transits with \textit{JWST} NIRISS-SOSS, NIRSpec-G395H, and MIRI-LRS configurations. \label{fig:jwst_atmo}
        }
\end{figure}

Figure\,\ref{fig:jwst_atmo} shows the synthetic transmission spectra for the atmospheric configurations described above. The H/He model atmospheres exhibit strong H$_2$O and CH$_4$ absorption features of $\gtrsim$100--1000 ppm (parts per million), depending on metallicity and haze, while the steam H$_2$O atmosphere has absorption features $\lesssim$100 ppm. The predicted error bars for a single transit observation are 60--273\,ppm (mean error 110 ppm) for NIRISS-SOSS, 86--256 ppm (mean error 123 ppm) for NIRSpec-G395H, and 125--190 ppm (mean error 144 ppm) for MIRI-LRS. Based on our simulations, a single transit observations with NIRSpec-G395H or NIRISS-SOSS is well suited to detect an H/He atmosphere. While at least four transit observations may be required to reveal features in the case of a steam atmosphere.

\section{Conclusion}
\label{conclusion}
In this paper, we present a photodynamical analysis of the TOI-2015 exoplanetary system including the \emph{TESS} and ground-based photometry, together with radial velocity measurements obtained with the \maroonx\ spectrograph (Section~\ref{sec:maroonx_obs}). 
To account for large-amplitude transit timing variations,
the gravitational interaction between the planets is also included in our analysis.
The host star is well characterized using optical and near-infrared spectroscopic observations obtained by \maroonx, Shane/KAST and Magellan/FIRE spectrographs, together with SED (section~\ref{SED_evol_analys}), BMA (section~\ref{BMA}) and stellar evolutionary models.
The target was observed by the \emph{TESS} mission in Sectors 24, 51 and 78 with a 2-min cadence. The intensive ground-based photometric observations were collected using SPECULOOS-North/Artemis-1m0, SAINT-EX-1m0, MuSCAT, MuSCAT2, LCOGT-2m0/MuSCAT3, LCOGT-1m0, TRAPPIST-North/-South-0.6m, TUBITAK-1m0, IAC80, OSN-1.5m and CAHA-1.23m telescopes (Section~\ref{ground_based_photo}). 
The photodynamical analysis was performed using the \pyttv pipeline developed by \cite{Korth2023,korth2024ApJ} (Section~\ref{Global_modelling}), and an independent analysis (Section~\ref{indep_fit}) was performed
the {\tt NbodyGradient} code \citep{Agol2021b}.
TOI-2015 is a $K=10.3$\,mag M4-type dwarf with a sub-solar metallicity of [Fe/H]\,$=-0.31\pm 0.16$, an effective temperature of $T_{\rm eff} = 3211\pm 51$\,K, and a surface gravity of $\log{g_\star}=5.04 \pm 0.04$~(cgs).

The TOI-2015 system is composed of two sub-Neptune exoplanets. The inner planet, TOI-2015\,b is a transiting sub-Neptune at $P_b = 3.35$~days, with  a radius of $R_p = 3.31\pm 0.01~R_\oplus$ and a mass of $M_p = 9.86\pm 0.37~M_\oplus$. 
To place TOI-2015\,b in context within the exoplanet population, we plot the radius-mass diagram (Figure\,\ref{Rp_Mp_diagram}) for transit exoplanets with mass measurements.  As a preliminary comparison, the planet has a degeneracy in its interior composition based on the mass-radius composition models from \cite{Lopez_2014}  and \cite{Aguichine_2021ApJ}. The measurement of a transmission spectrum from \emph{JWST} can help to break this degeneracy by providing the planet's atmospheric composition.

The transit events of TOI-2015\,b exhibit large transiting timing variations indicative of an outer companion in the system, TOI-2015\,c. Our best-fit solution places the planet near a  5:3 resonance with an orbital period of $P_c = 5.58$~days, a mass of $M_p = 9.20^{+0.32}_{-0.35}~M_\oplus$, and an eccentricity of $e_b = 0.0789^{+0.0018}_{-0.0016}$. Other scenarios for TOI-2015\,c might be possible, such as the 2:1, 5:2, 3:2 and 4:3 resonances. More photometric data and radial velocity measurements are required in order to break this degeneracy on the orbital period of TOI-2015\,c.

\section{Data availability}
The \emph{TESS} photometric observations that we used in this work are available via the Mikulski Archive for Space Telescopes (MAST) and the ExoFOP-\emph{TESS} platform. All ground-based photometric observations are also available via the ExoFOP-\emph{TESS} platform.
Radial velocity measurements collected with the \maroonx\ spectrograph are available in Table\,\ref{table_Maroonx_RVs_TOI2015}.

\bibliographystyle{aa}
\bibliography{aa.bib}

\onecolumn
\begin{appendix}

\section*{Acknowledgments}
The postdoctoral fellowship of KB is funded by F.R.S.-FNRS grant T.0109.20 and by the Francqui Foundation.
MG and EJ are F.R.S.-FNRS Research Directors. 
This publication benefits from the support of the French Community of Belgium in the context of the FRIA Doctoral Grant awarded to MT.
Funding for the TESS mission is provided by NASA's Science Mission Directorate. KAC and CNW acknowledge support from the TESS mission via subaward s3449 from MIT.
This paper made use of data collected by the TESS mission, obtained from the Mikulski Archive for Space Telescopes MAST data archive at the Space Telescope Science Institute (STScI). Funding for the TESS mission is provided by the NASA Explorer Program. STScI is operated by the Association of Universities for Research in Astronomy, Inc., under NASA contract NAS 5–26555. We acknowledge the use of public TESS data from pipelines at the TESS Science Office and at the TESS Science Processing Operations Center.  Resources supporting this work were provided by the NASA High-End Computing (HEC) Program through the NASA Advanced Supercomputing (NAS) Division at Ames Research Center for the production of the SPOC data products.
This research has made use of the Exoplanet Follow-up Observation Program (ExoFOP; DOI: 10.26134/ExoFOP5) website, which is operated by the California Institute of Technology, under contract with the National Aeronautics and Space Administration under the Exoplanet Exploration Program.
J.d.W. and MIT gratefully acknowledge financial support from the Heising-Simons Foundation, Dr. and Mrs. Colin Masson and Dr. Peter A. Gilman for Artemis, the first telescope of the SPECULOOS network situated in Tenerife, Spain. The ULiege's contribution to SPECULOOS has received funding from the European Research Council under the European Union's Seventh Framework Programme (FP/2007-2013) (grant Agreement n$^\circ$ 336480/SPECULOOS), from the Balzan Prize and Francqui Foundations, from the Belgian Scientific Research Foundation (F.R.S.-FNRS; grant n$^\circ$ T.0109.20), from the University of Liege, and from the ARC grant for Concerted Research Actions financed by the Wallonia-Brussels Federation. 
The research leading to these results has received funding from  the ARC grant for Concerted Research Actions, financed by the Wallonia-Brussels Federation. TRAPPIST is funded by the Belgian Fund for Scientific Research (Fond National de la Recherche Scientifique, FNRS) under the grant PDR T.0120.21. TRAPPIST-North is a project funded by the University of Liege (Belgium), in collaboration with Cadi Ayyad University of Marrakech (Morocco).
The ULiege's contribution to SPECULOOS has received funding from the European Research Council under the European Union's Seventh Framework Programme (FP/2007-2013) (grant Agreement n$^\circ$ 336480/SPECULOOS), from the Balzan Prize and Francqui Foundations, from the Belgian Scientific Research Foundation (F.R.S.-FNRS; grant n$^\circ$ T.0109.20), from the University of Liege, and from the ARC grant for Concerted Research Actions financed by the Wallonia-Brussels Federation. 
This work is supported by a grant from the Simons Foundation (PI Queloz, grant number 327127).
J.d.W. and MIT gratefully acknowledge financial support from the Heising-Simons Foundation, Dr. and Mrs. Colin Masson and Dr. Peter A. Gilman for Artemis, the first telescope of the SPECULOOS network situated in Tenerife, Spain.
This work is supported by the Swiss National Science Foundation (PP00P2-163967, PP00P2-190080 and the National Centre for Competence in Research PlanetS).
This work has received fund from the European Research Council (ERC)
 under the European Union's Horizon 2020 research and innovation programme (grant agreement n$^\circ$ 803193/BEBOP), from the MERAC foundation, and from the Science and Technology Facilities Council (STFC; grant n$^\circ$ ST/S00193X/1). 
This work is based upon observations carried out at the Observatorio Astron\'omico Nacional on the Sierra de San Pedro M\'artir (OAN-SPM), Baja California, M\'exico.
SAINT-EX observations and team were supported by the Swiss National Science Foundation (PP00P2-163967 and PP00P2-190080),
 the Centre for Space and Habitability (CSH) of the University of Bern,  the National Centre for Competence in Research PlanetS, supported by the Swiss National Science Foundation (SNSF). 
This work makes use of observations from the LCOGT network. Part of the LCOGT telescope time was granted by NOIRLab through the Mid-Scale Innovations Program (MSIP). MSIP is funded by NSF.
This work is partly supported by JSPS KAKENHI Grant Number JP21K13955, JP24H00017, JP24H00248, JP24K00689, JP24K17082 and JP24K17083, JSPS Grant-in-Aid for JSPS Fellows Grant Number JP24KJ0241, Astrobiology Center SATELLITE Research project AB022006, JSPS Bilateral Program Number JPJSBP120249910 and JST SPRING, Grant Number JPMJSP2108. This article is based on observations made with the MuSCAT2 instrument, developed by ABC, at Telescopio Carlos Sánchez operated on the island of Tenerife by the IAC in the Spanish Observatorio del Teide. This paper is based on observations made with the MuSCAT3 instrument, developed by the Astrobiology Center and under financial supports by JSPS KAKENHI (JP18H05439) and JST PRESTO (JPMJPR1775), at Faulkes Telescope North on Maui, HI, operated by the Las Cumbres Observatory.
We thank T\"urkiye National Observatories for the partial support in using T100 telescope with the project number 	22BT100-1958.
Visiting Astronomer at the Infrared Telescope Facility, which is operated by the University of Hawaii under contract 80HQTR24DA010 with the National Aeronautics and Space Administration.
L.M. acknowledges financial contribution from PRIN MUR 2022 project 2022J4H55R.
B.V.R. thanks the Heising-Simons Foundation for Support.
This material is based upon work supported by the National Aeronautics and Space Administration under Agreement No.\ 80NSSC21K0593 for the program ``Alien Earths''.
The results reported herein benefited from collaborations and/or information exchange within NASA’s Nexus for Exoplanet System Science (NExSS) research coordination network sponsored by NASA’s Science Mission Directorate.
BR-A acknowledges funding support ANID Basal project FB210003.
Support for this work was provided by NASA through the NASA Hubble Fellowship grant \#HST-HF2-51559.001-A awarded by the Space Telescope Science Institute, which is operated by the Association of Universities for Research in Astronomy, Inc., for NASA, under contract NAS5-26555.
VVG is a F.R.S.-FNRS Research Associate.
Y.G.M.C acknowledges support from UNAM PAPIIT-IG101224.
DR was supported by NASA under award number NNA16BD14C for NASA Academic Mission Services.
We acknowledge financial support from the Agencia Estatal de Investigaci\'on of the Ministerio de Ciencia e Innovaci\'on MCIN/AEI/10.13039/501100011033 and the ERDF “A way of making Europe” through project PID2021-125627OB-C32, and from the Centre of Excellence “Severo Ochoa” award to the Instituto de Astrofisica de Canarias.
M.S. acknowledges the support of the Italian National Institute of Astrophysics (INAF) through the project 'The HOT-ATMOS Project: characterizing the atmospheres of hot giant planets as a key to understand the exoplanet diversity' (1.05.01.85.04).
E. E-B. acknowledges financial support from the European Union and the State Agency of Investigation of the Spanish Ministry of Science and Innovation (MICINN) under the grant PRE2020-093107 of the Pre-Doc Program for the Training of Doctors (FPI-SO) through FSE funds.
J. S. acknowledges support from STFC under grant number ST/Y002563/1.

\section{TESS observations log for TOI-2015}
\begin{table}[h!]
 \begin{center}
 {\renewcommand{\arraystretch}{1.4}
 \resizebox{0.45\textwidth}{!}{
 \begin{tabular}{l c c c c}
 \toprule
Sector  &  Camera & CCD &  Observation date\\ 
 \hline
 24 &  1 & 4 & 2020 April 16 -- May 13 \\
 51 &  2 & 2 & 2022 April 22 -- May 18 \\
 78 &  1 & 3 & 2024 April 23 -- May 21 \\
\hline
 \end{tabular}}}
 \caption{TESS observations log for TOI-2015.}
 \label{TESS_obs_table}
 \end{center}
\end{table}

\section{Ground-based observations log for TOI-2015\,b.}
\begin{table*}[!]
\caption{Ground-based observations log for TOI-2015\,b: Telescope, size, observation date and filter.}
 \begin{center}
 {\renewcommand{\arraystretch}{0.83}
 \resizebox{0.84\textwidth}{!}{
 \begin{tabular}{l c c c c c ccc}
 \toprule
Observatory & Aperture [m] & Date (UT) & Filter &    Coverage  \\ 
 \hline
LCO-SAAO & 1.0 & 2020.07.05  & Sloan-$i'$  &  Ingress  \\
MuSCAT3 & 2.0 & 2021.06.12  & Sloan-$g',r',i'$, $z_s$  &  Full  \\
MuSCAT3 & 2.0 & 2021.07.28  & Sloan-$g',r',i'$, $z_s$  &  Full  \\
LCO-CTIO & 1.0 & 2021.08.04  & Sloan-$i'$  &  Egress  \\
MuSCAT3 & 2.0 & 2022.03.26  & Sloan-$i'$, $z_s$  &  Full  \\
MuSCAT3 & 2.0 & 2022.04.05  & Sloan-$i'$, $z_s$  &  Ingress  \\
MuSCAT & 1.88 & 2022.05.06  & Sloan-$r'$, $z_s$  &  Full  \\
SPECULOOS-North & 1.0 & 2022.05.19 & $I+z'$ & Full \\
MuSCAT2 & 1.5 & 2022.05.19  & Sloan-$g',r',i'$, $z_s$  &  Full  \\
MuSCAT3 & 2.0 & 2022.05.22  & Sloan-$r',i'$, $z_s$  &  Full  \\
SPECULOOS-North & 1.0 & 2022.05.29 & $I+z'$ & Full \\
MuSCAT2 & 1.5 & 2022.05.29  & Sloan-$g',r',i'$, $z_s$  &  Full  \\
TRAPPIST-South & 0.6 & 2022.06.09 & $I+z'$ & Full \\
SPECULOOS-North & 1.0 & 2022.06.19 & $I+z'$ & Full \\
SAINT-EX & 1.0 & 2022.06.19 & $I+z'$ & Full \\
SPECULOOS-North & 1.0 & 2022.06.25 & $I+z'$ & Egress \\
OSN & 1.5 & 2022.06.25 & $Ic$ & Egress \\
SPECULOOS-North & 1.0 & 2022.07.15 & $I+z'$ & Full \\
SPECULOOS-North & 1.0 & 2022.07.25 & $I+z'$ & Full \\
SPECULOOS-North & 1.0 & 2022.08.04 & $I+z'$ & Ingress \\
LCO-Teid & 1.0 & 2022.08.04  & Sloan-$i'$  &  Ingress  \\
SPECULOOS-North & 1.0 & 2023.04.02 & $I+z'$ & Full \\
TRAPPIST-North & 1.0 & 2023.04.02 & $I+z'$ & Full + Flip \\
SPECULOOS-North & 1.0 & 2023.04.12 & $I+z'$ & Ingress \\
LCO-CTIO & 1.0 & 2023.04.12  & Sloan-$i'$  &  Full  \\
LCO-McD & 1.0 & 2023.04.12  & Sloan-$i'$  &  Full  \\
LCO-McD & 1.0 & 2023.04.12  & Sloan-$i'$  &  Full  \\
MuSCAT3 & 2.0 & 2023.04.15  & Sloan-$g',r',i'$, $z_s$  &  Full  \\
LCO-CTIO & 1.0 & 2023.04.23  & Sloan-$i'$  &  Full  \\
LCO-SSO & 1.0 & 2023.04.26  & Sloan-$i'$  &  Ingress  \\
TRAPPIST-North & 1.0 & 2023.04.29 & $I+z'$ & Full + thin clouds  *\\ 
TRAPPIST-North & 1.0 & 2023.05.09 & $I+z'$ & Full + Flip \\
SPECULOOS-North & 1.0 & 2023.05.09 & $I+z'$ & Full \\
LCO-Teid & 1.0 & 2023.05.09  & Sloan-$i'$  &  Full  \\
IAC80 & 1.0 & 2023.05.09  & Sloan-$r'$  &  Full   *\\
LCO-HAL & 0.4 & 2023.05.12  & Sloan-$i'$  &  Full  \\
SPECULOOS-North & 1.0 & 2023.05.19 & $I+z'$ & Full \\
LCO-Teid & 1.0 & 2023.05.19  & Sloan-$i'$  &  Full  \\
SPECULOOS-North & 1.0 & 2023.05.29 & $I+z'$ & Full \\
LCO-Teid & 1.0 & 2023.05.29  & Sloan-$i'$  &  Full  \\
LCO-CTIO & 1.0 & 2023.05.29  & Sloan-$i'$  &  Full  \\
LCO-CTIO & 1.0 & 2023.06.08  & Sloan-$i'$  &  Full  \\
SPECULOOS-North & 1.0 & 2023.06.15 & $I+z'$ & Egress \\
CAHA-1.23m & 1.23 & 2023.06.15 & GG-495 & Full\\
SPECULOOS-North  & 1.0 & 2023.06.25 & $I+z'$ & Full \\
LCO-Teid & 1.0 & 2023.06.25 & $i'$ & Full \\
TUBITAK-1m0 & 1.0 & 2023.06.25 & Clear & Full \\
LCO-McD & 1.0 & 2023.06.29  & Sloan-$i'$  &  Full  \\
SPECULOOS-North & 1.0 & 2023.07.05 & $I+z'$ & Full \\
MuSCAT3 & 2.0 & 2024.03.19  & Sloan-$g', r',i'$, $z_s$  &  Ingress  \\
LCO-CTIO & 1.0 & 2024.03.26  & Sloan-$i'$  &  Full  \\
SPECULOOS-North & 1.0 & 2024.04.02 & $I+z'$ & Full \\
MuSCAT3 & 2.0 & 2024.04.05  & Sloan-$i'$, $z_s$  &  Full  \\
 \hline
 \end{tabular}}}
 \end{center}
 \tablefoot{* designs the transits that we excluded in our photodynamical analysis due to the low SNR.}
 \label{obs_table}
\end{table*}

\section{Mid-transit timings measured for TOI-2015\,b}
\begin{table*}[!]
\caption{Mid-transit timings measured for TOI-2015\,b in this work.}
\begin{center}
\label{tab3}%
\resizebox{0.8\textwidth}{!}{
{\renewcommand{\arraystretch}{0.98}
\begin{tabular}{llc}
 \toprule
 Epoch & Transit mid-time (BJD$_{TDB}$) & Telescope(s)  \\ 
 \hline
-226 & $ 2458956.03370946 \pm 0.00311000 $ &   \emph{TESS}  \\ 
-225 & $ 2458959.37366368 \pm 0.00517000 $ &   \emph{TESS}  \\ 
-224 & $ 2458962.72816271 \pm 0.00549000 $ &  \emph{TESS}   \\ 
-223 & $ 2458966.07238390 \pm 0.00337000 $ &   \emph{TESS}  \\ 
-222 & $ 2458969.41420846 \pm 0.00285000 $ &  \emph{TESS}   \\ 
-221 & $ 2458972.76329561 \pm 0.00335000 $ &  \emph{TESS}   \\ 
-220 & $ 2458976.11441643 \pm 0.00640000 $ &   \emph{TESS}  \\ 
-219 & $ 2458979.45588928 \pm 0.00344000 $ &  \emph{TESS}   \\ 
-202 & $ 2459036.39117005 \pm 0.00071500 $ &  LCOGT-1m0   \\ 
-100 & $ 2459377.94893691 \pm 0.00860149 $ &  MuSCAT3    \\ 
-86  & $ 2459424.78535186 \pm 0.00373972 $ &  MuSCAT3   \\ 
-84  & $ 2459431.48034340 \pm 0.00059700 $ &  LCOGT-1m0   \\ 
-14  & $ 2459666.06899834 \pm 0.00014800 $ &  MuSCAT3   \\ 
-11  & $ 2459676.12237684 \pm 0.00370000 $ &   MuSCAT3  \\ 
-4   & $ 2459699.53654669 \pm 0.00463000 $ &   \emph{TESS}  \\ 
-2   & $ 2459706.23337622 \pm 0.00048800 $ &  \emph{TESS} + MuSCAT   \\ 
0    & $ 2459712.92259750 \pm 0.00251000 $ &    \emph{TESS} \\ 
1    & $ 2459716.27075880 \pm 0.00178000 $ &    \emph{TESS} \\ 
2    & $ 2459719.61366655 \pm 0.00036900 $ &  SPECULOOS-North + MuSCAT2   \\ 
3    & $ 2459722.95871120 \pm 0.00014900 $ &   MuSCAT3  \\ 
5    & $ 2459729.64513527 \pm 0.00039600 $ &  SPECULOOS-North   \\ 
8    & $ 2459739.67809341 \pm 0.00063500 $ &  TRAPPIST-South   \\ 
11   & $ 2459749.70844875 \pm 0.00039500 $ &  SPECULOOS-North   \\ 
13   & $ 2459756.38869089 \pm 0.00426315 $ & SPECULOOS-North + OSN-1.5m    \\ 
16   & $ 2459766.42456143 \pm 0.00031500 $ &  SAINT-EX   \\ 
19   & $ 2459776.45123640 \pm 0.00039400 $ &  SPECULOOS-North   \\ 
22   & $ 2459786.48088223 \pm 0.00039300 $ &  SPECULOOS-North   \\ 
25   & $ 2459796.51408078 \pm 0.00298935 $ &  SPECULOOS-North + LCOGT-1m0   \\ 
97   & $ 2460037.68757200 \pm 0.00031000 $ &  SPECULOOS-North + TRAPPIST-North   \\ 
100  & $ 2460047.74237832 \pm 0.00023400 $ &  SPECULOOS-North + LCOGT-1m0   \\ 
101  & $ 2460051.09383990 \pm 0.00019800 $ &  MuSCAT3   \\ 
103  & $ 2460057.79748197 \pm 0.00040000 $ &  LCOGT-1m0   \\ 
104  & $ 2460061.14790259 \pm 0.00153000 $ &  LCOGT-1m0   \\ 
105  & $ 2460064.51864170 \pm 0.02000000 $ &   TRAPPIST-North  \\ 
108  & $ 2460074.55214637 \pm 0.00031000 $ &  SPECULOOS-North + LCOGT-1m0    \\ 
109  & $ 2460077.89791170 \pm 0.00330000 $ &   LCOGT-0.35m  \\ 
111  & $ 2460084.60438173 \pm 0.00303000 $ &  SPECULOOS-North + LCOGT-1m0   \\ 
114  & $ 2460094.65119102 \pm 0.00024000 $ &  SPECULOOS-North + LCOGT-1m0   \\ 
117  & $ 2460104.69785000 \pm 0.00040000 $ &   LCOGT-1m0  \\ 
119  & $ 2460111.39533000 \pm 0.00020000 $ &  SPECULOOS-North + CAHA-1.23m   \\ 
122  & $ 2460121.43747558 \pm 0.0002400 $ & SPECULOOS-North   \\
     &                                     & + LCOGT-1m0 + TUBITAK-1m0\\
125  & $ 2460131.47835115 \pm 0.00051600 $ & SPECULOOS-North   \\
202  & $ 2460389.14534935 \pm 0.00034517 $ &  MuSCAT3 \\
204  & $ 2460395.84795994 \pm 0.00095855 $ & LCOGT-1m0 \\
206 & $ 2460402.55465865 \pm 0.00039847 $ &    SPECULOOS-North \\
207 & $ 2460405.90637088 \pm 0.00024873 $ &  MuSCAT3 \\
216 & $2460436.083329 \pm 0.002400 $  & \emph{TESS} \\
217 & $ 2460439.440196 \pm 0.002300$  & \emph{TESS} \\
218 & $ 2460442.784055 \pm 0.002300$  & \emph{TESS} \\
220 & $ 2460449.479219 \pm 0.001500$  & \emph{TESS} \\
\hline
\end{tabular}} }
\end{center}
\label{timing_table}
\end{table*}

\section{Results for 2:1 near resonance scenario}
\begin{table*}[h!]
\caption{Physical parameters of the TOI-2015 system for the 2:1 near resonance scenario.}
	\begin{center}
		{\renewcommand{\arraystretch}{1.2}
				\resizebox{0.9\textwidth}{!}{
			\begin{tabular}{lllc}
				\hline
			        Parameter & Unit &  TOI-2015\,b  & TOI-2015\,c    \\
           \hline
            Orbital period $P$ & days             & $3.34649323^{+0.00004513}_{-0.00004738}$ & $6.698178^{+0.000195}_{-0.000131}$ \\ 
            Transit-timing $T_0$ & BJD$_{\rm TDB}$ &  $2459424.78461920^{+0.00019598}_{-0.00018066}$  &   --   \\
            Orbital semi-major axis $a$ & au      & $0.02931232 \pm 0.000003$ &  $0.046559 \pm 0.000005$   \\
            Impact parameter $b$ & $R_\star$      & $0.841^{+0.004}_{-0.004}$  &  $30.31^{+0.09}_{-0.14}$  \\
			Transit duration $W$ & hour           &  $1.118 \pm 0.007$  &  --  \\
            eccentricity $e$     & --             & $0.1777 ^{+0.0084}_{-0.0087}$   & $0.0099^{+0.0044}_{-0.0033}$ \\
            $\sqrt{e}\cos(w)$    &  --           &  $ -0.4136^{+0.0111}_{-0.0105}$  &  $0.0111^{+0.0021}_{-0.0024}$ \\
            $\sqrt{e}\sin(w)$    &  --         &  $0.0816^{+0.0087}_{-0.0092}$  &  $-0.099^{+0.0183}_{-0.0202}$ \\
            Mean anomaly  $M$    & deg          &  --  &   $6.246^{+0.022}_{-0.020}$     \\
			Orbital inclination $i$ & deg         &  $87.87 \pm 0.01$ &  $28.44^{+3.03}_{-2.00}$  \\
			Radius ratio $R_p /R_\star $       & $R_\star$ & $0.09849^{+0.00045}_{-0.00042}$ & -- \\
            Scaled semi-major axis  $a/R_\star$    &    &   $19.2758^{+0.0064}_{-0.0056}$   &   $ 30.6220 \pm 0.0102 $   \\
            Planet radius $R_p$ & $R_\oplus $    &  $3.5119^{+0.016}_{-0.015}$ &  -- \\
            Planet Mass $M_p$   & $M_\oplus$       &  $8.032^{+0.67}_{-0.44}$  &  $ 18.11^{+1.05}_{-1.08} $ \\
            Planet density $\rho_p$  & g/cm$^3$       & $  1.022^{+0.077}_{-0.064} $   &   -- \\
			Planet irradiation $S_p$  & $S_\oplus$     &  $ 13.19 \pm 0.38 $ &  $5.23 \pm 0.15$\\
        Planet surface gravity $\log (g_p) $ &   --  &   $ 2.91581 \pm 0.21 $   &   --   \\
   \hline
		\end{tabular}}}
	\end{center}
	\label{tois_mcmc_params_21}
\end{table*}

\section{Results for 5:2 near resonance scenario}
\begin{table*}[h!]
\caption{Physical parameters of the TOI-2015 system for the 5:2 near resonance scenario.}
	\begin{center}
		{\renewcommand{\arraystretch}{1.2}
				\resizebox{0.9\textwidth}{!}{
			\begin{tabular}{lllc}
				\hline
			        Parameter & Unit &  TOI-2015\,b  & TOI-2015\,c    \\
           \hline
            Orbital period $P$ & days             & $3.34800359^{+0.00005184}_{-0.00005563}$ & $8.374068^{+0.000574}_{-0.000504}$ \\ 
            Transit-timing $T_0$ & BJD$_{\rm TDB}$ &  $2459424.785375^{+0.000181}_{-0.000174}$  &   --   \\
            Orbital semi-major axis $a$ & au      & $0.029322 \pm 0.000003$ &  $0.0540318 \pm 0.000007$   \\
            Impact parameter $b$ & $R_\star$      & $0.8146^{+0.0040}_{-0.0041}$  &  $17.23^{+0.66}_{-0.67}$  \\
			Transit duration $W$ & hour           &  $1.094 \pm 0.006$  &  --  \\
            eccentricity $e$     & --             & $0.324 ^{+0.0072}_{-0.0073}$   & $0.211^{+0.006}_{-0.007}$ \\
            $\sqrt{e}\cos(w)$    &  --           &  $ -0.273^{+0.014}_{-0.014}$  &  $0.4107^{+0.009}_{-0.009}$ \\
            $\sqrt{e}\sin(w)$    &  --         &  $-0.499^{+0.011}_{-0.011}$  &  $-0.207^{+0.022}_{-0.016}$ \\
            Mean anomaly  $M$    & deg          &  --  &   $2.422^{+0.025}_{-0.022}$     \\
			Orbital inclination $i$ & deg         &  $87.715 \pm 0.016$ &  $48.54^{+1.18}_{-1.29}$  \\
			Radius ratio $R_p /R_\star $       & $R_\star$ & $0.09756^{+0.00042}_{-0.00040}$ & -- \\
            Scaled semi-major axis  $a/R_\star$ &    &   $19.2826^{+0.0062}_{-0.0060}$   &   $ 35.539 \pm 0.012 $   \\
            Planet radius $R_p$ & $R_\oplus $    &  $3.479^{+0.015}_{-0.014}$ &  -- \\
            Planet Mass $M_p$   & $M_\oplus$       &  $22.12^{+0.79}_{-0.88}$  &  $ 11.61^{+0.71}_{-0.74} $ \\
            Planet density $\rho_p$  & g/cm$^3$       & $ 2.89^{+0.11}_{-0.12} $   &   -- \\
			Planet irradiation $S_p$  & $S_\oplus$     &  $ 13.18 \pm 0.38 $ &  $3.88 \pm 0.11$\\
            Planet surface gravity $\log (g_p) $ &   --  &   $ 3.25 \pm 0.12 $   &   --   \\
   \hline
		\end{tabular}}}
	\end{center}
	\label{tois_mcmc_params_52}
\end{table*}

\newpage

\section{Model parameters and priors}

\begin{table}[!h]
\centering
\caption{\pyttv model parameters and priors.}
\label{table:pyttv_parameters}
\resizebox{0.8\textwidth}{!}{
{\renewcommand{\arraystretch}{1.1}
\begin{tabular*}{\columnwidth}{@{\extracolsep{\fill}} lll}
\toprule\toprule
Description & Units & Prior \\
\midrule     
\multicolumn{3}{l}{\emph{Stellar parameters}} \\
\midrule
Stellar mass & [$M_\odot$] & $\NP(0.33, 0.02)$ \\
Stellar radius & [$R_\odot$] & $\NP(0.339, 0.016)$ \\
Limb darkening q$_1^a$ & [-] & $\UP(0, 1)$ \\
Limb darkening q$_2^a$ & [-] & $\UP(0, 1)$ \\
\\
\multicolumn{3}{l}{\emph{Planet b parameters}} \\
\midrule
log$_{10}$ mass  & [$\log_{10}\, M_\odot$] & $\NP(-4.8, 0.1)^b$ \\
Radius ratio  & [$R_\star$] & $\NP(0.1, 0.01)$ \\
Transit centre & [BJD$_{\rm TDB}$] & $\NP(2459712.930, 0.005)$ \\
Orbital period & [d] & $\NP(3.35, 0.01)$\\
Impact parameter & [$R_\star$] & $\UP(0.5, 1.0)$ \\
$\sqrt{e} \cos\omega$ & [-]  & $\UP(-0.25, 0.25)$ \\
$\sqrt{e} \sin\omega$  & [-]  & $\UP(-0.25, 0.25)$ \\
$\Omega$ & [rad] & $\NP(\pi, 0.0001)$ \\
\\
\multicolumn{3}{l}{\emph{Planet c parameters}} \\
\midrule
log$_{10}$ mass & [$\log_{10}\, M_\odot$] & $\UP(-4.4, 0.1)^b$ \\
Radius ratio & [$R_\star$] & $\NP(0.1, 0.01)$ \\
Mean anomaly at $T_\mathrm{ref}$  & [rad] & $\UP(0, 2\pi)$ \\
Orbital period & [d] & Depends on the scenario\\
Impact parameter & [$R_\star$] & $\NP(0.0, 3.0)$ \\
$\sqrt{e} \cos\omega$ & [-]  & $\UP(-0.25, 0.25)$ \\
$\sqrt{e} \sin\omega$ & [-]  & $\UP(-0.25, 0.25)$ \\
$\Omega$  & [rad] & $\NP(\pi, 0.0001)$ \\
\\
\multicolumn{3}{l}{\emph{RV parameters}} \\
\midrule
Linear trend  & [m/s/d] & $\NP(0, 1.0)$ \\
Systemic velocity 1 & [m/s] & $\NP(-25855, 200)$ \\
Systemic velocity 2 & [m/s] & $\NP(-25416, 200)$ \\
log$_{10}$ jitter 1 & [$\log_{10}$ m/s] & $\NP(-1, 0.1)$ \\
log$_{10}$ jitter 2 & [$\log_{10}$ m/s] & $\NP(-1, 0.1)$ \\
\bottomrule
\end{tabular*}} }
\tablefoot{
    All the planetary parameters except the radius ratio and log$_{10}$ mass are defined at a reference time, $t_\mathrm{ref}=2459424.785$. \\
    \tablefoottext{a}{We use the triangular parameterization for quadratic limb darkening introduced by \citet{Kipping_2013MNRAS.435.2152K}.}
    \tablefoottext{b}{The log$_{10}$ planet mass priors are loosely based on two-planet RV analysis.}
    }
\end{table}

\section{Posterior probability distribution for TOI-2015 for the 5:3 scenario.}

\begin{figure*}[!h]
	\centering
	\includegraphics[scale=0.14]{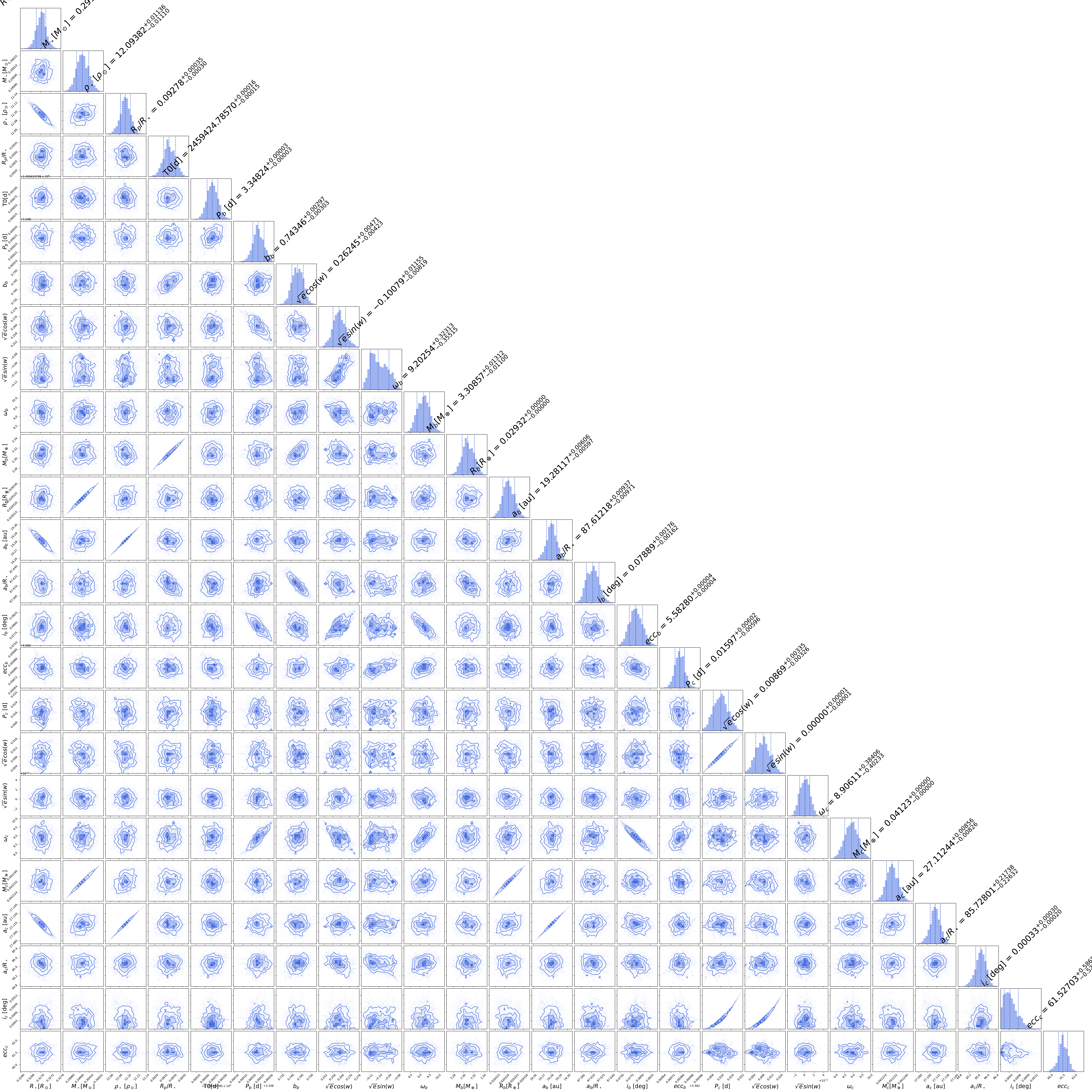}
	\caption{Corner plot of the posterior probability distribution from the PyTTV analysis for the parameters of the host star for the 5:3 scenario.}
	\label{fig:corner_TOI2015}
\end{figure*}

\end{appendix}
\end{document}

%% file: coauthors_list.tex
\newcommand{\orcidauthorA}{0000-0003-1464-9276} 
\newcommand{\orcidauthorB}{0000-0003-1572-7707}
\newcommand{\orcidauthorC}{}
\newcommand{\orcidauthorD}{0000-0002-0802-9145}
\newcommand{\orcidauthorE}{}
\newcommand{\orcidauthorF}{}
\newcommand{\orcidauthorG}{}
\newcommand{\orcidauthorH}{}
\newcommand{\orcidauthorI}{0000-0002-0149-1302}
\newcommand{\orcidauthorJ}{0000-0002-3627-1676}
\newcommand{\orcidauthorK}{0000-0003-2404-2427}
\newcommand{\orcidauthorL}{0000-0001-9204-8498}
\newcommand{\orcidauthorM}{0000-0003-1462-7739}
\newcommand{\orcidauthorN}{0000-0002-4671-2957}
\newcommand{\orcidauthorO}{0000-0001-8907-4775}
\newcommand{\orcidauthorP}{0000-0003-2196-6675}
\newcommand{\orcidauthorQ}{0009-0006-9361-9153}
\newcommand{\orcidauthorR}{0000-0003-4508-2436}
\newcommand{\orcidauthorS}{0000-0002-4262-5661}
\newcommand{\orcidauthorT}{0000-0003-0996-6402}
\newcommand{\orcidauthorU}{0000-0002-5370-7494}
\newcommand{\orcidauthorV}{0000-0003-1368-6593}
\newcommand{\orcidauthorW}{}
\newcommand{\orcidauthorY}{0000-0002-7486-6726}
\newcommand{\orcidauthorX}{}
\newcommand{\orcidauthorZ}{0000-0002-9350-830X}
\newcommand{\orcidauthora}{}
\newcommand{\orcidauthorb}{}
\newcommand{\orcidauthorc}{0000-0001-8879-7138}
\newcommand{\orcidauthord}{}
\newcommand{\orcidauthore}{}
\newcommand{\orcidauthorf}{0000-0001-9087-1245}
\newcommand{\orcidauthorg}{0000-0002-1787-3444}
\newcommand{\orcidauthorh}{}
\newcommand{\orcidauthori}{}
\newcommand{\orcidauthorj}{}
\newcommand{\orcidauthorl}{}
\newcommand{\orcidauthorm}{0000-0003-0030-332X}
\newcommand{\orcidauthorn}{}
\newcommand{\orcidauthoro}{}
\newcommand{\orcidauthorp}{}
\newcommand{\orcidauthorq}{0009-0006-9361-9153}
\newcommand{\orcidauthorr}{0009-0002-5545-3034}
\newcommand{\orcidauthors}{}
\newcommand{\orcidauthort}{}
\newcommand{\orcidauthoru}{}
\newcommand{\orcidauthorv}{}
\newcommand{\orcidauthorw}{}
\newcommand{\orcidauthory}{}
\newcommand{\orcidauthorx}{}
\newcommand{\orcidauthorz}{}
\newcommand{\orcidauthoraa}{}
\newcommand{\orcidauthorbb}{}
\newcommand{\orcidauthorcc}{}
\newcommand{\orcidauthordd}{}
\newcommand{\orcidauthoree}{}

\author{
K. Barkaoui\orcidA{}\inst{\ref{astro_liege},\ref{MIT},\ref{IAC_Laguna}}\thanks{E-mail: \color{blue}khalid.barkaoui@uliege.be}
\and J.~Korth\inst{\ref{lund}}  
\and E.~Gaidos\inst{\ref{Dep_Hawaii_USA},\ref{Univ_Vienna}} 
\and E.~Agol\orcidD{}\inst{\ref{Dept_Astro_Washington},\ref{NExSS_Washington}} 
\and H.~Parviainen\inst{\ref{Univ_LaLaguna},\ref{IAC_Laguna}}  
\and F.J.~Pozuelos\orcidB{}\inst{\ref{iaa}} 
\and E.~Palle\inst{\ref{IAC_Laguna},\ref{Univ_LaLaguna}} 
\and N.~Narita\inst{\ref{Univ_tokyo},\ref{Astro_tokyo},\ref{IAC_Laguna}} 
\and S.~Grimm\inst{\ref{Inst_Zurith_Switz},\ref{Dep_Zurith_Switz}}   
\and M.~Brady\orcidK{}\inst{\ref{Astro_Chicago}} 
\and J.~L.~Bean\inst{\ref{Astro_Chicago}} 
\and G.~Morello\orcidS{}\inst{\ref{iaa},\ref{INAF_Pa}} 
\and B.V.~Rackham\orcidJ{}\inst{\ref{MIT},\ref{Kavli_MIT}} 
\and A.~J.~Burgasser\inst{\ref{UCSDiego}} 
\and V.~Van~Grootel\inst{\ref{star_liege}} 
\and B.~Rojas-Ayala\orcidI\inst{\ref{tara}} 
\and A.~Seifahrt\inst{\ref{NOIRLab}} 
\and E.~Marfil\orcidO{}\inst{\ref{Univ_Madrid}} 
\and V.~M.~Passegger\inst{\ref{Sebaru_telescope},\ref{IAC_Laguna},\ref{Univ_LaLaguna},\ref{Ham_Germany}} 
\and M.~Stalport\orcidT{}\inst{\ref{star_liege}} 
\and M.~Gillon\orcidM{}\inst{\ref{astro_liege}}  
\and K.A.~Collins\inst{\ref{Harvard_USA}} 
\and A.~Shporer\inst{\ref{Kavli_MIT}}  
\and S.~Giacalone\inst{\ref{Dep_Astro_Califor_Inst}}  
\and S.~Yal\c{c}{\i}nkaya\inst{\ref{Dep_turkey},\ref{Univ_turkey},\ref{astro_liege}}    
\and E.~Ducrot\inst{\ref{Paris_Region},\ref{cea}} 
\and M.~Timmermans\inst{\ref{astro_liege}}  
\and A.~H.~M.~J.~Triaud\inst{\ref{birmingham}} 
\and J.~de~Wit\inst{\ref{MIT}} 
\and A.~Soubkiou\inst{\ref{ouka}} 
\and C.N.~Watkins\inst{\ref{Harvard_USA}}     
\and C.~Aganze\inst{\ref{Kavli_inst_Partic_Cosmo}} 
\and R.~Alonso\inst{\ref{IAC_Laguna},\ref{Univ_LaLaguna}} 
\and P.J.~Amado\inst{\ref{iaa}} 
\and R.~Basant\orcidR{}\inst{\ref{Astro_Chicago}} 
\and \"O.~Ba\c{s}t\"urk\inst{\ref{Dep_turkey},\ref{Univ_turkey}}  
\and Z.~Benkhaldoun\inst{\ref{ouka}} 
\and A.~Burdanov\inst{\ref{MIT}} 
\and Y.~Calatayud-Borras\inst{\ref{IAC_Laguna},\ref{Univ_LaLaguna}} 
\and J.~Chouqar\inst{\ref{ouka}} 
\and D.~M.~Conti\inst{\ref{American_Association}} 
\and K.I.~Collins\inst{\ref{George_Mason_Uni}} 
\and F.~Davoudi\orcidg{}\inst{\ref{astro_liege}} 
\and L.~Delrez\inst{{\ref{astro_liege},\ref{star_liege}}} 
\and C.D.~Dressing\inst{\ref{Dep_Astro_Califor_Univ}}  
\and J.~de~Leon\inst{\ref{Dep_Multi_Dsci_Japan}}  
\and  M.~D\'evora-Pajares\inst{\ref{ugr}} 
\and B.O.~Demory\inst{\ref{unibe}} 
\and G.~Dransfield\inst{\ref{birmingham}} 
\and E.~Esparza-Borges\inst{\ref{IAC_Laguna},\ref{Univ_LaLaguna}} 
\and G.~Fern\'andez-Rodriguez\inst{\ref{IAC_Laguna},\ref{Univ_LaLaguna}} 
\and I.~Fukuda\inst{\ref{Dep_Multi_Dsci_Japan}} 
\and A.~Fukui\inst{\ref{Univ_tokyo},\ref{IAC_Laguna}}    
\and P.P.M.~Gallardo\inst{\ref{IAC_Laguna},\ref{Univ_LaLaguna}} 
\and L.~Garcia\inst{\ref{Flatiron_instit},\ref{astro_liege}} 
\and N.A.~Garcia\inst{\ref{IAC_Laguna},\ref{Univ_LaLaguna}} 
\and M.~Ghachoui\inst{\ref{ouka},\ref{astro_liege}}    
\and S.~Gerald\'ia-González\orcidr{}\inst{\ref{IAC_Laguna},\ref{Univ_LaLaguna}} 
\and Y.~G\'omez~Maqueo~Chew\orcidY{}\inst{\ref{ciudad}} 
\and J.~Gonz\'alez-Rodr\'iguez\inst{\ref{IAC_Laguna},\ref{Univ_LaLaguna}} 
\and M.N.~G\"unther\inst{\ref{estec}} 
\and Y.~Hayashi\inst{\ref{Dep_Multi_Dsci_Japan}}      
\and K.~Horne\inst{\ref{SUPA_Physics_UK}}   
\and M.J.~Hooton\orcidm{}\inst{\ref{Cavendish}} 
\and C.C.~Hsu\orcidU{}\inst{\ref{CIERA}} 
\and K.~Ikuta\inst{\ref{Dep_Multi_Dsci_Japan}} 
\and K.~Isogai\inst{\ref{Okayama_Japan},\ref{Dep_Multi_Dsci_Japan}} 
\and E.~Jehin\inst{\ref{star_liege}} 
\and J.M.~Jenkins\inst{\ref{Ames_NASA}} 
\and K.~Kawauchi\inst{\ref{Ritsumeikan_Japan}}  
\and T.~Kagetani\inst{\ref{Dep_Multi_Dsci_Japan}}   
\and Y.~Kawai\inst{\ref{Dep_Multi_Dsci_Japan}}   
\and D.~Kasper\inst{\ref{Astro_Chicago}} 
\and J.F.~Kielkopf\inst{\ref{Dept_Louisville}} 
\and P.~Klagyivik\inst{\ref{Univ_Berlin}}  
\and G.~Lacedelli\inst{\ref{IAC_Laguna}}  
\and D.W.~Latham\inst{\ref{Harvard_USA}}  
\and F.~Libotte\inst{\ref{IAC_Laguna},\ref{Sabadell_Spain},\ref{Europlanet_Society_Belgium}} 
\and R.~Luque\orcidN{}\inst{\ref{Astro_Chicago},\ref{NHFP_Sagan_Fel}} 
\and J.H.~Livingston\inst{\ref{Astro_tokyo},\ref{National_Astro_Japan},\ref{Astro_Scien_Program_Japan}}  
\and L.~Mancini\inst{\ref{Univ_Rome},\ref{Univ_Turin},\ref{Max_planck_inst}} 
\and B.~Massey\orcidc{}\inst{\ref{Villa_Obs}}    
\and M.~Mori\orcidV\inst{\ref{Astro_tokyo},\ref{National_Astro_Japan}}     
\and S.~Muñoz~Torres\inst{\ref{IAC_Laguna},\ref{Univ_LaLaguna}} 
\and F.~Murgas\orcidf{}\inst{\ref{IAC_Laguna},\ref{Univ_LaLaguna}} 
\and P.~Niraula\inst{\ref{MIT}} 
\and J.~Orell-Miquel\inst{\ref{IAC_Laguna},\ref{Univ_LaLaguna}} 
\and David Rapetti\orcidP{}\inst{\ref{Ames_NASA},\ref{Research_inst_Washington}} 
\and R.~Rebolo-L\'opez\inst{\ref{IAC_Laguna},\ref{Univ_LaLaguna}} 
\and G.~Ricker\inst{\ref{Kavli_MIT}} 
\and R.~Papini\orcidQ{}\inst{\ref{Wild_Boar_Italy}} 
\and P.P.~Pedersen\inst{\ref{Cavendish},\ref{ETH_Zur_Queloz}} 
\and A.~Peláez-Torres\orcidL{}\inst{\ref{iaa}} 
\and J.A.~Pérez-Prieto\inst{\ref{IAC_Laguna}} 
\and E.~Poultourtzidis\inst{\ref{IAC_Laguna},\ref{Dept_Physic_Greece}} 
\and P.M.~Rodriguez\inst{\ref{IAC_Laguna},\ref{Univ_LaLaguna}} 
\and D.~Queloz\inst{\ref{Cavendish},\ref{ETH_Zur_Queloz}} 
\and A.B.~Savel\inst{\ref{Univ_Maryland}}  
\and N.~Schanche\inst{\ref{unibe}} 
\and M.~S\'anchez-Benavente\inst{\ref{IAC_Laguna},\ref{Univ_LaLaguna}} 
\and L.~Sibbald\inst{\ref{Citi_Canada}}     
\and R.~Sefako\inst{\ref{SAAO}}     
\and S.~Sohy\inst{\ref{star_liege}} 
\and A.~Sota\inst{\ref{iaa}}  
\and R.P.~Schwarz\inst{\ref{Harvard_USA}} 
\and S.~Seager\inst{\ref{UCSDiego},\ref{Univ_LaLaguna},\ref{Univ_Maryland}} 
\and D.~Sebastian\inst{\ref{birmingham}} 
\and J.~Southworth\inst{\ref{Univ_Keele}}  
\and M.~Stangret\inst{\ref{INAF_italy}}   
\and G.~Stef\'ansson\inst{\ref{Pannekoek_Insti_Nether}} 
\and J.~St{\"u}rmer\inst{\ref{Landessternwarte}} 
\and  G.~Srdoc\inst{\ref{Kotiza_Obs}}  
\and S.J.~Thompson\inst{\ref{Cavendish}} 
\and Y.~Terada\inst{\ref{Insti_Taiwan},\ref{Dep_Taiwan}}   
\and R.~Vanderspek\inst{\ref{Kavli_MIT}} 
\and G.~Wang\inst{\ref{Joh_Hopkins_univ}}     
\and N.~Watanabe\inst{\ref{Dep_Multi_Dsci_Japan}}   
\and F.P.~Wilkin\inst{\ref{Union_Sche}}   
\and J.~Winn\inst{\ref{Astro_Prin}}   
\and R.D.~Wells\inst{\ref{unibe}} 
\and C.~Ziegler\inst{\ref{Dep_Eng_Ast_Nacogdoches}} 
\and S.~Z\'u\~niga-Fern\'andez\orcidZ{}\inst{\ref{astro_liege}} 
}
	
\institute{
Astrobiology Research Unit, Universit\'e de Li\`ege, All\'ee du 6 Ao\^ut 19C, B-4000 Li\`ege, Belgium \label{astro_liege}
\and Department of Earth, Atmospheric and Planetary Science, Massachusetts Institute of Technology, 77 Massachusetts Avenue, Cambridge, MA 02139, USA \label{MIT}
\and Instituto de Astrof\'isica de Canarias (IAC), Calle V\'ia L\'actea s/n, 38200, La Laguna, Tenerife, Spain \label{IAC_Laguna} 
\and Lund Observatory, Division of Astrophysics, Department of Physics, Lund University, Box 118, 22100 Lund, Sweden \label{lund}
\and Department of Earth Sciences, University of Hawaii at Manoa, 1680 East-West Rd, Honolulu, HI 96822, USA \label{Dep_Hawaii_USA}
\and Institute for Astrophysics, University of Vienna, T\"{u}rkenschanzstrasse 17, A-1180 Vienna, Austria \label{Univ_Vienna}
\and Department of Astronomy and Astrobiology Program, University of Washington, Box 351580, Seattle, Washington 98195, USA \label{Dept_Astro_Washington}
\and NExSS Virtual Planetary Laboratory, Box 351580, University of Washington, Seattle, Washington 98195, USA \label{NExSS_Washington}
\and Departamento de Astrof\'isica, Universidad de La Laguna (ULL), E-38206 La Laguna, Tenerife, Spain \label{Univ_LaLaguna}
\and Instituto de Astrof\'isica de Andaluc\'ia (IAA-CSIC), Glorieta de la Astronom\'ia s/n, 18008 Granada, Spain \label{iaa}
\and INAF- Palermo Astronomical Observatory, Piazza del Parlamento, 1, 90134 Palermo, Italy \label{INAF_Pa}
\and Komaba Institute for Science, The University of Tokyo, 3-8-1 Komaba, Meguro, Tokyo 153-8902, Japan \label{Univ_tokyo}
\and Astrobiology Center, 2-21-1 Osawa, Mitaka, Tokyo 181-8588, Japan \label{Astro_tokyo}
\and Research Institute for Advanced Computer Science, Universities Space Research Association, Washington, DC 20024, USA \label{Research_inst_Washington}
\and Institute for Particle Physics and Astrophysics , ETH Z\"urich, OttoStern-Weg 5, 8093 Z\"urich, Switzerland \label{Inst_Zurith_Switz}
\and Department of Astrophysics, University of Z\"urich, Winterthurerstrasse 190 8057 Z\"rich, Switzerland  \label{Dep_Zurith_Switz}
\and Department of Astronomy \& Astrophysics, University of Chicago, Chicago, IL, USA \label{Astro_Chicago}
\and NHFP Sagan Fellow \label{NHFP_Sagan_Fel}
\and Department of Physics and Kavli Institute for Astrophysics and Space Research, Massachusetts Institute of Technology, Cambridge, MA 02139, USA \label{Kavli_MIT}
\and Space Sciences, Technologies and Astrophysics Research (STAR) Institute, Universit\'e de Li\`ege, All\'ee du 6 Ao\^ut 19C, B-4000 Li\`ege, Belgium \label{star_liege}
\and Instituto de Alta Investigaci\'on, Universidad de Tarapac\'a, Casilla 7D, Arica, Chile \label{tara}
\and Gemini Observatory/NSF NOIRLab, 670 N. A'ohoku Place, Hilo, HI 96720, USA \label{NOIRLab}
\and Departamento de Ingenier\'{i}a Topogr\'{a}fica y Cartograf\'{i}a, E.T.S.I. en Topograf\'{i}a, Geodesia y Cartograf\'{i}a, Universidad Polit\'{e}cnica de Madrid, 28031 Madrid, Spain \label{Univ_Madrid}
\and Subaru Telescope, National Astronomical Observatory of Japan, 650 North A‘ohoku Place, Hilo, HI 96720, USA \label{Sebaru_telescope}
\and Hamburger Sternwarte, Gojenbergsweg 112, D-21029 Hamburg, Germany \label{Ham_Germany}
\and Oukaimeden Observatory, High Energy Physics and Astrophysics Laboratory, Faculty of sciences Semlalia, Cadi Ayyad University, Marrakech, Morocco \label{ouka}
\and Center for Astrophysics \textbar \ Harvard \& Smithsonian, 60 Garden Street, Cambridge, MA 02138, USA \label{Harvard_USA}
\and Kavli Institute for Particle Astrophysics \& Cosmology, Stanford University, Stanford, CA 94305, USA \label{Kavli_inst_Partic_Cosmo}
\and Department of Astronomy, California Institute of Technology, Pasadena, CA 91125, USA \label{Dep_Astro_Califor_Inst}
\and Department of Astronomy \& Space Sciences, Faculty of Science, Ankara University, TR-06100, Ankara, T\"urkiye \label{Dep_turkey}
\and Ankara University, Astronomy and Space Sciences Research and Application Center (Kreiken Observatory), Incek Blvd., TR-06837, Ahlatlıbel, Ankara, T\"urkiye \label{Univ_turkey}
\and Paris Region Fellow, Marie Sklodowska-Curie Action \label{Paris_Region}
\and AIM, CEA, CNRS, Universit\'e Paris-Saclay, Universit\'e de Paris, F-91191 Gif-sur-Yvette, France \label{cea}
\and School of Physics \& Astronomy, University of Birmingham, Edgbaston, Birmingham B15 2TT, UK \label{birmingham}
\and Center for Astrophysics and Space Sciences, UC San Diego, UCSD Mail Code 0424, 9500 Gilman Drive, La Jolla, CA 92093-0424, USA \label{UCSDiego}
\and American Association of Variable Star Observers, 185 Alewife Brook Parkway, Suite 410, Cambridge, MA 02138, USA \label{American_Association}
\and George Mason University, 4400 University Drive, Fairfax, VA, 22030 USA \label{George_Mason_Uni}
\and Department of Astronomy, University of California Berkeley, Berkeley, CA 94720, USA \label{Dep_Astro_Califor_Univ}
\and Department of Multi-Disciplinary Sciences, Graduate School of Arts and Sciences, The University of Tokyo, 3-8-1 Komaba, Meguro, Tokyo 153-8902, Japan \label{Dep_Multi_Dsci_Japan}
\and Dpto. F\'isica Te\'orica y del Cosmos. Universidad de Granada. 18071. Granada, Spain \label{ugr}
\and Center for Space and Habitability, University of Bern, Gesellschaftsstrasse 6, 3012, Bern, Switzerland \label{unibe}
\and Center for Computational Astrophysics, Flatiron Institute, 162 Fifth Avenue, New York, New York 10010, USA \label{Flatiron_instit}
\and Universidad Nacional Aut\'onoma de M\'exico, Instituto de Astronom\'ia, AP 70-264, Ciudad de M\'exico,  04510, M\'exico \label{ciudad}
\and European Space Agency (ESA), European Space Research and Technology Centre (ESTEC), Keplerlaan 1, 2201 AZ Noordwijk, The Netherlands \label{estec}
\and SUPA Physics and Astronomy, University of St. Andrews, Fife, KY16 9SS Scotland, UK \label{SUPA_Physics_UK}
\and Cavendish Laboratory, JJ Thomson Avenue, Cambridge CB3 0HE, UK \label{Cavendish}
\and Center for Interdisciplinary Exploration and Research in Astrophysics (CIERA), Northwestern University,
1800 Sherman, Evanston, IL 60201, USA \label{CIERA}
\and Okayama Observatory, Kyoto University, 3037-5 Honjo, Kamogatacho, Asakuchi, Okayama 719-0232, Japan \label{Okayama_Japan}
\and NASA Ames Research Center, Moffett Field, CA 94035, USA \label{Ames_NASA}
\and Department of Physical Sciences, Ritsumeikan University, Kusatsu, Shiga 525-8577, Japan \label{Ritsumeikan_Japan}
\and Department of Physics and Astronomy, University of Louisville, Louisville, KY 40292, USA \label{Dept_Louisville}
\and Freie Universit\"at Berlin, Institute of Geological Sciences, Malteserstr. 74-100, 12249 Berlin, Germany \label{Univ_Berlin}
\and Sabadell Astronomical Society, 08206 Sabadell, Barcelona, Spain \label{Sabadell_Spain}
\and Europlanet Society, Department of Planetary Atmospheres of the Royal Belgian Institute for Space Aeronomy, B-1180 Brussels, Belgium \label{Europlanet_Society_Belgium}
\and National Astronomical Observatory of Japan, 2-21-1 Osawa, Mitaka, Tokyo 181-8588, Japan \label{National_Astro_Japan}
\and Astronomical Science Program, Graduate University for Advanced Studies, SOKENDAI, 2-21-1, Osawa, Mitaka, Tokyo, 181-8588, Japan \label{Astro_Scien_Program_Japan}
\and Department of Physics, University of Rome ``Tor Vergata'', Via della Ricerca Scientifica 1, I-00133, Rome, Italy \label{Univ_Rome}
\and INAF -- Astrophysical Observatory of Turin, via Osservatorio 20, I-10025, Pino Torinese, Italy  \label{Univ_Turin}
\and Max Planck Institute for Astronomy, K\"{o}nigstuhl 17, D-69117, Heidelberg, Germany \label{Max_planck_inst}
\and Villa '39 Observatory, Landers, CA 92285, USA \label{Villa_Obs}
\and Wild Boar Remote Observatory, San Casciano in val di Pesa, Firenze, 50026 Italy \label{Wild_Boar_Italy}
\and Institute for Particle Physics and Astrophysics , ETH Z\"urich, Wolfgang-Pauli-Strasse 2, 8093 Z\"urich, Switzeland \label{ETH_Zur_Queloz}
\and  Department of Physics, Aristotle University of Thessaloniki, University Campus, Thessaloniki, 54124, Greece \label{Dept_Physic_Greece}
\and Department of Astronomy, University of Maryland, College Park, College Park, MD 20742 USA \label{Univ_Maryland}
\and 7 Skies Observatory, Cypress County, Alberta, RASC (Royal Astronomical Society of Canada)\label{Citi_Canada}
\and South African Astronomical Observatory, P.O. Box 9, Observatory, Cape Town 7935, South Africa \label{SAAO}
\and Astrophysics Group, Keele University, Staffordshire, ST5 5BG, UK \label{Univ_Keele}
\and INAF -- Osservatorio Astronomico di Padova, Vicolo dell'Osservatorio 5, 35122, Padova, Italy \label{INAF_italy}
\and Anton Pannekoek Institute for Astronomy, University of Amsterdam, Science Park 904, 1098 XH Amsterdam, The Netherlands \label{Pannekoek_Insti_Nether}
\and Landessternwarte, Zentrum f{\"u}r Astronomie der Universit\"{a}t Heidelberg, K{\"o}nigstuhl 12, D-69117 Heidelberg, Germany \label{Landessternwarte}
\and Kotizarovci Observatory, Sarsoni 90, 51216 Viskovo, Croatia \label{Kotiza_Obs}
\and Institute of Astronomy and Astrophysics, Academia Sinica, P.O. Box 23-141, Taipei 10617, Taiwan, R.O.C. \label{Insti_Taiwan}
\and Department of Astrophysics, National Taiwan University, Taipei 10617, Taiwan, R.O.C. \label{Dep_Taiwan}
\and Department of Physics \& Astronomy, Johns Hopkins University, 3400 N. Charles Street, Baltimore, MD 21218, USA \label{Joh_Hopkins_univ}
\and Department of Physics and Astronomy, Union College, 807 Union St., Schenectady, NY 12308, USA \label{Union_Sche}
\and Department of Astrophysical Sciences, Princeton University, Princeton, NJ 08544, USA \label{Astro_Prin}
\and Department of Physics, Engineering and Astronomy, Stephen F. Austin State University, 1936 North St, Nacogdoches, TX 75962, USA \label{Dep_Eng_Ast_Nacogdoches} 
}